\documentclass[12pt]{article}
\pdfoutput=1
\usepackage{epsf}
\usepackage{amsmath,amssymb}

\usepackage{amssymb,amsmath,color}
\usepackage{mathtools,eqparbox,MnSymbol}

\usepackage{graphicx}

\usepackage{comment}
\usepackage{cite}

\allowdisplaybreaks

\begin{document}

\newcommand{\lsim}{\stackrel{<}{_\sim}}
\newcommand{\gsim}{\stackrel{>}{_\sim}}
\newcommand{\tr}{{\rm Tr\,}}
\newcommand{\Det}{{\mbox{Det}\,}}

\newcommand{\etaIII}{\eta_3}
\newcommand{\lambdaIII}{\lambda_3}

\mathchardef\mhyphen="2D

\newcommand\coolunder[3]{\smash{\underbrace{
\begin{matrix} \eqmakebox[#1]{$#3$} \end{matrix}
}_{\eqmakebox[#1]{$#2$}}}}

\newcommand\coolright[1]{
\left.\vphantom{\begin{matrix}#1\end{matrix}}\right\}#1}

\newcommand{\rem}[1]{{$\spadesuit$\bf #1$\spadesuit$}}

\renewcommand{\theequation}{\thesection.\arabic{equation}}

\renewcommand{\thefootnote}{\fnsymbol{footnote}}
\setcounter{footnote}{0}

\begin{titlepage}

\begin{center}

\hfill KEK-TH-2244\\
\hfill July, 2020\\

\vskip .75in

{\Large\bf
  Precise Calculation of the Decay Rate of False Vacuum \\[2mm]
  with Multi-Field Bounce
}

\vskip .5in

{\large
  So Chigusa$^{(a)}$, Takeo Moroi$^{(b)}$ and Yutaro Shoji$^{(c)}$
}

\vskip 0.25in

$^{\rm (a)}${\em
  KEK Theory Center, IPNS, KEK, Tsukuba, Ibaraki 305-0801, Japan}

\vskip 0.1in
$^{\rm (b)}${\em
  Department of Physics, The University of Tokyo, Tokyo 113-0033, Japan}

\vskip 0.1in
$^{\rm (c)}${\em
  Kobayashi-Maskawa Institute for the Origin of Particles and the Universe,
  Nagoya University, Nagoya, Aichi 464-8602, Japan}

\end{center}

\vskip .5in

\begin{abstract}

  We study the decay rate of a false vacuum in gauge theory at the
  one-loop level.  We pay particular attention to the case where the
  bounce consists of an arbitrary number of scalar fields.  With a
  multi-field bounce, which has a curved trajectory in the field
  space, the mixing among the gauge fields and the scalar fields
  evolves along the path of the bounce in the field space and the
  one-loop calculation of the vacuum decay rate becomes complicated.
  We consider the one-loop contribution to the decay rate with an
  arbitrary choice of the gauge parameter, and obtain a gauge
  invariant expression of the vacuum decay rate.  We also give
  proper treatments of gauge zero modes and renormalization.

\end{abstract}

\end{titlepage}

\setcounter{page}{1}
\renewcommand{\thefootnote}{\#\arabic{footnote}}
\setcounter{footnote}{0}

\section{Introduction}
\label{sec_intro}
\setcounter{equation}{0}

The decay of a false vacuum has attracted theoretical and
phenomenological interests in particle physics and cosmology. For
example, in the standard model (SM) and models beyond the SM, there
may exist a vacuum whose energy density is lower than that of the
electroweak (EW) vacuum. If this is the case, the EW vacuum becomes a
false vacuum and is not absolutely stable. Thus, the longevity of the
EW vacuum often provides an important constraint on model parameters.
In particular, assuming that the standard model is valid up to the
Planck scale, the EW vacuum decays within a timescale shorter than the
present cosmic age if the top-quark mass is too large or the Higgs
mass is too small \cite{Isidori:2001bm,Arnold:1991cv,Espinosa:1995se,
  ArkaniHamed:2008ym,
  EliasMiro:2011aa,Plascencia:2015pga,DiLuzio:2015iua,
  Espinosa:2016nld, Lalak:2016zlv, Andreassen:2017rzq,
  Chigusa:2017dux,Chigusa:2018uuj}.\footnote
{For discussion of the absolute stability of the EW vacuum in the SM,
  see \cite{Cabibbo:1979ay, Hung:1979dn, Lindner:1988ww, Ford:1992mv,
    Casas:1994qy, Casas:1996aq, Einhorn:2007rv, Ellis:2009tp,
    Degrassi:2012ry, Alekhin:2012py, Bezrukov:2012sa,
    Andreassen:2014gha, DiLuzio:2014bua, Bednyakov:2015sca}.}
In addition, the decay of the false vacuum is also important for the
studies of phase transitions in cosmological history, which may be
related to inflation or the baryon asymmetry of the Universe.  Thus,
the precise calculation of the decay rate of the false vacuum is of
great importance.

The calculation of a vacuum decay rate has been formulated in
\cite{Coleman:1977py,Callan:1977pt}, where the field configuration
called the bounce plays a central role.  The bounce is a saddle-point
solution of the Euclidean equation of motion, which dominates the path
integral for the decay process of the false vacuum.  With the bounce
configuration being given, the decay rate of a vacuum in unit volume
is expressed as
\begin{align}
  \gamma=\mathcal A e^{-\mathcal B},
\end{align}
where $\mathcal B$ is the action of the bounce and $\mathcal A$
contains the effects of the quantum corrections to the action. At the
one-loop level, $\mathcal A$ is obtained by evaluating the functional
determinants of the fluctuation operators around the false vacuum and
those around the bounce.  For the precise determination of a vacuum decay
rate, the calculation of $\mathcal A$ is necessary not only because it
fixes the overall factor but also because it cancels out the
renormalization scale dependence of $\mathcal B$ at the one-loop level
\cite{Endo:2015ixx}.\footnote
{The numerical impact of the prefactor is demonstrated in \cite{Endo:2015ixx,Chigusa:2018uuj,Oda:2019njo}.
}

If scalar fields responsible for the bounce couple to the gauge
fields, the fluctuation operator generally depends on the gauge-fixing
parameter (which we call $\xi$) and $\xi$ appears everywhere in the
calculation of the prefactor $\mathcal{A}$.  On the other hand, the
decay rate of the false vacuum should be independent of $\xi$ because
the effective action is gauge independent at its extrema
\cite{Nielsen:1975fs,Fukuda:1975di}. An explicit check of the gauge
invariance at the one-loop level is quite formidable and the
first calculation appeared only recently in \cite{Endo:2017gal,
Endo:2017tsz}. In these papers, a manifestly gauge-invariant
expression of the decay rate has been obtained for the case where the
bounce consists of a single field (single-field bounce).  They also
address another issue that arises when a gauge symmetry preserved in
the false vacuum is broken by the bounce configuration.  In such a
case, there appears a flat direction of the action corresponding to
the global part of the gauge symmetry; it can be seen as a gauge zero
mode in the calculation of the functional determinant. Since the
fluctuation toward such a flat direction cannot be treated with the
saddle point method in the path integral, we need special treatment; a
correct prescription for the gauge zero mode has been developed for
the single-field bounce \cite{Endo:2017tsz}.  The prescriptions to
calculate the decay rate of false vacuum given in \cite{Endo:2017gal,
Endo:2017tsz} are essential to perform a complete one-loop calculation
of the decay rate of the EW vacuum in the SM
\cite{Andreassen:2017rzq,Chigusa:2017dux,Chigusa:2018uuj}, which
update the previous result \cite{Isidori:2001bm}.\footnote
{When the true vacuum appears as a result of the renormalization group
  improvement of the potential, a careful treatment is required to
  avoid the double-counting of quantum corrections.  If the properties
  of the bounce are well described by the (renormalizable) Lagrangian
  with choosing relevant renormalization scale, however, the
  functional-determinant method can give the one-loop contribution to
  the prefactor without the double counting; examples include the
  standard model (see
  \cite{Andreassen:2017rzq,Chigusa:2017dux,Chigusa:2018uuj}).}
In addition, they are also applied to models beyond the SM
\cite{Chigusa:2018uuj,Oda:2019njo}.\footnote
{For other studies about the stability of the electroweak vacuum in
  models beyond the SM, see, for example, \cite{Branchina:2018qlf,
    Krauss:2018thf, Basler:2017nzu, Chakrabarty:2016smc,
    Bagnaschi:2015pwa, Chowdhury:2015yja, Ferreira:2015rha,
    Das:2015mwa, Chakrabarty:2014aya, EliasMiro:2012ay,
    Lebedev:2012zw, Pruna:2013bma, Costa:2014qga, Branchina:2013jra,
    Branchina:2014usa, Branchina:2014rva, Branchina:2016bws, Salvio:2016mvj, Joti:2017fwe}.}

In this paper, we extend the results of \cite{Endo:2017gal,Endo:2017tsz}
to the case where the bounce consists of more than one field
(multi-field bounce).  We give a prescription to obtain a
gauge-invariant expression of the vacuum decay rate, adopting two
different gauge-fixing conditions; one is the Fermi gauge and the other
is the background gauge. In the Fermi gauge, the treatment of the gauge
zero mode can be understood easily, but the numerical calculation
becomes difficult due to a severe cancellation.  On the other hand, in
the background gauge, the treatment of the gauge zero mode is
complicated, but the numerical calculation becomes easier because of
better behavior of fluctuation operators.  Thus,
we give a prescription to convert the result in the background gauge to
that in the Fermi gauge which is guaranteed to be gauge invariant.

This paper is organized as follows. In Section~\ref{sec_setup}, we
explain our basic setup for the calculation of the vacuum decay rate.
We show that the vacuum decay rate (in particular, the prefactor $\mathcal
A$) can be expressed by using solutions of a set of differential
equations.  In Section~\ref{sec_solutions}, we provide a decomposition
of the solutions.  In Section~\ref{sec_determinant}, we construct a
set of solutions and calculate their asymptotic behavior, which is
needed for the evaluation of $\mathcal A$.  In
Section~\ref{sec_zeromodes}, we provide a general method to treat the
zero modes and apply it to the zero modes in association with the
gauge and the translational symmetries.  In Section~\ref{sec_results}, we
summarize the analytic results.  In Section~\ref{sec_renormalization},
issues related to the renormalization are discussed.
Section~\ref{sec_conclusions} is devoted to conclusions and
discussion.

\section{Setup and Formulation}
\label{sec_setup}
\setcounter{equation}{0}

\subsection{Lagrangian and bounce}

We consider a Euclidean four-dimensional gauge theory with a
direct-product gauge group, $\mathfrak G$. We concentrate on the
contributions from scalar fields and gauge fields; the effects of
fermions, if they exist, can be taken into account separately.  The
Lagrangian is given by
\begin{eqnarray}
  \mathcal L_E&=&\frac{1}{4}F^a_{\mu\nu}F^a_{\mu\nu}+
  \frac{1}{2}(D_\mu \phi)_i(D_\mu \phi)_i+V(\phi)+
  \mathcal L^{\rm (GF)}+\mathcal L^{\rm (ghost)},
\end{eqnarray}
where $V$ is a scalar potential, while $\mathcal L^{\rm (GF)}$ and
$\mathcal L^{\rm (ghost)}$ include the gauge fixing terms and the
terms containing the ghosts, respectively, which will be defined
later.  In addition, $F^a_{\mu\nu}$ is the field strength of the gauge
field $A_\mu^a$ (with $a$ being the adjoint index of $\mathfrak G$),
while $\phi_i$ (with $i=1-n_\phi$ being the index distinguishing
scalars) are real scalar fields.\footnote
{In our convention, complex scalar fields are understood to be
  decomposed into pairs of real scalar fields.}
Here and hereafter, the summation over the repeated scalar indices is
implicit.  The covariant derivative of $\phi$ is defined as
\begin{align}
 D_\mu \phi=(\partial_\mu +g_a A_\mu^a T^a)\phi.
\end{align}
In the real basis we are working with, the generators satisfy
\begin{align}
 (T^a)^T&=-T^a,\\
 [T^a,T^b]&=-f^{abc}T^c,
\end{align}
where the superscript ``$T$'' denotes transpose and $f^{abc}$ is the
structure constant.  The gauge coupling constants, denoted as $g_a$,
can be different for different subgroups of $\mathfrak G$.  We assume
that the scalar potential, $V(\phi)$, has two minima, {\it i.e.}, the
false vacuum and the true vacuum (the false vacuum has higher
potential energy than the true vacuum).

For the process of the false vacuum decay, the path integral is
dominated by the field configuration called ``bounce,'' an $O(4)$
symmetric saddle-point solution of the Euclidean equations of motion
\cite{Coleman:1977th,Blum:2016ipp}.  Due to the gauge symmetry, there
exist an infinite number of solutions. We adopt one solution with
$A_\mu=0$. Then, the bounce configuration, which we denote as
$\bar\phi(r)$, satisfies
\begin{align}
  \partial_r^2\bar\phi_i+\frac{3}{r}\partial_r\bar\phi_i
  =
  \left. \frac{\partial V}{\partial \phi_i} \right|_{\phi\rightarrow\bar\phi},
  \label{eq_bounce_eom}
\end{align}
with the following boundary conditions:
\begin{align}
  \partial_r\bar\phi_i(0)=0,\label{eq_boundary}
  \\
  \bar\phi_i(\infty)=v_i,
\end{align}
where $v_i$ denotes the scalar amplitude at the false vacuum.  Here,
$r \equiv \sqrt{x_\mu x_\mu}$ is the radius from the center of the
bounce, and $\partial_r$ denotes the derivative with respect to $r$.

The bounce solution has the following properties.  First, since the
potential is symmetric under the infinitesimal gauge transformation
$\phi\to\phi+\delta\theta^aT^a\phi$ with an arbitrary choice of
transformation parameters $\delta\theta^a$,
\begin{equation}
  \frac{\partial V}{\partial\phi_i}T^a_{ij}\phi_j =0.
\end{equation}
Differentiating it with respect to $\phi_k$, we obtain
\begin{equation}
 T^a_{ki}\frac{\partial V}{\partial\phi_i}=\frac{\partial^2 V}{\partial\phi_k\partial\phi_i}T^a_{ij}\phi_j.
\end{equation}
Then, from Eq.~\eqref{eq_bounce_eom},
\begin{equation}
 \partial_r^2(T^a\bar\phi)+\frac{3}{r}\partial_r(T^a\bar\phi)=\Omega (T^a\bar\phi),\label{eq_eom_Tphi}
\end{equation}
where
\begin{align}
  \Omega_{ij} =
  \left. \frac{\partial^2V}{\partial\phi_i\partial\phi_j} \right|_{\phi\rightarrow\bar\phi}.
\end{align}
From Eq.~\eqref{eq_eom_Tphi}, we obtain
\begin{equation}
  \frac{1}{r^3}\partial_rr^3[(\partial_r\bar\phi)^T T^a\bar\phi]
  =\frac{1}{2}\bar\phi^T \left[T^a,\Omega\right]\bar\phi=0,
\end{equation}
because $\Omega$ is gauge invariant. It implies that
$(\partial_r\bar\phi)^T T^a\bar\phi$ should be proportional to $r^{-3}$
or zero.  Since it should vanish at the origin,
\begin{align}
  (\partial_r\bar\phi)^T T^a\bar\phi=0.
  \label{eq_bounce_traj}
\end{align}

For the later convenience, we define
\begin{align}
  M_{ia} (r) = - g_aT^a_{ik}\bar\phi_k (r),
\end{align}
which satisfies
\begin{align}
  \left( \partial_r^2 + \frac{3}{r} \partial_r - \Omega \right) M = 0.
  \label{eq_M_eom}
\end{align}
Using Eq.~\eqref{eq_bounce_traj}, the following relation holds:
\begin{align}
  M^TM' = (M^T)'M,
  \label{eq_M_deriv}
\end{align}
where $M'=\partial_r M$.  Notice that the gauge boson mass matrix in
the false vacuum is given by $M^TM(r\rightarrow\infty)$.

In our analysis, we concentrate on the case where the following
conditions hold:
\begin{itemize}
\item The rank of the matrix $M^TM(r)$ is unchanged for $r<\infty$.
  (At the false vacuum, some of the broken gauge symmetries may be
  recovered so that the ranks of $M^TM(r<\infty)$ and $M^TM(\infty)$
  may be different.)
\item There are no zero modes except for the gauge zero modes and the
  translational zero modes.
\item At a large $r$, $M^TM(r)$ approaches to $M^TM(\infty)$ exponentially.
\end{itemize}
The second condition is just for simplicity and our results can be
extended to the cases with additional zero modes. The third condition is
violated when the theory has the (approximate) scale invariance, which
has already been discussed for the single-field case
\cite{Andreassen:2017rzq,Chigusa:2017dux,Chigusa:2018uuj} and the
multi-field case \cite{Oda:2019njo}.

We define the subset of gauge fields that acquire masses from
the bounce at $r < \infty$.  Then, we define the
following numbers:
\begin{itemize}
\item $n_G$: the number of the gauge bosons which acquire
  masses from the bounce at $r < \infty$.
\item $n_B$: the number of the gauge bosons which remain massive at
  the false vacuum.
\item $n_U$: the number of the gauge bosons which become massless
  at the false vacuum (but are massive at $r < \infty$).
\item $n_\varphi$: the total number of scalar fields.
\item $n_H$: the number of the scalar fields
  that do not mix with the gauge bosons at $r<\infty$.
\end{itemize}
Notice that $n_\varphi=n_G+n_H$ and $n_G=n_B+n_U$.

Now we introduce the gauge fixing terms.  We consider the following
two gauge fixing conditions:
\begin{enumerate}
 \item Fermi gauge:
\begin{align}
  \mathcal L^{\rm (GF)}_{\rm Fermi}
  &=\frac{1}{2\xi}(\partial_\mu A_\mu^a)^2,\label{eq_Fermi}\\
  \mathcal L^{\rm (ghost)}_{\rm Fermi}
  &=-\bar{c}\partial_\mu D_\mu c.
\end{align}
\item Background gauge:
  \begin{align}
  \mathcal L^{\rm (GF)}_{\rm BG}
  &=\frac{1}{2\xi}\left(\partial_\mu A_\mu^a+\xi g_a\phi^T T^a\bar\phi\right)^2,
  \label{eq_BG}\\
  \mathcal L^{\rm (ghost)}_{\rm BG}
  &=\bar{c}^a\left[-\partial_\mu D_\mu^{ab}+\xi (g_aT^a\bar\phi)^T (g_bT^b\phi)\right]c^b.
\end{align}
\end{enumerate}
These gauge fixing terms are consistent with the bounce of our choice
since their contributions to the equation of motion vanish in the
limit of $A_\mu\to0$ and $\phi\to\bar\phi$.  They are also consistent
with the spacial translational invariance; it corresponds to the
fluctuation of $\delta\phi=\delta x_\mu\partial_\mu\bar\phi$. As for
the global gauge transformation,
$\delta\phi=\delta\theta^aT^a\bar\phi$, the gauge fixing term of the
Fermi gauge is invariant \cite{Kusenko:1996bv, Endo:2017tsz}.  Then,
in the Fermi gauge, a prescription to take care of the gauge zero modes
is available.  On the contrary, in the background gauge, the proper
treatment of the gauge zero modes becomes more complicated, as we will
discuss.  The background gauge is, however, useful for numerical
calculation especially with $\xi=1$.  Thus, we also discuss how the
prefactors based on these two gauge fixing conditions are related in
this paper.

\subsection{Fluctuation operators}

To evaluate the decay rate of the false vacuum, we integrate out the
fluctuations around the bounce configuration.  As is explicitly given
in the next subsection, the prefactor is expressed with the functional
determinant of the fluctuation operators, which are the second
derivative of the action with respect to the fluctuations.

In discussing the contributions to the prefactor in our setup, we
should consider the fluctuation operator for the bosons and that for
the Faddeev-Popov (FP) ghosts.  In the Fermi gauge, they are given by\footnote
{We do not put the subscript ``Fermi'' on the fluctuation operators in the
  Fermi gauge for notational simplicity while we put the subscript ``BG'' on
  those in the background gauge.}
\begin{align}
  \mathcal{M}^{(A_\mu\varphi)} &= \begin{pmatrix}
    (-\partial^2 + M^T M) \delta_{\mu\nu} + (1-\frac{1}{\xi}) \partial_\mu \partial_\nu
    & (\partial_\nu M)^T - M^T \partial_\nu\\
    2(\partial_\mu M) + M \partial_\mu & -\partial^2 + \Omega
  \end{pmatrix},
\end{align}
and
\begin{align}
  \mathcal{M}^{(c\bar c)} &= -\partial^2,
\end{align}
respectively.  Note that $\mathcal{M}^{(A_\mu\varphi)}$ is an $(n_G +
n_\varphi) \times (n_G + n_\varphi)$ object and $\mathcal{M}^{(c\bar c)}$ is
$n_G\times n_G$.  In addition, in the background gauge,
\begin{align}
  \mathcal{M}_{\mathrm{BG}}^{(A_\mu\varphi)} &= \begin{pmatrix}
    (-\partial^2 + M^T M) \delta_{\mu\nu} + (1-\frac{1}{\xi}) \partial_\mu \partial_\nu
    & 2(\partial_\nu M)^T\\
    2(\partial_\mu M) & -\partial^2 + \Omega + \xi M M^T
  \end{pmatrix},\\
  \mathcal{M}_{\mathrm{BG}}^{(c\bar c)} &= -\partial^2 + \xi M^T M.
\end{align}

In order for the following discussion, let us define
\begin{align}
  \widehat{M} \equiv &\, M (r\rightarrow\infty),
  \\
  \widehat{\Omega} \equiv &\, \Omega (r\rightarrow\infty).
\end{align}
Then, using Eq.\ \eqref{eq_M_eom} as well as
$M'(r\rightarrow\infty)=0$, we obtain
\begin{align}
  \widehat\Omega \widehat{M} = 0.
  \label{hatOmegahatM=0}
\end{align}
For later convenience, we choose the following basis of the gauge
bosons and the scalars:
\begin{align}
  \widehat{M} = &\,
  \begin{pmatrix}
    \eqmakebox[tagWhat]{$\widehat W$}&0\\
    \coolunder{tagWhat}{n_B}{0}  & \coolunder{tag0}{n_U}{0}
  \end{pmatrix}
  \begin{matrix*}[l]
    \coolright{n_B}\\
    \coolright{n_U+n_H}
  \end{matrix*},\label{eq_basis_M}\\ \nonumber\\
  \widehat{\Omega} = &\,
  \begin{pmatrix}
    0&0\\
    \coolunder{tag0}{n_B}{0}&\coolunder{tagm2hat}{n_U+n_H}{\widehat{m}^2}
  \end{pmatrix}
  \begin{matrix*}[l]
    \coolright{n_B}\\
    \coolright{n_U+n_H}
  \end{matrix*},\label{eq_basis_Omega}\\ \nonumber
\end{align}
where $\widehat W$ and $\widehat m^2$ are full rank diagonal matrices.\footnote
{We consider the case where there is no massless physical scalar in
  the false vacuum, and hence $\hat{m}^2$ does not have zero
  eigenvalues. }
Such a choice always exists because of Eq.\ \eqref{hatOmegahatM=0}.
In this basis, we also define submatrices of $M$ as
\begin{align}
 M (r) = \begin{pmatrix}
   \coolunder{}{n_B}{M_B (r)} & \coolunder{}{n_U}{M_U (r)}
 \end{pmatrix}.\\ \nonumber
\end{align}
In addition, we define
\begin{align}
  M_0 \equiv M(r=0).
\end{align}

The fluctuation around the bounce can be expanded using the
hyperspherical functions on $S^3$, which are functions of $\hat{x}_\mu
\equiv x_\mu / r$.  The rotational Lie algebra of the four dimensional
Euclidean space is equivalent to $SU(2)_L \times SU(2)_R$ and the
hyperspherical functions, $\mathcal{Y}_P (\hat{x}_\mu)$, are labeled by
indices $P=(\ell,m_A,m_B)$.  Here, $\ell = 0,1,2,\cdots$ is the
azimuthal quantum number and $m_A$ and $m_B$ are the magnetic quantum
numbers that take values of $-\ell/2, \ell/2+1, \cdots,
\ell/2$.\footnote
{The variable $J$ used in \cite{Endo:2017gal, Endo:2017tsz} and
$\ell$ are related as $\ell=2J$.}
Using mode functions that depend only on $r$ and the hyperspherical
functions, we expand fields around the bounce as
\begin{align}
  \phi(x)=&\bar\phi(r)+\alpha_{\varphi}^{P}(r)\mathcal Y_P,
  \\
  A_\mu^a(x)=&\alpha_S^{aP}(r)\frac{x_\mu}{r}\mathcal Y_P
  +\alpha_L^{aP}(r)\frac{r}{L}\partial_\mu\mathcal Y_P
  \nonumber\\ &
  +\alpha_{T1}^{aP}(r)i\epsilon_{\mu\nu\rho\sigma}V_\nu^{(1)}L_{\rho\sigma}\mathcal Y_P+\alpha_{T2}^{aP}(r)i\epsilon_{\mu\nu\rho\sigma}V_\nu^{(2)}L_{\rho\sigma}\mathcal Y_P,\\
 c^a(x)=&\alpha_c^{aP}(r)\mathcal Y_P,\\
 \bar c^a(x)=&\alpha_{\bar c}^{aP}(r)\mathcal Y_P,
\end{align}
where the summation over
$P$ is implicit.
Here, $V_\nu^{(1)}$ and $V_\nu^{(2)}$ are arbitrary two independent vectors and
\begin{align}
 L_{\rho\sigma}&=\frac{i}{\sqrt{2}}(x_\rho\partial_\sigma-x_\sigma\partial_\rho),\\
 L&=\sqrt{\ell(\ell+2)}.
\end{align}
Note that $(L)$ and $(T)$ modes do not have $\ell =0$ mode.  For the
following discussion, it is convenient to define the derivative
operator that corresponds to the Laplacian acting on the modes with
the azimuthal quantum number $\ell$:
\begin{align}
 \Delta_\ell&=\partial_r^2+\frac{3}{r}\partial_r-\frac{L^2}{r^2}.
\end{align}

Similarly, we define the fluctuation operators at the false vacuum,
which can be obtained by replacement $M\to\widehat M$ and
$\Omega\to\widehat\Omega$. We denote them as
$\widehat{\mathcal{M}}^{(A_\mu\varphi)}$ and
$\widehat{\mathcal{M}}^{(c\bar c)}$.

In the following, we will show the hyperspherical expansion of the
fluctuation operators around the bounce and around the false vacuum.

\subsubsection{FP ghosts}

The fluctuations of the FP ghosts can be expanded by using the
hyperspherical functions.  Correspondingly, the fluctuation operator
for the FP ghosts can be block-diagonalized as
\begin{align}
  \mathcal M^{(c\bar c)}&=
  \bigoplus_{\ell=0}^\infty\left(\mathcal M^{(c\bar c)}_\ell\right)^{2(\ell+1)^2},
\end{align}
where the power comes from $(\ell+1)^2$ different choices of
$(m_A,m_B)$ and the complexity of the FP ghosts.
In the Fermi gauge,
\begin{align}
 \mathcal M^{(c\bar c)}_{\ell}=-\Delta_\ell,
 \label{eq_M_ccbar_Fermi}
\end{align}
while in the background gauge,
\begin{align}
  \mathcal M^{(c\bar c)}_{\ell,{\rm BG}}=-\Delta_\ell+\xi M^TM.
 \label{eq_M_ccbar_BG}
\end{align}
Notice that the fluctuation operators for the ghosts are
$n_G\times n_G$ objects.

At the false vacuum, we have a similar block-diagonalization:
\begin{align}
  \widehat{\mathcal M}^{(c\bar c)}&=
  \bigoplus_{\ell=0}^\infty\left(\widehat{\mathcal M}^{(c\bar c)}_\ell\right)^{2(\ell+1)^2},
\end{align}
where, in the Fermi gauge,
\begin{align}
 \widehat{\mathcal M}_{\ell}^{(c\bar c)}&=-\Delta_\ell,
\end{align}
while in the background gauge,
\begin{align}
  \widehat{\mathcal M}_{\ell,{\rm BG}}^{(c\bar c)}&=-\Delta_\ell+\xi \widehat M^T\widehat M.
\end{align}

\subsubsection{Gauge bosons and scalars}

Due to the mixing between the gauge bosons and the scalars, we cannot
discuss their effects separately.  Since only the mode functions with
the same $P=(\ell,m_A,m_B)$ mix, the fluctuation operator
$\mathcal{M}^{(A_\mu\varphi)}$ can be block-diagonalized as
\begin{align}
  \mathcal M^{(A_\mu \varphi)}&=\mathcal M_0^{(S\varphi)}\oplus\left[\bigoplus_{\ell=1}^\infty \left(\mathcal M_\ell^{(SL\varphi)}\right)^{(\ell+1)^2}\right]\oplus\left[\bigoplus_{\ell=1}^\infty \left(\mathcal M_\ell^{(T)}\right)^{2(\ell+1)^2}\right].
\end{align}
Here, independently of the choice of the gauge fixing,
the fluctuation operator for the $(T)$ modes is given by
\begin{align}
  \mathcal M_\ell^{(T)}&=-\Delta_\ell+M^TM.
  \label{M^T(bounce)}
\end{align}
Meanwhile, the fluctuation operators of $(S)$, $(L)$, and $(\varphi)$ modes
depend on the gauge fixing.  In the Fermi gauge,
\begin{align}
\mathcal M_{0}^{(S\varphi)}&=
\begin{pmatrix}
 -\frac{1}{\xi}\Delta_1+M^TM&(M')^T-M^T\partial_r\\
 2M'+M\frac{1}{r^3}\partial_rr^3&-\Delta_0+\Omega
\end{pmatrix},\\
  \mathcal M_{\ell}^{(SL\varphi)}&=
\begin{pmatrix}
 -\Delta_\ell+\frac{3}{r^2}+M^TM&-\frac{2L}{r^2}&(M')^T-M^T\partial_r\\
 -\frac{2L}{r^2}&-\Delta_\ell-\frac{1}{r^2}+M^TM & -\frac{L}{r}M^T\\
 2M'+M\frac{1}{r^3}\partial_rr^3& -\frac{L}{r}M&-\Delta_\ell+\Omega
\end{pmatrix}\nonumber\\
&\hspace{3ex}+\left(1-\frac{1}{\xi}\right)
\begin{pmatrix}
 \partial_r^2+\frac{3}{r}\partial_r-\frac{3}{r^2}&-L\left(\frac{1}{r}\partial_r-\frac{1}{r^2}\right)&0\\
 L\left(\frac{1}{r}\partial_r+\frac{3}{r^2}\right)&-\frac{L^2}{r^2}&0 \\
 0&0 &0
\end{pmatrix}.
\end{align}
Note that $\mathcal M_0^{(S\varphi)}$ is $(n_G + n_\varphi) \times (n_G + n_\varphi)$, while $\mathcal M_\ell^{(SL\varphi)}$ is $(2n_G + n_\varphi) \times (2n_G + n_\varphi)$.

Meanwhile, in the background gauge,
\begin{align}
 \mathcal M_{0,{\rm BG}}^{(S\varphi)}&=
\begin{pmatrix}
 -\frac{1}{\xi}\Delta_1+M^TM&2(M')^T\\
 2M'&-\Delta_0+\Omega+\xi MM^T
\end{pmatrix},
\label{M0(BG)}\\
 \mathcal M_{\ell,{\rm BG}}^{(SL\varphi)}&=
\begin{pmatrix}
 -\Delta_\ell+\frac{3}{r^2}+M^TM&-\frac{2L}{r^2}&2(M')^T\\
 -\frac{2L}{r^2}&-\Delta_\ell-\frac{1}{r^2}+M^TM & 0\\
 2M'& 0&-\Delta_\ell+\Omega+\xi MM^T
\end{pmatrix}\nonumber\\
&\hspace{3ex}+\left(1-\frac{1}{\xi}\right)
\begin{pmatrix}
 \partial_r^2+\frac{3}{r}\partial_r-\frac{3}{r^2}&-L\left(\frac{1}{r}\partial_r-\frac{1}{r^2}\right)&0\\
 L\left(\frac{1}{r}\partial_r+\frac{3}{r^2}\right)&-\frac{L^2}{r^2}&0 \\
 0&0 &0
\end{pmatrix}.
\end{align}

At the false vacuum, similar block-diagonalizations hold.  For the
$(T)$ mode,
\begin{align}
  \widehat{\mathcal M}_\ell^{(T)}&=-\Delta_\ell+\widehat M^T\widehat M.
  \label{M^T(false)}
\end{align}

The fluctuation operator for $(SL\varphi)$ modes at the false vacuum
can be further block-diagonalized thanks to the choice of the basis of
Eqs.~\eqref{eq_basis_M} and \eqref{eq_basis_Omega}. For $\ell>0$, the
fluctuation operator can be expressed as
\begin{align}
  \widehat{\mathcal M}_\ell^{(SL\varphi)} =
  \left. {\mathcal M}_\ell^{(SL\varphi)} \right|_{\bar{\phi}\rightarrow v}
  =\widehat{\mathcal M}_\ell^{(B)}\oplus\widehat{\mathcal M}_\ell^{(U)}\oplus\widehat{\mathcal M}_\ell^{(\sigma)},
  \label{eq_block_ell}
\end{align}
corresponding to the contributions from massive gauge bosons and
Nambu-Goldstone (NG) bosons (due to broken gauge symmetry), massless
gauge bosons (in association with unbroken gauge symmetry), and
physical scalars.  Firstly, contributions to
$\widehat{\mathcal{M}}_\ell^{(B)}$ are from $(S)$ and $(L)$ modes of
gauge bosons corresponding to broken symmetries $(a = 1,\dots,n_B)$
and the corresponding NG modes $(i=1,\dots,n_B)$. Then, $\widehat{\mathcal
  M}_\ell^{(B)}$ is a $3n_B \times 3n_B$ object and is given by
\begin{align}
  \widehat{\mathcal M}_\ell^{(B)}
  = &\,
  \begin{pmatrix}
    -\Delta_\ell+\frac{3}{r^2}+\widehat W^T\widehat W&-\frac{2L}{r^2}&-\widehat W^T\partial_r\\
    -\frac{2L}{r^2}&-\Delta_\ell-\frac{1}{r^2}+\widehat W^T\widehat W & -\frac{L}{r}\widehat W^T\\
    \widehat W\frac{1}{r^3}\partial_rr^3& -\frac{L}{r}\widehat W&-\Delta_\ell
  \end{pmatrix}
  \notag\\ &\,
  +\left(1-\frac{1}{\xi}\right)
  \begin{pmatrix}
    \partial_r^2+\frac{3}{r}\partial_r-\frac{3}{r^2}&-L\left(\frac{1}{r}\partial_r-\frac{1}{r^2}\right)&0\\
    L\left(\frac{1}{r}\partial_r+\frac{3}{r^2}\right)&-\frac{L^2}{r^2}&0 \\
    0&0 &0
  \end{pmatrix}.
\end{align}
Secondly, contributions to $\widehat{\mathcal{M}}_\ell^{(U)}$ are from
massless gauge bosons in
$(S)$ and  $(L)$ modes. We obtain a $2n_U
\times 2n_U$ fluctuation operator as
\begin{align}
 \widehat{\mathcal M}_\ell^{(U)}&=
\begin{pmatrix}
 -\Delta_\ell+\frac{3}{r^2}&-\frac{2L}{r^2}\\
 -\frac{2L}{r^2}&-\Delta_\ell-\frac{1}{r^2}
\end{pmatrix}+\left(1-\frac{1}{\xi}\right)
\begin{pmatrix}
 \partial_r^2+\frac{3}{r}\partial_r-\frac{3}{r^2}&-L\left(\frac{1}{r}\partial_r-\frac{1}{r^2}\right)\\
 L\left(\frac{1}{r}\partial_r+\frac{3}{r^2}\right)&-\frac{L^2}{r^2}
\end{pmatrix}.
\end{align}
For the other massive scalars,
\begin{align}
 \widehat{\mathcal M}_\ell^{(\sigma)}&=-\Delta_\ell+\widehat{m}^2,
\end{align}
which is an $(n_U + n_H) \times (n_U + n_H)$ object.

Similarly, for $\ell=0$, we obtain
\begin{align}
  \widehat{\mathcal M}_0^{(S\varphi)}=\widehat{\mathcal M}_0^{(B)}\oplus\widehat{\mathcal M}_0^{(U)}\oplus\widehat{\mathcal M}_0^{(\sigma)},
  \label{eq_block_0}
\end{align}
where
\begin{align}
  \widehat{\mathcal M}_0^{(B)}&=
 \begin{pmatrix}
  -\frac{1}{\xi}\Delta_1+\widehat{W}^T\widehat{W}&-\widehat{W}^T\partial_r\\
  \widehat{W}\frac{1}{r^3}\partial_rr^3&-\Delta_0
 \end{pmatrix},\\
 \widehat{\mathcal M}_0^{(U)}&=-\frac{1}{\xi}\Delta_1,\\
 \widehat{\mathcal M}_0^{(\sigma)}&=-\Delta_0+\widehat{m}^2.
\end{align}

\subsection{Prefactor and functional determinant}

The prefactor $\mathcal A$ is expressed as
\begin{align}
  \mathcal A=
  \mathcal A^{(c\bar c)} \mathcal{A}^{(A_\mu\varphi)},
  \label{eq_prefactor}
\end{align}
where $\mathcal{A}^{(c\bar{c})}$ denotes the contributions from the FP
ghosts and $\mathcal{A}^{(A_\mu \varphi)}$ denotes those from the gauge
bosons and the scalars.  For the evaluation of the prefactor at the
one-loop level, the following quantities are necessary
\cite{Callan:1977pt}:
\begin{align}
  \mathcal A^{(c\bar c)}&=\frac{\Det\mathcal M^{(c\bar c)}}{\Det\widehat{\mathcal M}^{(c\bar c)}}=\prod_{\ell=0}^{\infty}\left(\frac{\Det\mathcal M_\ell^{(c\bar c)}}{\Det\widehat{\mathcal M}_\ell^{(c\bar c)}}\right)^{(\ell+1)^2},
  \label{eq_A_ccbar}
  \\
 \mathcal A'^{(A_\mu\varphi)}&=\left[\frac{\Det'\mathcal M^{(A_\mu\varphi)}}{\Det\widehat{\mathcal M}^{(A_\mu\varphi)}}\right]^{-1/2}\nonumber\\
 &=\left(\frac{\Det'\mathcal M_0^{(S\varphi)}}{\Det\widehat{\mathcal M}_0^{(S\varphi)}}\right)^{-1/2}\left(\frac{\Det'\mathcal M^{(SL\varphi)}_1}{\Det\widehat{\mathcal M}_1^{(SL\varphi)}}\right)^{-2}\nonumber\\
 &\hspace{3ex}\times\left[\prod_{\ell=2}^{\infty}\left(\frac{\Det\mathcal M_\ell^{(SL\varphi)}}{\Det\widehat{\mathcal M}_\ell^{(SL\varphi)}}\right)^{-(\ell+1)^2/2}\right]\left[\prod_{\ell=1}^{\infty}\left(\frac{\Det\mathcal M_\ell^{(T)}}{\Det\widehat{\mathcal M}_\ell^{(T)}}\right)^{-(\ell+1)^2}\right].
 \label{eq_A_Aphi}
\end{align}
Here and hereafter, the ``prime'' is used for quantities after the
proper subtraction of the zero modes if necessary.  In particular,
because of the translational invariance, the $\ell=1$ contribution
inevitably contains the effects of translational zero modes
\cite{Callan:1977pt}.  In addition, if $n_U>0$, both $\mathcal
M_{0}^{(S\varphi)}$ and $\mathcal M_{0,{\rm BG}}^{(S\varphi)}$
have zero eigenvalues.  More details about the zero mode subtraction and
the relation between $\mathcal A^{(A_\mu\varphi)}$ and $\mathcal
A'^{(A_\mu\varphi)}$ will be explained in Section \ref{sec_zeromodes}.

To evaluate the ratio of two functional determinants, we adopt the
method developed in
\cite{Gelfand:1959nq,Dashen:1974ci,Coleman:1985rnk,Kirsten:2003py,Kirsten:2004qv}.  Let $\mathcal
M_\ell^{(X)}$ and $\widehat{\mathcal M}_\ell^{(X)}$ be $n\times n$
fluctuation  operators.  We first prepare $n$ linearly independent
$n$-dimensional functions $\psi_{\ell}^{(X)(I)}(r)$ and
$\widehat{\psi}_{\ell}^{(X)(I)}(r)$ (with $I=1-n$) that satisfy
\begin{align}
  \mathcal M_\ell^{(X)} \psi_{\ell}^{(X)(I)}=0,
  \label{eq_MPsiX=0}
  \\
  \widehat{\mathcal M}_\ell^{(X)}\widehat\psi_{\ell}^{(X)(I)}=0,
  \label{eq_hatMPsiX=0}
\end{align}
and are regular at $r\to 0$.  Then, we define $n\times n$ objects
$\Psi_{\ell}^{(X)}(r)$ and $\widehat{\Psi}_{\ell}^{(X)}(r)$ as
\begin{align}
  \Psi_{\ell}^{(X)}(r) \equiv &\
  \left( \psi_{\ell}^{(X)(1)}(r) ~~\cdots~~ \psi_{\ell}^{(X)(n)}(r) \right),
  \\
  \widehat\Psi_{\ell}^{(X)}(r) \equiv &\
  \left( \widehat\psi_{\ell}^{(X)(1)}(r) ~~\cdots~~ \widehat\psi_{\ell}^{(X)(n)}(r) \right),
\end{align}
with which  the ratio of functional determinants can be
evaluated as
\begin{align}
  \frac{\Det\mathcal M_\ell^{(X)}}{\Det\widehat{\mathcal M}_\ell^{(X)}}=
  \left(\frac{\det\Psi_\ell^{(X)}(r_0)}{\det\widehat{\Psi}_\ell^{(X)}(r_0)}\right)^{-1}
  \left(\frac{\det\Psi_\ell^{(X)}(r_\infty)}{\det\widehat{\Psi}_\ell^{(X)}(r_\infty)}\right),
 \label{eq_det_ratio}
\end{align}
where $r_0$ and $r_\infty$ are abbreviations of $r\rightarrow 0$ and
$r\rightarrow\infty$, respectively.

Thus, for the evaluation of the prefactor $\mathcal A$, we need to
understand the asymptotic behavior of $\Psi_\ell^{(X)}$ and
$\widehat\Psi_\ell^{(X)}$ (and hence those of
$\det{\Psi}_\ell^{(X)}(r)$ and $\det\widehat{\Psi}_\ell^{(X)}(r)$) at
$r\rightarrow 0$ and $r\rightarrow\infty$.  In particular, the
behavior of $\det\Psi_\ell^{(SL\varphi)}$ and
$\det\Psi_0^{(S\varphi)}$ is non-trivial because of the mixing among
$(S)$, $(L)$, and $(\varphi)$ modes. In the following sections, we
discuss how we can evaluate those quantities.

\section{Decomposition of Solutions}
\label{sec_solutions}
\setcounter{equation}{0}

In this section, we provide a decomposition of the set of solutions
$\Psi_\ell^{(SL\varphi)}$ and $\Psi_0^{(S\varphi)}$, generalizing the
results of \cite{Endo:2017gal,Endo:2017tsz}.

\subsection{$\ell>0$}

Let us consider the solutions of the following equation in the Fermi
gauge:
\begin{align}
  \mathcal M_\ell^{(SL\varphi)}\Psi_\ell^{(SL\varphi)}=0.
  \label{M*Psi(SLphi)=0}
\end{align}
Because $\mathcal{M}_\ell^{(SL\varphi)}$ is $(2n_G + n_\varphi) \times (2n_G +
n_\varphi)$, there exist $(2n_G + n_\varphi)$ linearly
independent solutions of Eq.\ \eqref{M*Psi(SLphi)=0}.

Solutions of Eq.\ \eqref{M*Psi(SLphi)=0} can be decomposed by using
four functions, $\chi$, $\eta$, $\zeta$, and $\lambda$:
\begin{align}
 \Psi_\ell^{(SL\varphi)}=
\begin{pmatrix}
 \Psi_\ell^{\rm (top)}\\
 \Psi_\ell^{\rm (mid)}\\
 \Psi_\ell^{\rm (bot)}
\end{pmatrix}=
\begin{pmatrix}
 \partial_r\chi\\
 \frac{L}{r}\chi\\
 M\chi
\end{pmatrix}
+
\begin{pmatrix}
 (M^TM)^{-1}\left[\frac{L}{r}\eta-2(M')^T \lambda\right]\\
 (M^TM)^{-1}\frac{1}{r^2}\partial_rr^2\eta\\
  \lambda
\end{pmatrix}
+
\begin{pmatrix}
 [\partial_r(M^TM)^{-1}]\zeta\\
 0\\
 M(M^TM)^{-1}\zeta
\end{pmatrix}.
\label{eq_decomp}
\end{align}
Here, the functions $\chi$, $\eta$, $\zeta$, and $\lambda$ are
dependent on the azimuthal quantum number $\ell$; the subscript
``$\ell$'' is omitted from these functions for notational simplicity.
The shape of $\chi$, $\eta$, and $\zeta$ is $n_G \times (2n_G +
n_\varphi)$, while that of $\lambda$ is $n_\varphi \times (2n_G +
n_\varphi)$; here, the $(2n_G + n_\varphi)$ columns are linearly
independent and are distinguished by the boundary conditions at $r=0$.

The evolution of functions $\chi$ and $\zeta$ is governed by the
following differential equations:
\begin{align}
  \mathcal M^{(c\bar c)}_\ell\chi =&
           [\partial_r(M^TM)^{-1}]\frac{L}{r}\eta-\frac{2}{r^3}\partial_rr^3(M^TM)^{-1}(M')^T\lambda
  \nonumber\\ &
  -\mathcal M^{(c\bar c)}_0(M^TM)^{-1}\zeta-\frac{1}{r^3}\partial_rr^3(M^TM)^{-1}\partial_r\zeta+\xi\zeta,
  \label{eq_chi}\\
  \mathcal M^{(c\bar c)}_\ell\zeta =& 0,
  \label{eq_zeta}
\end{align}
while $\eta$ and $\lambda$ satisfy
\begin{align}
  \Delta_\ell \eta = &\,
  M^TM \left[ \eta - \{\partial_r(M^TM)^{-1}\} \frac{1}{r^2} \partial_r r^2 \eta \right]
  - \frac{2L}{r} M'^T \lambda
  + \frac{L}{r} M^TM [\partial_r(M^TM)^{-1}] \zeta,
  \label{eq_eta}
  \\
  \Delta_\ell \lambda = &\,
  \Omega \lambda
  - 4 M' (M^TM)^{-1} M'^T \lambda
  + \frac{2L}{r} M' (M^TM)^{-1} \eta
  - 2 M' (M^TM)^{-1} \zeta'
  \nonumber \\ &\,
  + M
  \left[
    -\frac{2}{r^3} \partial_r r^3 (M^TM)^{-1} M'^T \lambda
    + \frac{L}{r} \{\partial_r(M^TM)^{-1}\} \eta
    - \{\partial_r(M^TM)^{-1}\} \zeta'
    \right].
  \label{eq_lambda}
\end{align}
In addition, $\lambda$ satisfies the following constraint:
\begin{align}
  M^T \lambda = 0.
  \label{Mlambda=0}
\end{align}
The above constraint is consistent with the evolution equations; one can
derive
$\Delta_\ell (M^T\lambda)=0$,
so that Eq.\
\eqref{Mlambda=0} holds if it is satisfied at $r=0$.

So far, we have discussed in the Fermi gauge.  We note here that the
decomposition given in Eq.\ \eqref{eq_decomp} is also applicable in
the background gauge if we replace $\mathcal{M}_\ell^{(c\bar{c})}$ in
Eqs.\ \eqref{eq_chi} and \eqref{eq_zeta} by $\mathcal M^{(c\bar
  c)}_{\ell,{\rm BG}}$ given in Eq.\ \eqref{eq_M_ccbar_BG}.  (In the
background gauge,
the relation $\mathcal M^{(c\bar c)}_{\ell,{\rm BG}}
(M^T\lambda)=0$ holds.  Thus, the condition $M^T\lambda=0$ is
again consistent with the evolution equations.)

From Eqs.~\eqref{eq_decomp} -- \eqref{eq_lambda}, we can derive the
following equations:
\begin{align}
  \mathcal M_{\ell}^{(c\bar c)}\left(r\Psi^{({\rm mid})}_\ell\right)&=
  -\frac{1}{r^3}\partial_rr^4\eta
  + \xi L\zeta,
  \label{eq_mid}\\
  \Psi^{({\rm top})}_\ell&=\frac{1}{L}\partial_r\left(r\Psi^{({\rm mid})}_\ell\right)-\frac{r}{L}\eta,\label{eq_top}\\
  \Psi^{({\rm bot})}_\ell&=\frac{r}{L}M\Psi^{({\rm mid})}_\ell-M(M^TM)^{-1}\frac{1}{Lr}\partial_rr^2\eta+\lambda+M(M^TM)^{-1}\zeta.
  \label{eq_bot}
\end{align}
These expressions are useful to derive the asymptotic behaviors of the
solutions.  Notice that, in the background gauge, Eq.\ \eqref{eq_mid}
is replaced by
\begin{align}
  \mathcal M_{\ell,{\rm BG}}^{(c\bar c)}\left(r\Psi^{({\rm mid})}_{\ell,{\rm BG}}\right)&=
  -\frac{1}{r^3}\partial_rr^4\eta
  +
  \frac{\xi}{r}\partial_rr^2\eta,
  \label{eq_mid_BG}
\end{align}
while Eqs.\ \eqref{eq_top} and \eqref{eq_bot} are unchanged.

\subsection{$\ell=0$}

For $\ell=0$ mode, we need solutions of
\begin{align}
  \mathcal M_0^{(S\varphi)}\Psi_0^{(S\varphi)}=0.
  \label{M*Psi(Sphi)=0}
\end{align}
The solutions can be decomposed as
\begin{align}
  \Psi_0^{(S\varphi)}=
  \begin{pmatrix}
    \Psi_0^{\rm (top)}\\
    \Psi_0^{\rm (bot)}
  \end{pmatrix}=
  \begin{pmatrix}
    \partial_r\chi\\
    M\chi
  \end{pmatrix}
  +
  \begin{pmatrix}
    -2(M^TM)^{-1}(M')^T \lambda\\
    \lambda
  \end{pmatrix}
  +
  \begin{pmatrix}
    [\partial_r(M^TM)^{-1}]\zeta\\
    M(M^TM)^{-1}\zeta
  \end{pmatrix},
  \label{eq_decomp_0}
\end{align}
where $\chi$, $\zeta$, and $\lambda$ for $\ell=0$ satisfy Eqs.\
\eqref{eq_chi}, \eqref{eq_zeta}, and \eqref{eq_lambda} with replacing
$\ell\rightarrow 0$ and $\eta\rightarrow 0$.  Notice that now $\chi$ and
$\zeta$ are $n_G \times (n_G + n_\varphi)$ objects and $\lambda$ is
$n_\varphi \times (n_G + n_\varphi)$; the columns correspond to
$(n_G+n_\varphi)$ independent choices of the boundary conditions at
$r=0$.  Thus, $\Psi_0^{(S\varphi)}$ is an
$(n_G+n_\varphi)\times(n_G+n_\varphi)$ object.

\section{Functional Determinants}
\label{sec_determinant}
\setcounter{equation}{0}

In this section, we study the behavior of the functional determinants
of the fluctuation operators in the Fermi gauge.  Equivalence to the
results in the background gauge is also discussed.

\subsection{$\ell>0$}

Let us consider the case with $\ell>0$. For the study of the
functional determinants, we should first understand the behavior of the
solutions of Eq.\ \eqref{M*Psi(SLphi)=0}.  For this purpose, we define
an $r$-dependent $n_\varphi\times n_H$ matrix $V_H$, which is given by
\begin{align}
 V_H&=
 \begin{pmatrix}
 u_1&\cdots&u_{n_H}
 \end{pmatrix},
\end{align}
with $u_p$ ($p=1,\cdots,n_H$) being zero eigenvectors of $M^T$, {\it i.e.},
\begin{align}
  M^Tu_p=0,~~~
  u_p^T u_{p'} =\delta_{pp'}.
\end{align}

The columns of the solution of Eq.\ \eqref{M*Psi(SLphi)=0} can be
classified into the following three types:
\begin{itemize}
\item {\bf Type 1}: $\eta^{(1)}=\lambda^{(1)}=\zeta^{(1)}=0$.
\item {\bf Type 2}: $\zeta^{(2)}\neq 0$.
\item {\bf Type 3}: $\zeta^{(3)}=0$, with non-vanishing
  $\eta^{(3)}$ and $\lambda^{(3)}$,
\end{itemize}
where the superscripts ``$(1)$,'' ``$(2)$,'' and ``$(3)$'' indicate
solutions for the Type $1$, $2$, and $3$, respectively.  Note that the
Type 1, 2, and 3 solutions include $n_G$, $n_G$, and $n_\varphi$
independent solutions, respectively.  The full matrix of solutions is
constructed as
\begin{align}
  \eta = \begin{pmatrix}
    \coolunder{tageta1}{n_G}{\eta^{(1)}} & \coolunder{tageta1}{n_G}{\eta^{(2)}} & \coolunder{tageta1}{n_\varphi}{\eta^{(3)}}
  \end{pmatrix},\\[-2ex]\nonumber
\end{align}
and similar for $\chi$, $\lambda$, and $\zeta$.
For each type of solutions, the boundary conditions at $r\to 0$ are
imposed as follows.
\begin{itemize}
\item {\bf Type 1}:
  We take
  \begin{align}
    \chi^{(1)}(r) = r^\ell I_G,
  \end{align}
  where $I_G$ is the $n_G\times n_G$ unit matrix.  Then, from Eq.~\eqref{eq_decomp}, we
  obtain
  \begin{align}
    \Psi_{\ell}^{(SL\varphi)(1)}(r\to0)\simeq
    \begin{pmatrix}
      \eqmakebox[tagPhi1orig]{$\ell r^{\ell-1} I_G$}\\
      \eqmakebox[tagPhi1orig]{$Lr^{\ell-1} I_G$}\\
      \coolunder{tagPhi1orig}{n_G}{M r^\ell}
%
    \end{pmatrix}
    \begin{matrix*}[l]
      \coolright{n_G}\\
      \coolright{n_G}\\
      \coolright{n_\varphi}
    \end{matrix*}.\label{BC_slphi1}\\[-2ex]\nonumber
  \end{align}

\item {\bf Type 2}: We take
\begin{align}
 \zeta^{(2)} (r) = r^\ell I_G.
\end{align}
At $r\to0$, $M'\propto r$ and $V_HV_H^TM'\propto r^2$. Thus,
from Eq.~\eqref{eq_eta} and Eq.~\eqref{eq_lambda}, $\eta^{(2)} (r\to0)=\mathcal O(r^{\ell+2})$ and $\lambda^{(2)} (r\to0)=\mathcal O(r^{\ell+3})$.
Consequently, using Eqs.~\eqref{eq_mid} -- \eqref{eq_bot}, we obtain
\begin{align}
  \Psi_{\ell}^{(SL\varphi)(2)}(r\to0)&\simeq
\begin{pmatrix}
 \eqmakebox[tagPhi2origF]{$-\frac{\xi}{4}r^{\ell+1}I_G$}\\
 \eqmakebox[tagPhi2origF]{$-\frac{\xi\ell}{4L}r^{\ell+1}I_G$}\\
 \coolunder{tagPhi2origF}{n_G}{M(M^TM)^{-1}r^\ell}
\end{pmatrix}
\begin{matrix*}[l]
    \coolright{n_G}\\
    \coolright{n_G}\\
    \coolright{n_\varphi}
\end{matrix*}.\label{BC_slphi2}\\[-2ex]\nonumber
\end{align}

\item {\bf Type 3}: This type contains two classes of solutions.  One
  is with $\eta^{(3\eta)}$ and $\lambda^{(3\eta)}$ which obey the
  following boundary conditions:
  \begin{align}
    \eta^{(3\eta)} (r\to0) \simeq r^\ell I_G, ~~~
    \lambda^{(3\eta)} (r\to0) \simeq 0.
    \label{Type3-1}
  \end{align}
  The other is with $\eta^{(3\lambda)}$ and $\lambda^{(3\lambda)}$
  whose boundary conditions are given by
  \begin{align}
    \eta^{(3\lambda)} (r\to0) \simeq 0,~~~
    \lambda^{(3\lambda)} (r\to0) \simeq r^\ell V_H(r\rightarrow 0).
    \label{Type3-2}
  \end{align}
  Combining two classes of solutions, we define $(2n_G+n_H)\times
  (n_G+n_H)$ object $\Psi_{\ell}^{(SL\varphi)(3)}$ whose behavior at
  $r\to 0$ is given by
  \begin{align}
    \Psi_{\ell}^{(SL\varphi)(3)}(r\to0) \simeq
    \begin{pmatrix}
      \eqmakebox[tagPhi3origF]{$\frac{\ell}{4L}r^{\ell+1}I_G$}&0\\
      \eqmakebox[tagPhi3origF]{$\frac{\ell+4}{4(\ell+2)}r^{\ell+1}I_G$}&0\\
      \coolunder{tagPhi3origF}{n_G}{-\frac{\ell+2}{L}M(M^T M)^{-1}r^\ell}
      &\coolunder{tagPhi3origFH}{n_H}{r^\ell V_H}
    \end{pmatrix}
    \begin{matrix*}[l]
      \coolright{n_G}\\
      \coolright{n_G}\\
      \coolright{n_\varphi}
    \end{matrix*}.
    \label{BC_slphi3}
    \\[-2ex]\nonumber
  \end{align}

  In our convention, because $M^T\bar{\phi}'=0$, one of $u_p$ is chosen
      to be proportional to
  $\bar{\phi}'$, which
  we denote as $u^{\rm (tr)}$:
  \begin{align}
    u^{\rm (tr)} \equiv \frac{\bar{\phi}'}{\sqrt{\bar{\phi}'^T\bar{\phi}'}}.
  \end{align}
  For $\ell=1$, it is related to the translational zero mode, and is
  important in eliminating the zero eigenvalues from the functional
  determinant.  The column of $\Psi_{1}^{(SL\varphi)(3)}$ in
  association with $u^{\rm (tr)}$ is obtained by $\chi^{\rm
    (tr)}=\eta^{\rm (tr)}=0$ and $\lambda^{\rm
    (tr)}\propto\bar{\phi}'$.  We will come back to this issue in
  Section~\ref{sec_zeromodes}.
\end{itemize}

The asymptotic behavior at $r\rightarrow\infty$ can be understood by
carefully observing the differential equation.  We leave the precise
discussion to Appendix~\ref{apx_calc} and, in this section, only show
the results.  Neglecting terms irrelevant for the calculation of
the functional determinant of our interest, the columns of
$\Psi_{\ell}^{(SL\varphi)}(r\rightarrow\infty)$ can be obtained by
linear combinations of those of the following objects:
\begin{align}
  \tilde{\psi}^{(1)} = &\,
  \begin{pmatrix}
    \eqmakebox[tagPhi1infF]{$\ell r^{\ell-1} I_G$}\\
    \eqmakebox[tagPhi1infF]{$Lr^{\ell-1} I_G$}\\
    \coolunder{tagPhi1infF}{n_G}{M r^\ell}
  \end{pmatrix}
  \begin{matrix*}[l]
    \coolright{n_G}\\
    \coolright{n_G}\\
    \coolright{n_\varphi}
  \end{matrix*}\hspace{2ex},
  \\[5ex]
  \tilde{\psi}^{(2)} = &\,
  \begin{pmatrix}
    \eqmakebox[tagPhi2infF]{$-\frac{\xi}{4}r^{\ell+1}I_B$}
    &\eqmakebox[tagPhi2infF2]{$0$}\\
    \eqmakebox[tagPhi2infF]{$0$}&\eqmakebox[tagPhi2infF2]{$\frac{\ell-(\ell+2)\xi}{4(\ell+2)}r^{\ell+1}I_U$}\\
    \eqmakebox[tagPhi2infF]{$-\frac{\xi\ell}{4L}r^{\ell+1}I_B$}&\eqmakebox[tagPhi2infF2]{$0$}\\
    \eqmakebox[tagPhi2infF]{$0$}&\eqmakebox[tagPhi2infF2]{$\frac{\ell[(\ell+4)-(\ell+2)\xi]}{4L(\ell+2)} r^{\ell+1}I_U$}\\
    \coolunder{tagPhi2infF}{n_B}{-\frac{\xi}{4(\ell+2)}r^{\ell+2}M_B}&\coolunder{tagPhi2infF2}{n_U}{0}
  \end{pmatrix}
  \begin{matrix*}[l]
    \coolright{n_B}\\
    \coolright{n_U}\\
    \coolright{n_B}\\
    \coolright{n_U}\\
    \coolright{n_\varphi}
  \end{matrix*}\hspace{2ex},
%
  %
  \\[5ex]
  \tilde{\psi}^{(3B)} = &
  \begin{pmatrix}
    \eqmakebox[tagPhi3-1infF]{$0$}\\
    \eqmakebox[tagPhi3-1infF]{$(M^TM)^{-1} \partial_r \tilde{\eta}^{(B)}$}\\
    \coolunder{tagPhi3-1infF}{n_B}{0}
  \end{pmatrix}
  \begin{matrix*}[l]
    \coolright{n_G}\\
    \coolright{n_G}\\
    \coolright{n_\varphi}
  \end{matrix*}\hspace{2ex},
  \\[5ex]
  \tilde{\psi}^{(3U)} = &
  \begin{pmatrix}
    \eqmakebox[tagPhi3-1pinfF]{$0$}\\
    \eqmakebox[tagPhi3-1pinfF]{$0$}\\
    \coolunder{tagPhi3-1pinfF}{n_U}{LM[\partial_r (M^TM)^{-1}]r^{-1} \tilde{\eta}^{(U)}}
  \end{pmatrix}
  \begin{matrix*}[l]
    \coolright{n_G}\\
    \coolright{n_G}\\
    \coolright{n_\varphi}
  \end{matrix*}\hspace{2ex},
  \\[5ex]
  \tilde{\psi}^{(3\lambda)} = &
  \begin{pmatrix}
    \eqmakebox[tagPhi3-2infF]{$0$}\\
    \eqmakebox[tagPhi3-2infF]{$0$}\\
    \coolunder{tagPhi3-2infF}{n_H}{\tilde{\lambda}}
  \end{pmatrix}.
  \\[-2ex] \nonumber
\end{align}
Here, $I_B$ and $I_U$ are the $n_B \times n_B$ and $n_U \times n_U$
unit matrices, respectively, while $\tilde{\eta}^{(B)}$ and
$\tilde{\eta}^{(U)}$ are $n_G\times n_B$ and $n_G\times n_U$ objects
both of which satisfy
\begin{align}
  \left\{ -\Delta_\ell + (M^T M) -
  (M^T M) [\partial_r (M^T M)^{-1}] \frac{1}{r^2} \partial_r r^2
  \right\} \tilde{\eta} = 0.
\end{align}
Among the solutions, $\tilde{\eta}^{(B)}$ corresponds to $n_B$
solutions that exponentially grow at $r\to\infty$, while
$\tilde{\eta}^{(U)}$ corresponds to $n_U$ solutions that behave as
$r^{-2}$ at $r\to\infty$.  In addition, $\tilde{\lambda}$ is
$n_\varphi\times n_H$, satisfying
\begin{align}
  \left( -\Delta_\ell + \Omega \right) \tilde{\lambda} = 0,
  \label{eq_lambda_tilde}
\end{align}
and
\begin{align}
  M^T \tilde{\lambda} = 0.
\end{align}

Then, $\Psi_{\ell}^{(SL\varphi)} (r\rightarrow\infty)$
can be expressed as
\begin{align}
  \Psi_{\ell}^{(SL\varphi)} (r\rightarrow\infty) =
  \left(
  \tilde{\psi}^{(1)} ~
  \tilde{\psi}^{(2)} ~
  \tilde{\psi}^{(3B)} ~
  \tilde{\psi}^{(3U)} ~
  \tilde{\psi}^{(3\lambda)}
  \right)
  \left(
  \begin{array}{ccc}
    I_G & 0 & 0 \\
    0 & I_G & 0 \\
    0 & 0 & \tau^{(\eta\lambda)}
  \end{array}
  \right)+(\text{irrelevant}),
\end{align}
with $\tau^{(\eta\lambda)}$ being an $n_\varphi\times n_\varphi$
orthogonal matrix.  In the above expression, elements irrelevant for
the following discussion are neglected.

In order to calculate the prefactor $\mathcal A$, we should also
consider the fluctuation operator around the false vacuum.  In
particular, we need to derive solutions of $\widehat{\mathcal
  M}_\ell^{(SL\varphi)}\widehat{\Psi}_\ell^{(SL\varphi)}=0$.  For this
purpose, we can use the fact that the fluctuation operator around the
false vacuum can be block-diagonalized as Eq.~\eqref{eq_block_ell};
the blocks are for the fluctuations of massive gauge bosons and NG
bosons, for massless gauge bosons, and for physical scalars.  Thus, we
can discuss their contributions separately.  Similarly to the
discussion above, we define $\widehat{\Psi}_\ell^{(X)} (X=B,U,\sigma)$
obeying $\widehat{\mathcal{M}}_\ell^{(X)}
\widehat{\Psi}_\ell^{(X)}=0$, which describe independent solutions of
the differential equation.  Note that $\widehat{\Psi}_\ell^{(B)}$,
$\widehat{\Psi}_\ell^{(U)}$, and $\widehat{\Psi}_\ell^{(\sigma)}$ are
$3n_B\times 3n_B$, $2n_U\times 2n_U$, and $(n_U+n_H) \times
(n_U+n_H)$, respectively.

Effects of the fluctuations of massive gauge boson and NG bosons
around the false vacuum are embedded in $\widehat{\Psi}_\ell^{(B)}$,
which behaves as
\begin{align}
  \widehat{\Psi}_\ell^{(B)} (r\rightarrow 0) =
  \begin{pmatrix}
    \eqmakebox[tagPhiB-1r0]{$\ell r^{\ell-1} I_B$}
    & \eqmakebox[tagPhiB-2r0]{$-\frac{\xi}{4}r^{\ell+1}I_B$}
    & \eqmakebox[tagPhiB-3r0]{$\frac{\ell}{4L}r^{\ell+1}I_B$}\\
    \eqmakebox[tagPhiB-1r0]{$Lr^{\ell-1} I_B$}
    & \eqmakebox[tagPhiB-2r0]{$-\frac{\xi\ell}{4L}r^{\ell+1}I_B$}
    & \eqmakebox[tagPhiB-3r0]{$\frac{\ell+4}{4(\ell+2)}r^{\ell+1}I_B$}\\
    \coolunder{tagPhiB-1r0}{n_B}{\widehat{W} r^\ell}
    & \coolunder{tagPhiB-2r0}{n_B}{\widehat{W}^{-1}r^\ell}
    & \coolunder{tagPhiB-3r0}{n_B}{-\frac{\ell+2}{L}\widehat{W}^{-1}r^\ell}
  \end{pmatrix}
  \begin{matrix*}[l]
    \coolright{n_B}\\
    \coolright{n_B}\\
    \coolright{n_B}
  \end{matrix*},
  \\[-2ex] \nonumber
\end{align}
and
\begin{align}
  \widehat{\Psi}_\ell^{(B)} (r\rightarrow\infty) =
    \left(
    \begin{array}{ccc}
      \ell r^{\ell-1} I_B & -\frac{\xi}{4}r^{\ell+1}I_B & 0\\
      Lr^{\ell-1} I_B  & -\frac{\xi\ell}{4L}r^{\ell+1}I_B & \widehat{W}^{-1} \widehat{\Psi}^{(\eta)}_\ell\\
      \widehat{W} r^\ell & -\frac{\xi}{4(\ell+2)}r^{\ell+2}\widehat{W} & 0
    \end{array}
    \right),
\end{align}
where $\widehat{\Psi}^{(\eta)}_\ell$ is an $n_B\times n_B$ function
satisfying
\begin{align}
  \left( -\Delta_\ell + \widehat{W}^T\widehat{W} \right)
  \widehat{\Psi}^{(\eta)}_\ell = 0,
  \label{eq_eta_hat}
\end{align}
with
\begin{align}
  \widehat{\Psi}^{(\eta)}_\ell (r\rightarrow 0) = I_B r^{\ell}.
  \label{bc_eta_hat}
\end{align}
Next, we consider the fluctuations of massless gauge bosons.  For
those, we can obtain the solutions in the following form:
\begin{align}
  \widehat{\Psi}_\ell^{(U)} =
  \begin{pmatrix}
    \eqmakebox[tagPhiU-1r0]{$\ell r^{\ell-1} I_U$}
    & \eqmakebox[tagPhiU-1r0B]{$\frac{(\ell+2)\xi-\ell}{4\ell(\ell+2)}r^{\ell+1} I_U$} \\
    \coolunder{tagPhiU-1r0}{n_U}{Lr^{\ell-1} I_U}
    & \coolunder{tagPhiU-1r0B}{n_U}{\frac{(\ell+2)\xi-(\ell+4)}{4L(\ell+2)} r^{\ell+1} I_U}
  \end{pmatrix}
  \begin{matrix*}[l]
    \coolright{n_U}\\
    \coolright{n_U}
  \end{matrix*}.
  \\\nonumber
\end{align}
Furthermore, the solutions related to the physical scalars are given
by an $(n_U + n_H) \times (n_U + n_H)$ object,
$\widehat{\Psi}_\ell^{(\sigma)}$, whose evolution is governed by
\begin{align}
  \left( -\Delta_\ell + \widehat{m}^2 \right) \widehat{\Psi}_\ell^{(\sigma)} = 0,
  \label{eq_lambda_hat}
\end{align}
with
\begin{align}
  \widehat{\Psi}_\ell^{(\sigma)}(r\to0) =
  \begin{pmatrix}
    I_U & 0\\
    0 & I_H
  \end{pmatrix}
  r^\ell,
  \label{bc_lambda_hat}
\end{align}
where $I_H$ is the $n_H \times n_H$ unit matrix.  Then,
$\det\widehat{\Psi}_\ell^{(SL\varphi)}(r)$ can be calculated as
\begin{align}
  \det\widehat{\Psi}_\ell^{(SL\varphi)} (r) =
  \det \left( \widehat{\Psi}_\ell^{(B)} (r) \right)
  \det \left( \widehat{\Psi}_\ell^{(U)} (r) \right)
  \det \left( \widehat{\Psi}_\ell^{(\sigma)} (r) \right).
\end{align}

By using $\det\Psi_\ell^{(SL\varphi)}$ and
$\det\widehat{\Psi}_\ell^{(SL\varphi)}$, the functional determinant of
our interest can be expressed as
\begin{align}
  \frac{\Det\mathcal M_\ell^{(SL\varphi)}}
       {\Det\widehat{\mathcal M}_\ell^{(SL\varphi)}} =
       \left(
       \frac{\det\Psi_\ell^{(SL\varphi)}(r_0)}
            {\det\widehat{\Psi}_\ell^{(SL\varphi)}(r_0)}
       \right)^{-1}
       \left(
       \frac{\det\Psi_\ell^{(SL\varphi)}(r_\infty)}
            {\det\widehat{\Psi}_\ell^{(SL\varphi)}(r_\infty)}
       \right).
\end{align}
In order to evaluate the above quantity, we define
\begin{align}
  \Psi_\ell^{(\eta\lambda)} =
  \left(
  \begin{array}{cc}
    \eta^{\rm (3\eta)} & \eta^{\rm (3\eta)} \\
    V_H^T \lambda^{\rm (3\lambda)} & V_H^T \lambda^{\rm (3\lambda)}
  \end{array} \right),
\label{Psi(etalambda)}
\end{align}
which can be obtained by solving Eqs.\ \eqref{eq_eta} and
\eqref{eq_lambda} with taking the boundary conditions given in
Eq.\ \eqref{Type3-1} or \eqref{Type3-2}.  Then, by using the following
relation:
\begin{align}
  \det \Psi_\ell^{(\eta\lambda)} (r_\infty) =
  \det \tilde{\eta} (r_\infty)
  \det \left[ V_H^T (r_\infty) \tilde{\lambda} (r_\infty) \right],
\end{align}
we obtain
\begin{align}
  \frac{\Det\mathcal M_\ell^{(SL\varphi)}}
       {\Det\widehat{\mathcal M}_\ell^{(SL\varphi)}} =&
  \left( \frac{\ell}{r_\infty} \right)^{n_U}
  \left( \det M^T M \right)_\infty^{1/2}
  \left[ \det \left(M_U^T M_U \right)' \right]^{-1}_\infty
  \notag \\ &
  \left( \det \Psi_\ell^{(\eta\lambda)} (r_\infty) \right)
  \left( \det \widehat{\Psi}^{(\eta)}_\ell (r_\infty) \right)^{-1}
  \left( \det \widehat{\Psi}_\ell^{(\sigma)} (r_\infty) \right)^{-1}
  \notag \\ &
  \left( \det M_0^T M_0 \right)^{1/2}
  \left( \det \widehat{W}^T \widehat{W} \right)^{-1},
  \label{det_l>1}
\end{align}
where the subscript ``$\infty$'' implies that the quantity should be
evaluated at $r=r_\infty$; one can check that $r_0$ dependence and
$r_\infty$ dependence cancel out after taking $r_0\to 0$ and
$r_\infty\to \infty$. It is important to notice that Eq.~\eqref{det_l>1}
does not contain the gauge parameter, $\xi$.

\subsection{$\ell= 0$}

Formally, the functional determinant of $\mathcal M_0^{(S\varphi)}$
can be obtained by a similar procedure as in the case of $\ell >0$.
The solutions of $\mathcal M_0^{(S\varphi)}\Psi_0^{(S\varphi)(I)}=0$, corresponding to
the Type 1, 2, and 3 solutions, are given in the following form:
\begin{itemize}

\item {\bf Type 1}:
  \begin{align}
    \Psi_0^{(S\varphi)(1)} =
    \begin{pmatrix}
      \eqmakebox[tagPhi0-1orig]{$0$}\\
      \coolunder{tagPhi0-1orig}{n_G}{M}
    \end{pmatrix}
    \begin{matrix*}[l]
      \coolright{n_G}\\
      \coolright{n_\varphi}
    \end{matrix*}.
    \label{psi1(l=0)}
    \\[-2ex]\nonumber
  \end{align}

\item {\bf Type 2}:
  \begin{align}
    \Psi_0^{(S\varphi)(2)} =
    \begin{pmatrix}
      \eqmakebox[tagPhi0-2orig]{$-\frac{\xi}{4} r I_G$}\\
      \coolunder{tagPhi0-2orig}{n_G}{0}
    \end{pmatrix}
    \begin{matrix*}[l]
      \coolright{n_G}\\
      \coolright{n_\varphi}
    \end{matrix*}
    + \mbox{(irrelevant)}\label{psi2(l=0)},
    \\[-2ex]\nonumber
  \end{align}
  where we neglect terms that are irrelevant for the calculation of
  the functional determinant of our interest.

\item {\bf Type 3}:
  \begin{align}
    \Psi_0^{(S\varphi)(3)} =
    \begin{pmatrix}
      \eqmakebox[tagPhi0-3orig]{$0$}\\
      \coolunder{tagPhi0-3orig}{n_H}{\tilde\lambda}
    \end{pmatrix}
    \begin{matrix*}[l]
      \coolright{n_G}\\
      \coolright{n_\varphi}
    \end{matrix*}
    + \mbox{(irrelevant)},
    \label{psi3(l=0)}
    \\[-2ex] \nonumber
  \end{align}
  where the function $\tilde\lambda$ satisfies Eq.\ \eqref{eq_lambda}
  with taking $\eta\rightarrow 0$ and $\zeta\rightarrow 0$, as well as
  $M^T \tilde\lambda=0$.

\end{itemize}
Then, we define
\begin{align}
  \Psi_0^{(S\varphi)} \equiv
  \left( \Psi_0^{(S\varphi)(1)} ~ \Psi_0^{(S\varphi)(2)} ~ \Psi_0^{(S\varphi)(3)} \right),
\end{align}
which will be related to the functional determinant of the fluctuation
operator around the bounce.

Recalling the block-diagonalization shown in
Eq.~\eqref{eq_block_0}, we can solve $\widehat{\mathcal{M}}_0^{(X)}
\widehat{\Psi}_0^{(X)} =0$ ($X=B$ and $U$) to obtain
\begin{align}
  \widehat{\Psi}_0^{(B)}(r) &=
  \begin{pmatrix}
    \eqmakebox[tagPhil=0B-1]{$0$} & \eqmakebox[tagPhil=0B-2]{$-\frac{1}{4}\xi r I_B$}
    \\
    \coolunder{tagPhil=0B-1}{n_B}{\widehat{W}} &
    \coolunder{tagPhil=0B-2}{n_B}{0}
  \end{pmatrix}
  \begin{matrix*}[l]
    \coolright{n_B}\\
    \coolright{n_B}\\
  \end{matrix*},
 \\[-2ex]\notag
\end{align}
and
\begin{align}
\widehat{\Psi}_0^{(U)}(r)=\frac{\xi}{4} r I_U.
\end{align}
We also define $\widehat{\Psi}_0^{(\sigma)}(r)$ that satisfies
\begin{align}
  (\Delta_0 - \widehat{m}^2) \widehat{\Psi}_0^{(\sigma)} = 0,
\end{align}
with
\begin{align}
  \widehat{\Psi}_0^{(\sigma)} (r\to 0) = \begin{pmatrix}
    I_U & 0 \\
    0 & I_H
  \end{pmatrix}.
\end{align}
Using quantities defined above, the determinant is calculated as
\begin{align}
 \det\widehat{\Psi}_0^{(S\varphi)}(r)=
 \det\left(\widehat{\Psi}_0^{(B)}(r)\right)
 \det\left(\widehat{\Psi}_0^{(U)}(r)\right)
 \det\left(\widehat{\Psi}_0^{(\sigma)}(r)\right).
\end{align}

We
obtain the following expression:
\begin{align}
  \frac{\Det \mathcal M_0^{(S\varphi)}}
       {\Det \widehat{\mathcal M}_0^{(S\varphi)}} =
       \left(
       \frac{\det \Psi_{0}^{(S\varphi)}(r_0)}
            {\det \widehat\Psi_{0}^{(S\varphi)}(r_0)}
       \right)^{-1}
       \left(
       \frac{\det \Psi_{0}^{(S\varphi)}(r_\infty)}
            {\det \widehat\Psi_{0}^{(S\varphi)}(r_\infty)}
            \right).
            \label{detM0}
\end{align}
Importantly, however, extra treatment is needed when there exists an
unbroken gauge symmetry at the false vacuum.  If it exists, there shows
up a gauge zero mode, which makes Eq.~\eqref{detM0} vanish.  Indeed, it
is easy to see the existence of zero modes explicitly.  For $\ell=0$,
the Type 1 solution given in Eq.~\eqref{psi1(l=0)} has $n_G$ columns
distinguished by the adjoint index of $M$.  Because
$M_{ia}(r\rightarrow\infty)=0$ if the adjoint index $a$ is for unbroken
generators at the false vacuum, $\det\mathcal M_0^{(S\varphi)}$ vanishes
if there exists an unbroken gauge symmetry at the false vacuum.  In our
setup, the number of gauge zero modes is $n_U$.

If there exist zero modes, a naive calculation of the prefactor ${\cal
  A}$ makes it divergent.  Such a divergence is an artifact arising from
the flat directions of the Euclidean action, implying the break down
of the saddle point method in the path integral.  The proper
treatments of the zero modes will be discussed in
Section~\ref{sec_zeromodes}.

\subsection{Background gauge}

Although it is convenient to use the Fermi gauge to discuss the gauge
invariance of the decay rate, the background gauge is useful in
performing the numerical calculation of the decay rate.  Here, we show
that the calculations based on the Fermi and the background gauges give
the same result for $\ell>0$.  (For $\ell=0$, the zero mode
subtraction is non-trivial in the background gauge.  The treatments of
the zero modes will be discussed in the next section.)

In order to discuss the functional determinants in the background
gauge, we first define the function $\Psi_{\ell,{\rm
    BG}}^{(c\bar{c})}$ which is $n_G\times n_G$.  It obeys the
following differential equation:
\begin{align}
  \left( -\Delta_\ell+\xi M^TM \right) \Psi_{\ell,{\rm BG}}^{(c\bar{c})} = 0,
  \label{Eq_for_fFP}
\end{align}
with the boundary condition
\begin{align}
  \Psi_{\ell,{\rm BG}}^{(c\bar{c})} (r\rightarrow 0) = r^\ell I_G.
  \label{BC_for_fFP}
\end{align}
In our convention, because $M^TM(r\rightarrow\infty)$ reduces to the block-diagonal form
(see Eq.\ \eqref{eq_basis_M}), the relevant part of $\Psi_{\ell,{\rm BG}}^{(c\bar{c})}$ for
our analysis behaves as
\begin{align}
  \Psi_{\ell,{\rm BG}}^{(c\bar{c})} (r\rightarrow \infty) \simeq
  \left( \begin{array}{cc}
    f^{(c\bar{c})}_{\ell,B} & 0 \\
    0 & f^{(c\bar{c})}_{\ell,U}
  \end{array} \right) \tau^{(c\bar{c})},
\end{align}
where $f^{(c\bar{c})}_{\ell,B}$ and $f^{(c\bar{c})}_{\ell,U}$ are $n_B\times n_B$ and
$n_U\times n_U$, respectively, while $\tau^{(c\bar{c})}$ is an
$n_G\times n_G$ orthogonal matrix.  Notice that $f^{(c\bar{c})}_{\ell,B}(r\rightarrow\infty)$ exponentially grows while $f^{(c\bar{c})}_{\ell,U}(r\rightarrow\infty)$ is approximately proportional to
$r^{\ell}$.

Similarly, we also define the function $\widehat{\Psi}_{\ell,{\rm BG}}^{(c\bar{c})}$, obeying
\begin{align}
  \left( -\Delta_\ell+\xi \widehat{M}^T \widehat{M} \right)
  \widehat{\Psi}_{\ell,{\rm BG}}^{(c\bar{c})} = 0,
  \label{Eq_for_fFPhat}
\end{align}
and
\begin{align}
  \widehat{\Psi}_{\ell,{\rm BG}}^{(c\bar{c})} (r\rightarrow 0) = r^\ell I_G.
  \label{BC_for_fFPhat}
\end{align}
Because of the block-diagonal nature of $\widehat{M}^T \widehat{M}$ in
our convention, $\widehat{\Psi}_{\ell,{\rm BG}}^{(c\bar{c})}$ can be expressed as
\begin{align}
  \widehat{\Psi}_{\ell,{\rm BG}}^{(c\bar{c})} =
  \left( \begin{array}{cc}
    \widehat{f}^{(c\bar{c})}_{\ell,B} & 0 \\
    0 & \widehat{f}^{(c\bar{c})}_{\ell,U}
  \end{array} \right),
\end{align}
where $\widehat{f}^{(c\bar{c})}_{\ell,B}$ is an $n_B\times n_B$ object, while
$\widehat{f}^{(c\bar{c})}_{\ell,U}=r^\ell I_U$.  The
functional determinant of the fluctuation operator of the FP ghosts is
then given by
\begin{align}
  \frac{\Det{\cal M}_{\ell,{\rm BG}}^{(c\bar{c})}}{\Det\widehat{\cal M}_{\ell,{\rm BG}}^{(c\bar{c})}}
  =
  \frac{\det \Psi_{\ell,{\rm BG}}^{(c\bar{c})} (r_\infty)}{\det\widehat{\Psi}_{\ell,{\rm BG}}^{(c\bar{c})} (r_\infty)}.
\end{align}

For the calculation of the functional determinant of $\mathcal{M}_{\ell,{\rm
    BG}}^{(SL\varphi)}$, we should derive the solutions of the
following differential equation:
\begin{align}
  {\cal M}_{\ell,{\rm BG}}^{(SL\varphi)} \Psi_{\ell,{\rm BG}}^{(SL\varphi)} = 0.
\end{align}
As in the case of the Fermi gauge, we can find a set of functions that
are relevant for the determinant:
\begin{align}
  \tilde{\psi}_{\rm BG}^{(1)} = &\,
  \begin{pmatrix}
    \eqmakebox[tagpsi1BBG]{$\partial_r f^{(c\bar{c})}_{\ell,B}$}
    &\eqmakebox[tagpsi1UBG]{$0$}\\
    \eqmakebox[tagpsi1BBG]{$0$}
    &\eqmakebox[tagpsi1UBG]{$\ell f^{(c\bar{c})}_{\ell,U}$}\\
    \eqmakebox[tagpsi1BBG]{$0$}
    &\eqmakebox[tagpsi1UBG]{$0$}\\
    \eqmakebox[tagpsi1BBG]{$0$}
    &\eqmakebox[tagpsi1UBG]{$\frac{L}{r}f^{(c\bar{c})}_{\ell,U}$}\\
    \eqmakebox[tagpsi1BBG]{$0$}&\eqmakebox[tagpsi1UBG]{$0$}\\
    \coolunder{tagpsi1BBG}{n_B}{0}&\coolunder{tagpsi1UBG}{n_U}{0}
  \end{pmatrix}
  \begin{matrix*}[l]
    \coolright{n_B}\\
    \coolright{n_U}\\
    \coolright{n_B}\\
    \coolright{n_U}\\
    \coolright{n_B}\\
    \coolright{n_\varphi}
  \end{matrix*},
  \\[5ex]
  \tilde{\psi}_{\rm BG}^{(2)} = &\,
    \begin{pmatrix}
    \eqmakebox[tagpsi2BBG]{$0$}&\eqmakebox[tagpsi2UBG]{$0$}\\
    \eqmakebox[tagpsi2BBG]{$0$}
    &\eqmakebox[tagpsi2UBG]{$\frac{\ell-(\ell+2)\xi}{4(\ell+2)}
      rf^{(c\bar{c})}_{\ell,U}$}\\
    \eqmakebox[tagpsi2BBG]{$0$}&\eqmakebox[tagpsi2UBG]{$0$}\\
    \eqmakebox[tagpsi2BBG]{$0$}&
    \eqmakebox[tagpsi2UBG]{$\frac{\ell[(\ell+4)-(\ell+2)\xi]}{4L(\ell+2)} rf^{(c\bar{c})}_{\ell,U}$}\\
    \coolunder{tagpsi2BBG}{n_B}{M_B (M_B^T M_B)^{-1}f^{(c\bar{c})}_{\ell,B}}&\coolunder{tagpsi2UBG}{n_U}{0}
  \end{pmatrix}
  \begin{matrix*}[l]
    \coolright{n_B}\\
    \coolright{n_U}\\
    \coolright{n_B}\\
    \coolright{n_U}\\
    \coolright{n_\varphi}
  \end{matrix*},
  \\[5ex]
  \tilde{\psi}_{\rm BG}^{(3)} = &\,
  \left( \tilde{\psi}^{(3B)} ~ \tilde{\psi}^{(3U)} ~ \tilde{\psi}^{(3\lambda)} \right),
\end{align}
where terms irrelevant for our discussion are neglected.  With these
functions, $\Psi_{\ell,{\rm BG}}^{(SL\varphi)}(r\rightarrow\infty)$ can
be expressed as
\begin{align}
  \Psi_{\ell,{\rm BG}}^{(SL\varphi)} (r\rightarrow\infty) =
  \left(
  \tilde{\psi}_{\rm BG}^{(1)} ~ \tilde{\psi}_{\rm BG}^{(2)} ~ \tilde{\psi}_{\rm BG}^{(3)}
  \right)
  \left(
  \begin{array}{ccc}
    \tau_{\rm BG}^{(c\bar{c})} & 0 & 0 \\
    0 & \tau_{\rm BG}^{(c\bar{c})} & 0 \\
    0 & 0 & I_\varphi
  \end{array}
  \right)
  \left(
  \begin{array}{ccc}
    I_G & 0 & 0 \\
    0 & I_G & 0 \\
    0 & 0 & \tau_{\rm BG}^{(\eta\lambda)}
  \end{array}
  \right)+(\text{irrelevant}),
\end{align}
with $\tau_{\rm BG}^{(\eta\lambda)}$ being an $n_\varphi\times n_\varphi$
orthogonal matrix.

Meanwhile, the behavior of $\Psi_{\ell,{\rm BG}}^{(SL\varphi)}$ around
$r=0$ is given by
\begin{align}
  \Psi_{\ell,{\rm BG}}^{(SL\varphi)} (r\rightarrow0) \simeq
\begin{pmatrix}
\eqmakebox[tagPhi1origB]{$\ell r^{\ell-1} I_G$}&\eqmakebox[tagPhi2origB]{$0$}&\eqmakebox[tagPhi3origB]{$\frac{\ell-\xi(\ell+2)}{4L}r^{\ell+1}I_G$}&\eqmakebox[tagPhi3origBH]{$0$}\\
 \eqmakebox[tagPhi1origB]{$Lr^{\ell-1} I_G$}&\eqmakebox[tagPhi2origB]{$0$}&\eqmakebox[tagPhi3origB]{$\frac{(\ell+4)-\xi(\ell+2)}{4(\ell+2)}r^{\ell+1}I_G$}&\eqmakebox[tagPhi3origBH]{$0$}\\
 \coolunder{tagPhi1origB}{n_G}{M r^\ell}&\coolunder{tagPhi2origB}{n_G}{M_0(M_0^TM_0)^{-1}r^\ell}&\coolunder{tagPhi3origB}{n_G}{-\frac{\ell+2}{L}M_0(M_0^TM_0)^{-1}r^\ell}&\coolunder{tagPhi3origBH}{n_H}{V_Hr^\ell}
\end{pmatrix}
  \begin{matrix*}[l]
    \coolright{n_G}\\
    \coolright{n_G}\\
    \coolright{n_\varphi}
  \end{matrix*}.
\nonumber\\
\end{align}

Asymptotic behavior of the function $\widehat{\Psi}_{\ell,{\rm
    BG}}^{(SL\varphi)}$, which obeys
\begin{align}
  \widehat{\cal M}_{\ell,{\rm BG}}^{(SL\varphi)}
  \widehat\Psi_{\ell,{\rm BG}}^{(SL\varphi)} = 0,
\end{align}
can be again obtained using the block-diagonalization given in
Eq.~\eqref{eq_block_ell}.  We define
$\widehat{\Psi}_{\ell,\mathrm{BG}}^{(X)} (X=B,U,\sigma)$ through
$\widehat{\mathcal{M}}_{\ell,\mathrm{BG}}^{(X)}
\widehat{\Psi}_{\ell,\mathrm{BG}}^{(X)} = 0$, which describe
independent solutions of the differential equation.  For
$\widehat\Psi_{\ell,{\rm BG}}^{(B)}$, we obtain
\begin{align}
  \widehat\Psi_{\ell,{\rm BG}}^{(B)} (r) =
  \begin{pmatrix}
    \eqmakebox[tagPhiB-1BG]{$\partial_r \widehat{f}_{\ell,B}^{(c\bar{c})}$}
    & \eqmakebox[tagPhiB-2BG]{0}
    & \eqmakebox[tagPhiB-3BG]{0} \\
    \eqmakebox[tagPhiB-1BG]{0}
    & \eqmakebox[tagPhiB-2BG]{0}
    & \eqmakebox[tagPhiB-3BG]{$\widehat{W}^{-2} \partial_r \widehat{\Psi}^{(\eta)}_\ell$} \\
    \coolunder{tagPhiB-1BG}{n_B}{0}
    & \coolunder{tagPhiB-2BG}{n_B}{\widehat{W}^{-1}\widehat{f}_{\ell,B}^{(c\bar{c})}}
    & \coolunder{tagPhiB-3BG}{n_B}{0}
  \end{pmatrix}
  \begin{matrix*}[l]
    \coolright{n_B}\\
    \coolright{n_B}\\
    \coolright{n_B}
  \end{matrix*}.
  \\[-2ex]\nonumber
\end{align}
For the others, we can find
$\widehat{\Psi}_{\ell,\mathrm{BG}}^{(U)} =
\widehat{\Psi}_{\ell}^{(U)}$ and
$\widehat{\Psi}_{\ell,\mathrm{BG}}^{(\sigma)} =
\widehat{\Psi}_{\ell}^{(\sigma)}$.  Combining all of them,
$\det\widehat{\Psi}_{\ell,\mathrm{BG}}^{(SL\varphi)}(r)$ is
given by
\begin{align}
  \det\widehat{\Psi}_{\ell,\mathrm{BG}}^{(SL\varphi)} (r) =
  \det \left( \widehat{\Psi}_{\ell,\mathrm{BG}}^{(B)} (r) \right)
  \det \left( \widehat{\Psi}_{\ell,\mathrm{BG}}^{(U)} (r) \right)
  \det \left( \widehat{\Psi}_{\ell,\mathrm{BG}}^{(\sigma)} (r) \right).
\end{align}

Finally, we can find
\begin{align}
  \frac{\det\Psi_{\ell,{\rm BG}}^{(SL\varphi)}(r_\infty)}{\det\widehat{\Psi}_{\ell,{\rm BG}}^{(SL\varphi)}(r_\infty)}
  = \left(
  \frac{\det \Psi_{\ell,{\rm BG}}^{(c\bar{c})} (r_\infty)}{\det\widehat{\Psi}_{\ell,{\rm BG}}^{(c\bar{c})} (r_\infty)}
  \right)^2
  \frac{\det\Psi_{\ell}^{(SL\varphi)}(r_\infty)}{\det\widehat{\Psi}_{\ell}^{(SL\varphi)}(r_\infty)},
\end{align}
and consequently,
\begin{align}
  \frac{\Det{\cal M}_{\ell,{\rm BG}}^{(SL\varphi)}}{\Det\widehat{\cal M}_{\ell,{\rm BG}}^{(SL\varphi)}}
  = \left(
  \frac{\det \Psi_{\ell,{\rm BG}}^{(c\bar{c})} (r_\infty)}{\det\widehat{\Psi}_{\ell,{\rm BG}}^{(c\bar{c})} (r_\infty)}
  \right)^{2}
  \frac{\Det{\cal M}_{\ell}^{(SL\varphi)}}{\Det\widehat{\cal M}_{\ell}^{(SL\varphi)}}.
\end{align}

In the Fermi and the background gauges, the functional determinants of
the fluctuation operators of the $(SL\varphi)$ modes differ from each
other.  In the calculation of the prefactor ${\cal A}$, the difference
is compensated by the contributions from the FP ghosts.  Indeed, we can
find
\begin{align}
  \left(
  \frac{\Det{\cal M}_{\ell,{\rm BG}}^{(SL\varphi)}}{\Det\widehat{\cal M}_{\ell,{\rm BG}}^{(SL\varphi)}}
  \right)^{-1/2}
  \frac{\Det{\cal M}_{\ell,{\rm BG}}^{(c\bar{c})}}{\Det\widehat{\cal M}_{\ell,{\rm BG}}^{(c\bar{c})}}
  =
  \left(
  \frac{\Det{\cal M}_{\ell}^{(SL\varphi)}}{\Det\widehat{\cal M}_{\ell}^{(SL\varphi)}}
  \right)^{-1/2}
  \frac{\Det{\cal M}_{\ell}^{(c\bar{c})}}{\Det\widehat{\cal M}_{\ell}^{(c\bar{c})}}. \label{eq_gauge_inv_ell}
\end{align}
The above relation guarantees the equivalence of the total functional
determinants (on which the prefactor ${\cal A}$ depends) in the
Fermi and the background gauges for $\ell\geq 1$.

In performing the calculation in the background gauge, we should also study
the $\ell=0$ modes.
The boundary condition for $\ell=0$ (for the study of $\mathcal{M}_{0,{\rm
    BG}}^{(S\varphi)}$) are also obtained as
\begin{align}
  \Psi_{0,\rm BG}^{(S\varphi)(1)}(r\to0) \simeq &\,
  \begin{pmatrix}
    \eqmakebox[tagPhi1origL0B]{$\frac{\xi}{4}r (M_0^TM_0)$}\\
    \coolunder{tagPhi1origL0B}{n_G}{M_0}
  \end{pmatrix}
  \begin{matrix*}[l]
    \coolright{n_G}\\
    \coolright{n_\varphi}
  \end{matrix*},\\
  \nonumber \\
  \Psi_{0,\rm BG}^{(S\varphi)(2)}(r\to0) \simeq &\,
  \begin{pmatrix}
    \eqmakebox[tagPhi2origL0B]{$0$}\\
    \coolunder{tagPhi2origL0B}{n_G}{M_0(M_0^TM_0)^{-1}}
  \end{pmatrix}
  \begin{matrix*}[l]
    \coolright{n_G}\\
    \coolright{n_\varphi}
  \end{matrix*},\\
  \nonumber \\
  \Psi_{0,\rm BG}^{(S\varphi)(3)}(r\to0) \simeq &\,
  \begin{pmatrix}
    \eqmakebox[tagPhi3origL0B]{$0$}\\
    \coolunder{tagPhi3origL0B}{n_G}{V_H}
  \end{pmatrix}
  \begin{matrix*}[l]
    \coolright{n_G}\\
    \coolright{n_\varphi}
  \end{matrix*}.\\ \nonumber
\end{align}
For $\ell=0$, special care is necessary when there
exist gauge zero modes.  In such a case, both $\Det{\cal
  M}_{0}^{(S\varphi)}$ and $\Det{\cal M}_{0,{\rm BG}}^{(S\varphi)}$
vanish and the comparison between the calculations
in the Fermi and the background gauges is non-trivial
(see the discussion in the next section).

\section{Zero Modes}
\label{sec_zeromodes}
\setcounter{equation}{0}

When the action has flat directions around the bounce, mode functions
corresponding to those directions become zero modes of fluctuation
operators.  Here, we consider the zero modes in association with
continuous symmetries of the theory.  In particular, in the case of
our interest, the translational and the gauge symmetries result in
zero modes and their proper treatments are essential to calculate the
prefactor, $\mathcal A$.  In this section, we discuss how we can deal
with these zero modes.

\subsection{General issues}

Let us first discuss the treatment of zero modes in general.  As we have
mentioned in the previous section, the ratio of the functional
determinants of $n\times n$ fluctuation operators, ${\cal M}$ and
$\widehat{\cal M}$,\footnote
{Here, we omit the superscripts and the subscript for notational
  simplicity.}
is evaluated as
\begin{align}
  \frac{\Det{\cal M}}{\Det\widehat{\cal M}}
  =
  \left( \frac{\det\Psi (r_0)}{\det\widehat\Psi (r_0)} \right)^{-1}
  \left(
  \frac{\det\Psi (r_\infty)}{\det\widehat\Psi (r_\infty)}
  \right),
\end{align}
where $\Psi(r)$ and $\widehat\Psi(r)$ satisfy ${\cal M} \Psi = 0$ and
$\widehat{\cal M} \widehat\Psi = 0$, respectively.  Notice that $\Psi$
and $\widehat\Psi$ are $n\times n$ matrices, whose columns are
linearly independent solutions, {\it i.e.},
\begin{align}
  \Psi(r) = &\,
  \left( \psi^{(1)} (r) ~~ \cdots ~~ \psi^{(n)} (r) \right),
  \\
  \widehat\Psi(r) = &\,
  \left( \widehat\psi^{(1)} (r) ~~ \cdots ~~ \widehat\psi^{(n)} (r) \right),
\end{align}
with ${\cal M}\psi^{(I)}=0$ and $\widehat{\cal
  M}\widehat\psi^{(I)}=0$.  When ${\cal M}$ has zero eigenvalues,
$\det\Psi(r_\infty)=0$ and some of $\psi^{(I)}(r_\infty)$'s are not
independent. Here, the number of the dependent columns
of $\Psi(r_\infty)$
matches that of
the zero modes, which we denote as $n_{\rm zero}$. In this
subsection,
we take a basis in which
\begin{align}
  \psi^{(I \leq n_{\rm zero})} (r_\infty) =  0,~
  \psi^{(I > n_{\rm zero})} (r_\infty) \neq 0.
\end{align}
(In the following subsections, we may use a different convention.)

The functional determinant is due to the use of the saddle
point method in the path integral.  To evaluate the path integral, we
first expand the fields around the bounce configuration.  Let us denote
the fields as $\Phi$; in the case of our interest, $\Phi$ contains the
gauge and scalar fields, {\it i.e.}, $\Phi\ni(A_\mu,\phi)$.  Then, we
expand $\Phi$ around the bounce configuration (which is denoted as
$\bar{\Phi}$) as
\begin{align}
  \Phi
  =
  \bar{\Phi}
  + \sum_a c_a \mathcal G_a(x),
\end{align}
where $c_a$'s are expansion coefficients and $\mathcal G_a(x)$'s denote
the eigenfunctions of the fluctuation operator:
\begin{align}
  \mathcal M \mathcal G_a(x) = \omega_a\mathcal G_a(x),
\end{align}
with $\omega_a$'s being the corresponding eigenvalues.  The
eigenfunctions should satisfy $\mathcal G_a(r_\infty)=0$, and are
normalized as\footnote
{The inner product is defined as
  \begin{align*}
    \langle\mathcal G_a|\mathcal G_b\rangle
    =\int d^4x\mathcal G_a\cdot\mathcal G_b.
  \end{align*}
}
\begin{align}
 \langle\mathcal G_a|\mathcal G_b\rangle = \delta_{ab}.
\end{align}
Then, the path integral is evaluated as
\begin{align}
  \int \mathcal D \Phi e^{-S[\Phi]}
  \simeq
  e^{-S[\bar\Phi]}
  \int\left(\prod_a\frac{dc_a}{\sqrt{2\pi}}
  e^{-\frac{1}{2}\omega_a c_a^2}
  \right).
\end{align}
By performing the Gaussian integrals, we obtain
\begin{align}
  \int \mathcal D \Phi e^{-S[\Phi]}
  \simeq
  e^{-S[\bar\Phi]}
  \left( \prod_a \omega_a \right)^{-1/2}
  \equiv
  e^{-S[\bar\Phi]}
  \left(\Det \mathcal M \right)^{-1/2}.
  \label{eq_path_integ}
\end{align}
If there exist zero modes, some of $\omega_a$'s vanish and the above
result diverges.  In such a case, we cannot use the naive saddle point
method to evaluate the path integral.

When the zero mode is related to the symmetry of the theory, we may
properly eliminate the zero modes and avoid the divergence of the
transition amplitude mentioned above, as discussed in
\cite{Callan:1977pt} for the case of translational zero modes.  Let us
denote a generic symmetry transformation of the bounce configuration as
\begin{align}
  \bar{\Phi}
  \to
  \bar{\Phi}+\sum_{A}z_A\mathcal F_A(x)+\mathcal O(z^2),
\end{align}
where $z_A$ denotes the transformation parameter.  With
the symmetry transformation, the bounce action is invariant so that
$\mathcal F_A$ satisfies
\begin{align}
  \mathcal M \mathcal F_A =0,
\end{align}
with
\begin{align}
  \lim_{r\to\infty} \mathcal F_A(x)=0.
\end{align}
Thus, the following relation holds
\begin{align}
  \sum_{a\in {\cal I}_0} c_a \mathcal G_a(x)
  = \sum_{A}z_A\mathcal F_A(x),
\end{align}
where ${\cal I}_0$ denotes the set of indices for the zero modes,
resulting in
\begin{align}
  c_a = \sum_A \langle {\cal G}_a | {\cal F_A} \rangle z_A
  ~:~~~
  a\in {\cal I}_0.
\end{align}
Using
\begin{align}
  \langle {\cal F}_A | {\cal F}_B \rangle =
  \sum_{a\in {\cal I}_0}
  \langle {\cal F}_A | {\cal G}_a \rangle
  \langle {\cal G}_a | {\cal F}_B \rangle,
\end{align}
the Jacobian to convert the variable $c_a$ to $z_A$ is found to be
\begin{align}
  {\cal J} = \sqrt{\det_{AB} \langle {\cal F}_A | {\cal F}_B \rangle},
\end{align}
where $\det_{AB}$ denotes the determinant of the matrix with
the indices $A$ and $B$.

Based on the above argument, we reinterpret the path integral
containing zero modes related to the symmetry as
\begin{align}
  \int \mathcal D \Phi e^{-S[\Phi]}
  \rightarrow
  e^{-S[\bar\Phi]}
  \int
  \left( \prod_A \frac{dz_A}{\sqrt{2\pi}} \right)
  {\cal J}
  \left( \Det' {\cal M} \right)^{-1/2},
\end{align}
where $\Det' {\cal M}$ denotes the minor determinant of
${\cal M}$ with its zero eigenvalues eliminated:
\begin{align}
  \Det' {\cal M} \equiv \prod_{a\not\in {\mathcal I_0}} \omega_a.
\end{align}
With properly interpreting the integrals over $z_A$'s,
the divergence originating from the zero modes may be avoided.

The minor determinant, $\Det' {\cal M}$, can be calculated by
regulating the fluctuation operator as
\begin{align}
  \mathcal M_{\rm reg} \equiv \mathcal M + \nu,\label{eq_regularize_zero}
\end{align}
where $\nu$ is a constant.  With $\nu$ being small enough,
non-zero eigenvalues of
$\mathcal M$ are (almost) unchanged while the
zero eigenvalues are lifted by $\nu$.  Consequently,
\begin{align}
  \Det' {\cal M} =
  \lim_{\nu\rightarrow 0} \frac{1}{\nu^{n_{\rm zero}}}
  \Det \mathcal M_{\rm reg}.
\end{align}

We can calculate $\Det \mathcal M_{\rm reg}$ by using the procedure
mentioned above.  We can obtain a solution of $\mathcal M_{\rm
reg}\Psi_{\rm reg}=0$, where $\Psi_{\rm reg}$ is $n\times n$ and has $n$
linearly independent columns:
\begin{align}
  \Psi_{\rm reg} (r) =
  \left( \psi_{\rm reg}^{(1)} (r) ~~ \cdots ~~ \psi_{\rm reg}^{(n)} (r) \right).
\end{align}
Then, $\Det \mathcal M_{\rm reg}$ can be obtained by using
$\Psi_{\rm reg}$.

Because our purpose is to evaluate $\det\Psi_{\rm reg}
(r_\infty)$ up to $O(\nu^{n_{\rm zero}})$, we only need to calculate
the columns in association with the zero eigenvalues up to $O(\nu)$;
for the other columns, we can take $\psi_{\rm reg}^{(I)}\simeq\psi^{(I)}$
($I>n_{\rm zero}$).  We can calculate $\psi_{\rm reg}^{(I)}$ related
to the zero modes by treating $\nu$ as a perturbation.  We introduce
the function $\check{\psi}^{(I)}$ as
\begin{align}
  \psi_{\rm reg}^{(I\leq n_{\rm zero})}
  =
  \psi^{(I\leq n_{\rm zero})} + \nu \check{\psi}^{(I\leq n_{\rm zero})} + O(\nu^2),
\end{align}
where the superscript ``$(I\leq n_{\rm zero})$'' indicates the columns
in association with the zero modes.  Then, $\check{\psi}^{(I\leq
  n_{\rm zero})}$ should satisfy
\begin{align}
  {\cal M} \check\psi^{(I\leq n_{\rm zero})} =
  - \psi^{(I\leq n_{\rm zero})},
\end{align}
with $\psi^{(I\leq n_{\rm zero})}(r_0)=0$.
With $\check\psi^{(I\leq n_{\rm zero})}$ being obtained by solving the
above equation, we can take care of the
zero modes as
\begin{align}
  \left[ \frac{\Det{\cal M}}{\Det\widehat{\cal M}} \right]^{-1/2}
  \rightarrow
  \int
  \left( \prod_A \frac{dz_A}{\sqrt{2\pi}} \right)
  {\cal J}
  \left[ \frac{\Det'{\cal M}}{\Det\widehat{\cal M}} \right]^{-1/2},
\end{align}
where
\begin{align}
  \frac{\Det'{\cal M}}{\Det\widehat{\cal M}} =
  \left[ \frac{\det\Psi (r_0)}{\det\widehat\Psi (r_0)} \right]^{-1}
  \lim_{\nu\rightarrow 0} \frac{1}{\nu^{n_{\rm zero}}}
  \left[
    \frac{\det \{\Psi (r_\infty)+\nu\check\Psi (r_\infty) \}}{\det\widehat\Psi (r_\infty)}
  \right],
\end{align}
with $\check\Psi$ being $n\times n$ function
containing $\check{\psi}^{(I\leq n_{\rm zero})}$:
\begin{align}
  \check\Psi \equiv
  \left(
  \check\psi^{(1)} ~~ \cdots ~~ \check\psi^{(n_{\rm zero})} ~~
  0 ~~ \cdots ~~ 0
  \right).
\end{align}
Then, using the fact that
\begin{align}
  \Psi (r_\infty)+\nu\check\Psi (r_\infty) =
  \left(
  \nu\check\psi^{(1)} (r_\infty) ~~ \cdots ~~ \nu\check\psi^{(n_{\rm zero})} (r_\infty
) ~~
  \psi^{(n_{\rm zero}+1)} (r_\infty) ~~ \cdots ~~ \psi^{(n)} (r_\infty)
  \right),
\end{align}
we obtain
\begin{align}
  \frac{\Det'{\cal M}}{\Det\widehat{\cal M}} =
  \left[ \frac{\det\Psi (r_0)}{\det\widehat\Psi (r_0)} \right]^{-1}
  \left[
    \frac{\det \{\Psi (r_\infty)+\check\Psi (r_\infty) \}}{\det\widehat\Psi (r_\infty)}
  \right].
\end{align}
The actual calculations of $\Det' \mathcal M$ for the gauge and
translational zero modes will be discussed in the following
subsections.

\subsection{Gauge zero modes}

We first consider the gauge zero modes.  As we have mentioned in the
previous section, if a gauge symmetry, which is broken by the bounce, is
restored at the false vacuum, there show up zero modes in $\ell=0$
fluctuation operators.  In this subsection, we apply the discussion in
the previous subsection to the gauge zero modes.

The gauge zero modes can be given in the following form (see Eq.\
\eqref{psi1(l=0)}):
\begin{align}
  \psi^{(A)} =
  \begin{pmatrix}
    0 \\ M
  \end{pmatrix}
  \mathcal U^{(A)},
\end{align}
where $\mathcal U^{(A)}$'s are defined as
\begin{align}
  \begin{pmatrix}
    \mathcal U^{(1)}&\cdots&\mathcal U^{(n_U)}
  \end{pmatrix} \equiv
  \begin{pmatrix}
    0 \\ I_U
  \end{pmatrix}
  \begin{matrix*}[l]
    \coolright{n_B}\\
    \coolright{n_U}
  \end{matrix*}.
\end{align}
Notice that $\psi^{(A)}$ is a column of $\Psi^{(S\varphi)(1)}_0$ and
hence satisfy $\mathcal M_0^{(S\varphi)}\psi^{(A)}=0$.  Since the last
$n_U$ columns of $M$ vanish at $r\rightarrow r_\infty$, $\psi^{(A)}(r_\infty)=0$
and hence $\psi^{(A)}$ are eigenfunctions of $\mathcal
M_0^{(S\varphi)}$ with zero eigenvalues.

In order to follow the prescription given in the previous subsection,
we first calculate the function $\check\psi^{(A)}$ that satisfies
\begin{align}
  \mathcal M_0^{(S\varphi)} \check\psi^{(A)} = - \psi^{(A)},
\end{align}
with the boundary condition of
\begin{equation}
 \check\psi^{(A)}(r_0)=0.
\end{equation}
The function $\check\psi^{(A)}$ can be decomposed as
\begin{align}
  \check\psi^{(A)} =
  \begin{pmatrix}
    \partial_r\check\chi^{(A)}\\
    M\check\chi^{(A)}
  \end{pmatrix}
  +
  \begin{pmatrix}
    -2(M^TM)^{-1}(M')^T \check\lambda^{(A)}\\
    \check\lambda^{(A)}
  \end{pmatrix}
  +
  \begin{pmatrix}
    [\partial_r(M^TM)^{-1}]\check\zeta^{(A)}\\
    M(M^TM)^{-1}\check\zeta^{(A)}
  \end{pmatrix},\label{decomp_gauge_zero}
\end{align}
where
$\check{\chi}^{(A)}$,
$\check{\zeta}^{(A)}$, and
$\check{\lambda}^{(A)}$ evolve as
\begin{align}
  \Delta_0 \check{\chi}^{(A)} =&\,
  \left. \Delta_0 \chi \right|_{
    \chi\rightarrow\check{\chi}^{(A)},\
    \zeta\rightarrow\check\zeta^{(A)},\
    \lambda\rightarrow\check{\lambda}^{(A)}
  },
  \\
  \Delta_0 \check{\zeta}^{(A)} =&\,
  M^T M\mathcal U^{(A)},
  \label{Delta0checkzeta}
  \\
  \Delta_0 \check{\lambda}^{(A)} =&\,
  \left. \Delta_0 \lambda \right|_{
    \chi\rightarrow\check{\chi}^{(A)},\
    \zeta\rightarrow\check\zeta^{(A)},\
    \lambda\rightarrow\check{\lambda}^{(A)}
  },
  \label{Delta0checklambda}
\end{align}
and $\check{\lambda}$ satisfies
\begin{align}
  M^T \check\lambda = 0.
\end{align}

One can solve Eq.\ \eqref{Delta0checkzeta} to obtain
\begin{align}
  \check\zeta^{(A)} (r) = \int_0^r dr_1 r_1^{-3}
  \int_0^{r_1} dr_2 r_2^3 M^T M (r_2)\mathcal U^{(A)}
  + \mathrm{(irrelevant)}.
\end{align}
As $r$ becomes large, $\check\zeta^{(A)} (r)$ becomes constant
because $M^T M\mathcal U^{(A)}$ approaches to zero exponentially.
Then, the relevant part of $\check\psi^{(A)}$ for the calculation of
the prefactor $\mathcal A$ is given by
\begin{align}
  \check\psi^{(A)} (r) =
  \begin{pmatrix}
    0 \\
    M_U {\cal I}_U {\cal X}_U \mathcal U^{(A)}
  \end{pmatrix},
\end{align}
where ${\cal I}_{U}$ and ${\cal X}_{U}^{(A,B)}$ are $n_U\times n_U$
objects whose $(A,B)$ elements are given by
\begin{align}
  {\cal I}_{U}^{(A,B)} (r) \equiv
  \int^r dr r^{-3} \, {\mathcal U^{(A)T}} (M^T M)^{-1} \mathcal U^{(B)},
\end{align}
and
\begin{align}
  {\cal X}_{U}^{(A,B)} \equiv
  \frac{1}{2\pi^2}
  \int d^4 x \, {\mathcal U^{(A)T}} M^T M \mathcal U^{(B)}.
\end{align}
Thus, because
\begin{align}
  {\rm Det}'\mathcal M_0^{(S\varphi)} \propto
  \left.
  \det
  \left( \begin{array}{cccc}
    0 & 0 & -\frac{1}{4}\xi r I_G & 0 \\
    M_B & M_U {\cal I}_U {\cal X}_U & 0 & \tilde\lambda
  \end{array} \right)
  \right|_{r\rightarrow\infty},
\end{align}
with $\tilde\lambda$ being the function introduced in
Eq.\ \eqref{psi3(l=0)},
we can find
\begin{align}
  \frac{{\rm Det}'\mathcal M_0^{(S\varphi)}}
       {{\rm Det} \widehat{\mathcal M}_0^{(S\varphi)}}
  = &\,
  \det_{AB} {\cal X}_U^{(A,B)}
  \det_{AB} {\cal I}_U^{(A,B)}
  \left[ \frac{\det \widehat{\Psi}_0^{(\sigma)}(r_\infty)}{\det \widehat{\Psi}_0^{(\sigma)}(r_0)} \right]^{-1}
  \left[ \frac{\det {\Psi}_0^{(\lambda)}(r_\infty)}{\det {\Psi}_0^{(\lambda)}(r_0)} \right]
  \left[ \frac{\det (M^T M)_\infty}{\det (M_0^T M_0)} \right]^{1/2},
  \label{Det'M0/DetM0hat}
\end{align}
with
\begin{align}
  \Psi_0^{(\lambda)} \equiv V_H^T \tilde{\lambda}.
  \label{Psi0^lambda}
\end{align}

The gauge zero modes are associated with the gauge symmetries that are
restored at the false vacuum; the NG bosons in association with the
global gauge transformations are not eaten by the gauge bosons and
appear as the zero modes.  Thus, the path integration over the gauge
zero modes can be replaced by the integration over the gauge volume of
the unbroken gauge symmetry at the false vacuum.

We parameterize such a global gauge transformation of the bounce
configuration as
\begin{align}
  \bar{\phi} \rightarrow \bar{\phi}
  + \sum_A \theta^A \widetilde T^A \bar{\phi} + \mathcal O(\theta^2),
  \label{eq_global_sym}
\end{align}
where $\widetilde T^A$'s are generators of unbroken gauge symmetry at
the false vacuum with $\widetilde T^A\bar\phi$ being required to be
orthogonal, {\it i.e.},
\begin{equation}
  \int d^4x
  (\widetilde T^A\bar\phi)^T\widetilde T^B\bar\phi = 0
  ~~~(\mbox{for $A\neq B$}).
\end{equation}
Here, the generators $\widetilde T^A$ are introduced to set a
diagonal basis for the integration over the gauge volume, and are given
by linear combinations of the generators of our original choice, {\it
  i.e.},
\begin{align}
  \widetilde T^A = \kappa_{AB} T^B.
  \label{tildeT}
\end{align}
Note that the ambiguity in $\kappa$ is absorbed into the normalization
of the gauge volume ${\cal V}_U$ (see the following discussion).

Since the fluctuation operator in the Fermi gauge is invariant under the
transformation given in Eq.~\eqref{eq_global_sym}, the path integral
over the gauge zero modes can be interpreted as
\begin{align}
  \int \prod_{a} \frac{dc_a^{\rm (gauge)}}{\sqrt{2\pi}}
  \rightarrow
  \mathcal J^{\rm (gauge)}
  \int \prod_A \frac{d\theta^A}{\sqrt{2\pi}}
  \equiv
  \frac{\mathcal J^{\rm (gauge)}}{(2\pi)^{n_U/2}}
  {\cal V}_U,
\end{align}
where $c_a^{\rm (gauge)}$'s are the expansion coefficients in association
with the gauge zero modes, ${\cal V}_U$ is the volume of the moduli
space arising from the spontaneous symmetry breaking, and
\begin{align}
  \mathcal J^{\rm (gauge)} \equiv
  \det \kappa
  \left[ \det_{AB} \int d^4 x (T^A\bar\phi)^T T^B\bar\phi \right]^{1/2}
  =
  \left( \prod_A \frac{g_A^2}{2\pi^2} \right)^{-1/2}
  \det \kappa
  \left( \det_{AB} {\cal X}_U^{(A,B)} \right)^{1/2}.
\end{align}
Then,
using
$\det\widehat{\Psi}_0^{(\sigma)}(r_0)=\det {\Psi}_0^{(\lambda)}(r_0)=1$,
the right-hand side of
Eq.\ \eqref{Det'M0/DetM0hat} becomes
\begin{align}
  \frac{{\rm Det}'\mathcal M_0^{(S\varphi)}}
       {{\rm Det} \widehat{\mathcal M}_0^{(S\varphi)}}
  = &\,
  \left(\frac{1}{2r_\infty^3}\right)^{n_U}
  \left(\prod_A \frac{g_A^2}{2\pi^2} \right)
  \left( \frac{\mathcal J^{(\rm gauge)}}{\det\kappa} \right)^2
  \left( \det W'_U \right)^{-1}_\infty
  \frac{\det\Psi_0^{(\lambda)}(r_\infty)}{\det\widehat\Psi_0^{(\sigma)}(r_\infty)}\left[\frac{\det\widehat W^T\widehat W}{\det(M_0^T M_0)}\right]^{1/2},
\end{align}
and hence
the $\ell=0$ contribution, containing the path integral over the gauge
zero modes, can be written as
\begin{align}
  \int \prod_a \frac{dc_a^{\rm (gauge)}}{\sqrt{2\pi}}
  \left(
  \frac{{\rm Det}'\mathcal M_0^{(S\varphi)}}{{\rm Det} \widehat{\mathcal M}_0^{(S\varphi)}}
  \right)^{-1/2}
  \rightarrow & \,
  \left(\frac{r_\infty^3}{\pi}\right)^{n_U/2}
  \left(\prod_A \frac{g_A^2}{2\pi^2} \right)^{-1/2}
  \mathcal V_U\det \kappa
  \nonumber \\ & \,
  \left( \det W'_U \right)^{1/2}
  \left[\frac{\det\Psi_0^{(\lambda)}(r_\infty)}{\det\widehat\Psi_0^{(\sigma)}(r_\infty)}\right]^{-1/2}
  \left[\frac{\det\widehat W^T\widehat W}{\det(M_0^T M_0)}\right]^{-1/4}.
  \label{gaugezeromode}
\end{align}

Next, we consider the background gauge. The fluctuation operator in
the background gauge, {\it i.e.}, $\mathcal {\cal M}_{0,{\rm
    BG}}^{(S\varphi)}$ given in Eq.\ \eqref{M0(BG)}, also has zero
eigenvalues in association with the unbroken gauge symmetry in the
false vacuum.  The existence of the zero eigenvalues of
$\mathcal {\cal M}_{0,{\rm BG}}^{(S\varphi)}$ can be understood from
the fact that the following relation holds:
\begin{align}
  {\cal M}_{0,{\rm BG}}^{(S\varphi)}
  \begin{pmatrix}
    \partial_r \Psi_0^{(c\bar{c})} \\
    M \Psi_0^{(c\bar{c})}
  \end{pmatrix} = 0,
\end{align}
and that some of the columns of $\Psi_0^{(c\bar{c})}$ become constant at
$r\rightarrow\infty$ if there exists an unbroken gauge symmetry at the
false vacuum (see Eq.\ \eqref{Eq_for_fFP}). The proper treatment of such
zero modes is, however, complicated because the relation between the
path integral over such zero modes and the integration over the gauge
parameter is non-trivial in the background gauge.  Here, we consider a
way to reconstruct the result in the Fermi gauge from that in the
background gauge.  Even though the study in the Fermi gauge is enough
for the gauge invariant formulation of the decay rate, the background
gauge is advantageous for numerically calculating the decay rate.  This
is because the numerical calculation in the background gauge shows a
better convergence at $r\rightarrow\infty$.  Thus, it is desirable to
understand the procedure to transform the results in the background
gauge to those in the Fermi gauge. We will explicitly show such a
transformation in the following.

In discussing the treatment of the gauge zero modes in the background
gauge, we first define
\begin{align}
  \mathcal M_{0,{\rm BG},{\rm reg}}^{(S\varphi)}=&\,
  \mathcal M_{0,{\rm BG}}^{(S\varphi)}+\nu
  \begin{pmatrix}
    \xi^{-1}I_G&0\\
    0&I_\varphi
  \end{pmatrix},
  \label{M0_BG_reg}
  \\
  \widehat{\mathcal M}_{0,{\rm BG},{\rm reg}}^{(S\varphi)}=&\,
  \widehat{\mathcal M}_{0,{\rm BG}}^{(S\varphi)}+\nu
  \begin{pmatrix}
    \xi^{-1}I_G&0\\
    0&I_\varphi
  \end{pmatrix}.
  \label{hatM0_BG_reg}
\end{align}
Here, the regulator is chosen so that the final result becomes simple.
Then, we can find (see Appendix~\ref{apx_calc})
\begin{equation}
  \frac{{\rm Det}'\mathcal M_0^{(S\varphi)}}{{\rm Det} \widehat{\mathcal M}_0^{(S\varphi)}}=
  \frac{1}{\det_{AB} \mathcal K_{AB}}
  \left(\prod_A \frac{g_A^2}{2\pi^2} \right)
  \left( \frac{\mathcal J^{(\rm gauge)}}{\det\kappa} \right)^2
\left(\frac{{\rm Det}\mathcal M_{0,{\rm BG}}^{(c\bar c)}}{{\rm Det}\widehat{\mathcal M}_{0,{\rm BG}}^{(c\bar c)}}\right)^{-2}
\lim_{\nu\to0}\frac{1}{\nu^{n_U}}\frac{{\rm Det}\mathcal M_{0,{\rm BG},{\rm reg}}^{(S\varphi)}}{{\rm Det}\widehat{\mathcal M}_{0,{\rm BG},{\rm reg}}^{(S\varphi)}},
\label{Det'M0(BG)}
\end{equation}
where
\begin{equation}
  \mathcal K_{AB}=
  \lim_{r\to\infty}
  \frac{r^3}{\xi}\mathcal U^{(A)T}
  \left(\partial_r \Psi_0^{(c\bar{c})} \right)
  \left( \Psi_0^{(c\bar{c})} \right)^{-1}\mathcal U^{(B)}.
  \label{K_AB}
\end{equation}

In numerical calculations, Eq.\ \eqref{Det'M0(BG)} can be used to
obtain the Fermi gauge result from the calculation with the fluctuation
operators in the background gauge.  Here, one should note that, even
though our regulator for the background gauge becomes equal to that of
Eq.~\eqref{eq_regularize_zero} when $\xi=1$,
\begin{equation}
  \left.\left(\frac{{\rm Det}'\mathcal M_{0,{\rm BG}}^{(S\varphi)}}{{\rm Det}\widehat{\mathcal M}_{0,{\rm BG}}^{(S\varphi)}}\right)^{-1/2}\frac{{\rm Det}\mathcal M_{0,{\rm BG}}^{(c\bar c)}}{{\rm Det}\widehat{\mathcal M}_{0,{\rm BG}}^{(c\bar c)}}\right|_{\xi=1}
  \neq
  \left(\frac{{\rm Det}'\mathcal M_0^{(S\varphi)}}
       {{\rm Det} \widehat{\mathcal M}_0^{(S\varphi)}}\right)^{{-1/2}}\frac{{\rm Det}\mathcal M_{0}^{(c\bar c)}}{{\rm Det}\widehat{\mathcal M}_{0}^{(c\bar c)}}.
       \label{eq_gauge_inv_0}
\end{equation}
The above inequality is not surprising because the gauge
transformations relating different bounce configurations differ in two
choices of the gauge fixing.  In particular, in the background gauge,
different bounces are related  by a local, not global, gauge transformation
\cite{Endo:2017tsz}.  Thus, if we worked only in the background gauge,
it would become very non-trivial to relate the path integral over the
zero modes to the integration over the gauge volume.

\subsection{Translational zero modes}

Next, let us consider the zero modes due to the translational
invariance.  These zero modes are related to
\begin{align}
  \psi^{\rm (tr)} \equiv
  \frac{r_0}{\sqrt{\bar{\phi}'^T(r_0) \bar{\phi}'(r_0)}}
  \left( \begin{array}{c}
    0 \\ 0 \\ \bar{\phi}'
  \end{array} \right).
  \label{psi^tr}
\end{align}
One can see that $\psi^{\rm (tr)}$ satisfies ${\mathcal
  M}_1^{(SL\varphi)}\psi^{\rm (tr)}=0$ and $\psi^{\rm
  (tr)}(r\rightarrow\infty)=0$.
We note here that $\psi^{\rm (tr)}$ belongs to
the Type 3 solutions (see Eq.\ \eqref{Type3-2}).  It corresponds to
the solution obtained with $\chi^{\rm (tr)}=\eta^{\rm (tr)}=0$ and
$\lambda^{\rm (tr)}\propto\bar{\phi}'$; the relation
$(\Delta_1-\Omega)\bar{\phi}'=0$ holds and that $\lambda^{\rm
  (tr)}\propto\bar{\phi}'$ satisfies the constraint imposed on the function
$\lambda$, {\it i.e.}, $M^T\lambda^{\rm (tr)}=0$ (see
Eq.\ \eqref{eq_bounce_traj}).  The normalization of
$\psi^{\rm (tr)}$ is determined  so that
$\psi^{\rm (tr)}$ can be embedded into $\Psi_{\ell}^{(SL\varphi)(3)}$
(see Eq.\ \eqref{BC_slphi3}).

We need to eliminate the zero
eigenvalues from $\Det{\mathcal M}_1^{(SL\varphi)}$.
For the calculation of $\Det'{\mathcal M}_1^{(SL\varphi)}$, we
should obtain the function satisfying
\begin{align}
  {\mathcal M}_1^{(SL\varphi)} \check\psi^{\rm (tr)} = - \psi^{\rm (tr)}.
  \label{eq_checkpsitr}
\end{align}
The solution of the above equation can be expressed as
\begin{align}
  \check\psi^{\rm (tr)} =
  \begin{pmatrix}
    \partial_r\check{\chi}^{\rm (tr)}\\
    \frac{L}{r}\check{\chi}^{\rm (tr)}\\
    M\check{\chi}^{\rm (tr)}
  \end{pmatrix}
  +
  \begin{pmatrix}
    (M^TM)^{-1}\left[\frac{L}{r}\check{\eta}^{\rm (tr)}-2(M')^T \check{\lambda}^{\rm (tr)}\right]\\
    (M^TM)^{-1}\frac{1}{r^2}\partial_rr^2\check{\eta}^{\rm (tr)}\\
    \check{\lambda}^{\rm (tr)}
  \end{pmatrix},
\end{align}
where $\check{\chi}^{\rm (tr)}$, $\check{\eta}^{\rm (tr)}$, and
$\check{\lambda}^{\rm (tr)}$ are functions satisfying
\begin{align}
  \Delta_1 \check{\chi}^{\rm (tr)} =&\,
  \left. \Delta_1 \chi \right|_{
    \chi\rightarrow\check{\chi}^{\rm (tr)},\
    \eta\rightarrow\check{\eta}^{\rm (tr)},\
    \lambda\rightarrow\check{\lambda}^{\rm (tr)},\
    \zeta\rightarrow0},
  \\
  \Delta_1 \check{\eta}^{\mathrm{(tr)}} =&\,
  \left. \Delta_1 \eta \right|_{
    \chi\rightarrow\check{\chi}^{\rm (tr)},\
    \eta\rightarrow\check{\eta}^{\rm (tr)},\
    \lambda\rightarrow\check{\lambda}^{\rm (tr)},\
    \zeta\rightarrow0},
  \\
  \Delta_1 \check{\lambda}^{\mathrm{(tr)}} =&\,
  \left. \Delta_1 \lambda \right|_{
    \chi\rightarrow\check{\chi}^{\rm (tr)},\
    \eta\rightarrow\check{\eta}^{\rm (tr)},\
    \lambda\rightarrow\check{\lambda}^{\rm (tr)},\
    \zeta\rightarrow0}
  - \frac{r_0}{\sqrt{\bar{\phi}'^T(r_0) \bar{\phi}'(r_0)}}\bar{\phi}'.
\end{align}
By solving the above equations, $\check{\chi}^{\rm (tr)}$,
$\check{\eta}^{\rm (tr)}$, and $\check{\lambda}^{\rm (tr)}$ can be
obtained.  Then, we can evaluate $\check\psi^{\rm
  (tr)}(r\rightarrow\infty)$, with which ${\rm Det}'{\mathcal
  M}_1^{(SL\varphi)}$ can be calculated.

The bounce is localized in space-time, and the shift of the position
of the bounce does not change the action.  Thus, in four space-time
dimensions, there are four translational zero modes which can be
parameterized as
\begin{align}
  \begin{pmatrix}
    \bar A_\mu\\
    \bar\phi(x)
  \end{pmatrix}
  \to
  \begin{pmatrix}
    \bar A_\mu\\
    \bar\phi(x)
  \end{pmatrix}
  + dx_\mu
  \begin{pmatrix}
    0\\
    \partial_\mu\bar\phi
  \end{pmatrix}
  +\mathcal O(dx^2),
\end{align}
or as
\begin{align}
  \begin{pmatrix}
    \bar A_\mu\\
    \bar\phi(x)
  \end{pmatrix}
  \to
  \begin{pmatrix}
    \bar A_\mu\\
    \bar\phi(x)
  \end{pmatrix} +
  c_{(1,m_A,m_B)}^{\rm (tr)}
  \begin{pmatrix}
    0\\
    {\cal N}^{\rm (tr)} \bar{\phi}' {\cal Y}_{(1,m_A,m_B)}
  \end{pmatrix}
  +\mathcal O(c^2),
\end{align}
with ${\cal N}^{\rm (tr)}$ being the normalization constant:
\begin{align}
  {\cal N}^{\rm (tr)} =
  \frac{1}{\sqrt{\int dr r^3 \bar{\phi}'^2}}.
\end{align}
The path integral over the translational zero modes can be understood
as the integration over the position of the center of the bounce as
\cite{Callan:1977pt}:
\begin{align}
  \prod_{m_A, m_B} \frac{dc_{(1,m_A,m_B)}^{\rm (tr)}}{\sqrt{2\pi}}
  \rightarrow
  \frac{{\cal B}^2}{4\pi^2} \prod_\mu dx_\mu.
\end{align}
Consequently, the functional determinant from the path integral is
interpreted as
\begin{align}
  \left[ \frac{{\rm Det} {\mathcal M}_1^{(SL\varphi)}}{{\rm Det} \widehat{\mathcal M}_1^{(SL\varphi)}} \right]^{-1/2}
  \rightarrow
  \frac{{\cal B}^2}{4\pi^2}
  \left[ \frac{{\rm Det}' {\mathcal M}_1^{(SL\varphi)}}{{\rm Det} \widehat{\mathcal M}_1^{(SL\varphi)}} \right]^{-1/2}
  \prod_\mu dx_\mu.
  \label{trzeromode}
\end{align}
Integration over the space-time volume will disappear from the
expression of the decay rate per unit volume ({\it i.e.}, the
transition probability per unit time and unit volume).  We note
that, in the background gauge, we can obtain
Eq.\ \eqref{trzeromode} with the fluctuation operators in the Fermi
gauge being replaced by those in the background gauge.

\section{Semi-Analytic Expression of the Decay Rate}
\label{sec_results}
\setcounter{equation}{0}

Now, we summarize the semi-analytic expression of the decay rate of the
 false vacuum.  We are particularly interested in the expression of the
 prefactor ${\cal A}$ and its gauge invariance.  The prefactor ${\cal
 A}$ can be given as a product of the contributions of various fluctuations
 as
\begin{align}
  {\cal A} =
  {\cal A}^{(c\bar{c})}
  {\cal A}^{(T)}
  {\cal A}^{(SL\varphi)},
\end{align}
where ${\cal A}^{(c\bar{c})}$, ${\cal A}^{(T)}$, and ${\cal
  A}^{(SL\varphi)}$ denote the contributions of the FP ghosts,
  transverse mode of the gauge fields, and other bosonic fluctuations,
  respectively.  (If there exist fermions coupled to the bounce, their
  contributions should be also taken into account.)  Furthermore,
  because of the spherical nature of the bounce configuration, each
  contribution can be decomposed into those with different $(\ell, m_A, m_B)$:
\begin{align}
  {\cal A}^{(X)} = \prod_\ell \left[ {\cal A}^{(X)}_\ell \right]^{(\ell+1)^2},
\end{align}
with $(X)=(c\bar{c})$, $(T)$, or $(SL\varphi)$.  (Here, ${\cal A}^{(SL\varphi)}_0$
  should be understood as ${\cal A}^{(S\varphi)}_0$.)  For ${\cal
  A}^{(c\bar{c})}$ and ${\cal A}^{(SL\varphi)}$, $0\leq\ell<\infty$ and
  for ${\cal A}^{(T)}$, $1\leq\ell<\infty$.  The final result ${\cal A}$
  does not depend on the choice of the gauge fixing (as far as the gauge
  zero modes are irrelevant), although ${\cal A}^{(c\bar{c})}$ and
  ${\cal A}^{(SL\varphi)}$ in the Fermi and the background gauges differ
  from each other.  In the following, we first summarize the results in the
  Fermi gauge and then discuss those in the background gauge.

\subsection{Contributions of FP ghosts and transverse modes}
\label{sec_FP_and_T}

First, we consider the contributions of the FP ghosts.  In the Fermi
gauge, the FP ghosts do not couple to the bounce, and hence the
fluctuation operator of FP ghosts does not contain $\bar{\phi}$.
Consequently,
\begin{align}
  {\cal A}^{(c\bar{c})} = 1.
\end{align}

For the contributions of the transverse modes of the gauge fields,
we should solve
\begin{align}
  \mathcal M_\ell^{(T)} \Psi^{(T)}_\ell=0,\\
  \widehat{\mathcal M}_\ell^{(T)} \widehat\Psi^{(T)}_\ell=0,
\end{align}
where $\mathcal M_\ell^{(T)}$ and $\widehat{\mathcal M}_\ell^{(T)}$
are the fluctuation operators given in Eqs.\ \eqref{M^T(bounce)} and
\eqref{M^T(false)}, respectively.  In addition, $\Psi^{(T)}_\ell$ and
$\widehat\Psi^{(T)}_\ell$ are $n_G\times n_G$ objects, and their boundary
conditions are given by
\begin{align}
  \Psi^{(T)}_\ell(r\rightarrow 0) \simeq r^\ell I_G,
  \\
  \widehat\Psi^{(T)}_\ell(r\rightarrow 0) \simeq r^\ell I_G.
\end{align}
With these quantities,
\begin{align}
  \frac{\det \mathcal M_{\ell}^{(T)}}{\det \widehat{\mathcal M}_{\ell}^{(T)}}
  =
  \frac{\det\Psi_\ell^{(T)}(r_\infty)}{\det\widehat\Psi_\ell^{(T)}(r_\infty)}.
\end{align}
Then,
\begin{align}
  {\cal A}^{(T)} = \prod_{\ell=1}^{\infty}
  \left[
    \frac{\det\Psi_\ell^{(T)}(r_\infty)}{\det\widehat\Psi_\ell^{(T)}(r_\infty)}
    \right]^{-(\ell+1)^2/2}.
\end{align}

\subsection{Contributions of $(SL\varphi)$ modes}

Contributions of the $(SL\varphi)$ modes are complicated especially
because of the zero modes.  The prefactor originating from the
$(SL\varphi)$ modes can be written as
\begin{align}
  {\cal A}^{(SL\varphi)} =
  {\cal A}^{(S\varphi)}_0
  \prod_{\ell=1}^{\infty}
       \left[ {\cal A}^{(SL\varphi)}_\ell \right]^{(\ell+1)^2}.
\end{align}

For $\ell>1$, we do not expect zero modes.  Based on
Eq.\ \eqref{det_l>1}, we obtain
\begin{align}
  {\cal A}^{(SL\varphi)}_{\ell>1}
  = &\,
  \left( \frac{r_\infty}{\ell} \right)^{n_U/2}
  \left[ \det (M^T M)_\infty \det (M_0^T M_0) \right]^{-1/4}
  \left[ \det \left(1/W_U^T W_U \right)'_\infty \right]^{-1/2}
  \left[ \det \widehat{W}^T \widehat{W} \right]^{1/4}
  \nonumber \\ &\,
  \left[ \det \Psi_\ell^{(\eta\lambda)} (r_\infty) \right]^{-1/2}
  \left[ \det \widehat{\Psi}^{(\eta)}_\ell (r_\infty) \right]^{1/2}
  \left[ \det \widehat{\Psi}^{(\sigma)}_\ell (r_\infty) \right]^{1/2},
\end{align}
where $\Psi_\ell^{(\eta\lambda)}$ is give by Eq.\
\eqref{Psi(etalambda)}, while $\widehat{\Psi}^{(\eta)}_\ell$ and
$\widehat{\Psi}^{(\sigma)}_\ell$ are defined by Eqs.\ \eqref{eq_eta_hat}
and \eqref{bc_eta_hat} and Eqs.\ \eqref{eq_lambda_hat} and
\eqref{bc_lambda_hat}, respectively.

For $\ell=1$, the translational zero modes exist.
The zero eigenvalues from the translational zero modes are removed by
replacing $\eta^{\rm (tr)}$ and $\lambda^{\rm (tr)}$ by
$\check\eta^{\rm (tr)}$ and $\check\lambda^{\rm (tr)}$, respectively,
in $\det\Psi_1^{(SL\varphi)}(r\rightarrow\infty)$, with taking
account of the Jacobian (see Eq.\ \eqref{trzeromode}).  As a result,
${\cal A}^{(SL\varphi)}_{1}$ is obtained as
\begin{align}
  {\cal A}^{(SL\varphi)}_{1}
  = &\,
  \frac{{\cal B}^2}{4\pi^2}
  r_\infty^{n_U/2}
  \left[ \det (M^T M)_\infty \det (M_0^T M_0) \right]^{-1/4}
  \left[ \det \left(1/W_U^T W_U \right)'_\infty \right]^{-1/2}
  \left[ \det \widehat{W}^T \widehat{W} \right]^{1/4}
  \nonumber \\ &\,
  \left[ \det \left( \Psi_1^{(\eta\lambda)} (r_\infty) + \check\Psi_1^{(\eta\lambda)} (r_\infty) \right) \right]^{-1/2}
  \left[ \det \widehat{\Psi}^{(\eta)}_1 (r_\infty) \right]^{1/2}
  \left[ \det \widehat{\Psi}^{(\sigma)}_1 (r_\infty) \right]^{1/2},
\end{align}
where
\begin{align}
  \Psi_1^{(\eta\lambda)} + \check\Psi_1^{(\eta\lambda)}
  =
  \left.
  \Psi_1^{(\eta\lambda)}
  \right|_{\psi^{\rm (tr)}\rightarrow\check\psi^{\rm (tr)}}.
\end{align}
Here, the definitions of $\psi^{\rm (tr)}$ and $\check\psi^{\rm (tr)}$
are given in Eqs.\ \eqref{psi^tr} and \eqref{eq_checkpsitr},
respectively.  Notice that the space-time integral in
Eq.\ \eqref{trzeromode} is eliminated in calculating the decay rate per
unit volume.

For $\ell=0$, there may exist gauge zero modes.  Based on
Eq.\ \eqref{gaugezeromode}, we obtain
\begin{align}
  {\cal A}^{(S\varphi)}_0 = &\,
    \left(\frac{r_\infty^3}{\pi}\right)^{n_U/2}
  \left(\prod_A \frac{g_A^2}{2\pi^2} \right)^{-1/2}
  \mathcal V_U\det \kappa
  \left( \det W'_U \right)^{1/2}
  \left[\frac{\det\Psi_0^{(\lambda)}(r_\infty)}{\det\widehat\Psi_0^{(\sigma)}(r_\infty)}\right]^{-1/2}
  \left[\frac{\det\widehat W^T\widehat W}{\det(M_0^TM_0)}\right]^{-1/4},
\end{align}
where $\mathcal V_U$ is the gauge volume and $\kappa$ is the matrix
orthogonalizing $T^A\bar\phi$ (see Eq.\ \eqref{tildeT}).  In addition,
$\Psi_0^{(\lambda)}$ is given in Eq.\ \eqref{Psi0^lambda}.

\subsection{Background gauge}

Here, we summarize the relation between the results in the Fermi gauge and
in the background gauge.  In the background gauge, the FP contribution
becomes
\begin{equation}
  \mathcal A^{(c\bar c)}_{\rm BG}=
  \prod_{\ell=0}^{\infty}\left[\mathcal A_{\ell,{\rm BG}}^{(c\bar c)}\right]^{(\ell+1)^2}
  =
  \prod_{\ell=0}^{\infty}\left[ \frac{\det \Psi_{\ell,{\rm BG}}^{(c\bar{c})}}{\det \widehat \Psi_{\ell,{\rm BG}}^{(c\bar{c})}} \right]^{(\ell+1)^2}.
\end{equation}
The functions $\Psi_{\ell,{\rm BG}}^{(c\bar{c})}$ and $\widehat
\Psi_{\ell,{\rm BG}}^{(c\bar{c})}$ are defined in
Eqs.\ \eqref{Eq_for_fFP} and \eqref{BC_for_fFP} and
Eqs.\ \eqref{Eq_for_fFPhat} and \eqref{BC_for_fFPhat}, respectively.

Since the fluctuation operators for the $(T)$ modes are identical to
those in the Fermi gauge,
\begin{align}
  \mathcal A^{(T)}_{\ell,{\rm BG}} = \mathcal A^{(T)}_\ell.
\end{align}

As we have seen in Eq.~\eqref{eq_gauge_inv_ell}, the following
relation holds for $\ell>0$:
\begin{equation}
  \mathcal A^{(SL\varphi)}_{\ell,{\rm BG}}\mathcal A_{\ell,{\rm BG}}^{(c\bar c)}
  =
  \mathcal A^{(SL\varphi)}_\ell\mathcal A^{(c\bar c)}_\ell.
\end{equation}
Thus, we can explicitly see that the Fermi and the background gauges give
the same decay rate as far as the modes with $\ell>0$ are concerned.

For $\ell=0$, the proper treatment of the gauge zero modes in the
background gauge is non-trivial, as we have explained in
Section~\ref{sec_zeromodes}.  Thus, in the background gauge, we just
introduce a convenient way of regulating the fluctuation operator with
gauge zero modes and relate its functional determinant with that in
the Fermi gauge.  With the regulated fluctuation operators given in
Eqs.\ \eqref{M0_BG_reg} and \eqref{hatM0_BG_reg}, we can obtain
\begin{align}
  {\cal A}_0^{(S\varphi)} =
  \mathcal V_U
  \left( \det_{AB} {\cal K}_{AB} \right)^{1/2}
  \det \kappa
  \left(\prod_A \frac{g_A^2}{\pi} \right)^{-1/2}
  \left( \frac{{\rm Det}\mathcal M_{0,{\rm BG}}^{(c\bar c)}}{{\rm Det}\widehat{\mathcal M}_{0,{\rm BG}}^{(c\bar c)}} \right)
  \left(
  \lim_{\nu\to0}\frac{1}{\nu^{n_U}}\frac{{\rm Det}\mathcal M_{0,{\rm BG},{\rm reg}}^{(S\varphi)}}{{\rm Det}\widehat{\mathcal M}_{0,{\rm BG},{\rm reg}}^{(S\varphi)}}
  \right)^{-1/2},
  \label{A_0fromBG}
\end{align}
with ${\cal K}_{AB}$ being given in Eq.\ \eqref{K_AB}.
The above equality is useful for numerical calculations: the right-hand
side of Eq.\ \eqref{A_0fromBG} is written only with the quantities
in the background gauge and hence we can calculate them numerically.

\section{Renormalization}\label{sec_renormalization}
\setcounter{equation}{0}

The functional determinants diverge once the contributions of all $\ell$ are
taken into account, and the renormalization is necessary.  Here, we
discuss how the divergences are removed, adopting the $\overline{\rm
  MS}$ scheme.

In calculating the one-loop contribution to the decay rate of the
false vacuum, we are interested in the ratio of the functional
determinants in the following form:
\begin{align}
  \mathcal A^{(X)}=
  \left(\frac{\Det\mathcal M^{(X)}}{\Det\widehat{\mathcal M}^{(X)}}\right)^{-1/2}
  =
  \prod_{\ell=0}^\infty
  \left(\frac{\Det\mathcal M^{(X)}_\ell}{\Det\widehat{\mathcal M}^{(X)}_\ell}\right)^{-(\ell+1)^2/2},
\end{align}
where $\mathcal M^{(X)}$ and $\widehat{\mathcal M}^{(X)}$ are
fluctuation operators around the bounce and that around the false
vacuum, respectively.  Our procedure of the renormalization is as
follows \cite{Isidori:2001bm}:
\begin{itemize}
\item[(i)] We first identify terms which contain the divergence in
  $\ln \mathcal A^{(X)}$ (which we call $\delta\ln\mathcal A^{(X)}_{\rm
  div}$).
\item[(ii)] Next, we perform the $\overline{\rm MS}$ subtraction of
  $\delta\ln\mathcal A^{(X)}_{\rm div}$. The result is denoted as
  $\delta\ln\mathcal A^{(X)}_{\rm div,\overline{\rm MS}}$.
\end{itemize}
Then,  the divergence is removed as \cite{Isidori:2001bm}
\begin{align}
  \ln\mathcal A^{(X)} \rightarrow
  \ln\mathcal A^{(X)}_{\overline{\rm MS}} \equiv
  \ln\mathcal A^{(X)} - \delta \ln\mathcal A^{(X)}_{\rm div}
  + \delta \ln\mathcal A^{(X)}_{\rm div,\overline{\rm MS}}.
\end{align}
Notice that, in the above expression, $\ln\mathcal A^{(X)}$ and
$\delta\ln\mathcal A^{(X)}_{\rm div}$ are both divergent, while
$\ln\mathcal A^{(X)} - \delta\ln\mathcal A^{(X)}_{\rm div}$ and
$\delta\ln\mathcal A^{(X)}_{\rm div,\overline{\rm MS}}$ are finite.

The divergent part of the functional determinant can be obtained by a
diagrammatic consideration.  Let us define
\begin{align}
  \delta \mathcal M \equiv \mathcal M^{(X)}-\widehat{\mathcal M}^{(X)},
\end{align}
and $\delta\mathcal M_\ell\equiv\mathcal
M^{(X)}_\ell-\widehat{\mathcal M}^{(X)}_\ell$.  Then, treating $\delta
\mathcal M$ as a perturbation, $\ln\mathcal A^{(X)}$ can be expressed
as
\begin{align}
 \ln\mathcal A^{(X)} = -\sum_{I=1}^{\infty}s^{(X)}_I,
\end{align}
where $s^{(X)}_I$ indicates the contribution of diagrams with $I$ being
the number of insertions of $\delta \mathcal M$.

Before proceeding further, we comment on the choice of the gauge
fixing which affects the calculation of $\Det{\cal
  M}^{(SL\varphi)}_\ell$.  The calculation can be performed in any
gauge; as we have discussed in Section~\ref{sec_determinant}, the
Fermi and the background gauge calculations give the same result of
$\Det{\cal M}^{(SL\varphi)}_\ell$ for $\ell\geq 1$ and hence the
calculation of the divergent part is possible in either gauge.  In
practice, however, it is convenient to work in the background gauge
because the number of diagrams necessary to calculate the divergent
part is reduced.  This is because, in the background
gauge, $\delta \mathcal M$ does not contain derivative operators; thus, only $s^{(SL\varphi)}_1$ and $s^{(SL\varphi)}_2$
are divergent.  (In the Fermi gauge, in general,
$s^{(SL\varphi)}_{I\leq 4}$ are divergent.)  Thus, we adopt the
background gauge to calculate the divergent part.  In the following,
we consider the case where only $s^{(X)}_1$ and $s^{(X)}_2$ are
divergent while $s^{(X)}_{I\geq 3}$'s are finite.

We have calculated the functional determinants of fluctuation
operators by using the functions satisfying $\mathcal
M^{(X)}_\ell\Psi^{(X)}_\ell=0$ and $\widehat{\mathcal
  M}^{(X)}_\ell\widehat\Psi^{(X)}_\ell=0$.  We can express
$\Psi_\ell^{(X)}$ in the following form:
\begin{align}
  \Psi_\ell^{(X)} =
  \widehat\Psi_\ell^{(X)} +
  \sum_{I=1}^\infty\delta\Psi_{\ell,I}^{(X)},
\end{align}
where the following equations are satisfied:
\begin{align}
  \widehat{\mathcal M}_\ell^{(X)}\delta\Psi_{\ell,1}^{(X)}&=
  -\delta \mathcal M_\ell \widehat{\Psi}_{\ell}^{(X)},\\
  \widehat{\mathcal M}^{(X)}_\ell\delta\Psi_{\ell,I+1}^{(X)}&
  =-\delta \mathcal M_\ell \delta\Psi_{\ell,I}^{(X)} ~~~ (I\geq 1),
\end{align}
with $\delta\Psi_{\ell,I}^{(X)}(r\rightarrow 0)\simeq 0$.
Notice that $\delta\Psi_{\ell,I}^{(X)}$ can be understood as the
term which is of the order of $(\delta \mathcal M)^I$.
Then,
$\ln\mathcal A^{(X)}$ becomes
\begin{align}
  \ln\mathcal A^{(X)}
  &=-\lim_{r\to\infty}\sum_{\ell=0}^\infty \frac{(\ell+1)^2}{2}\left[\tr\ln\left(\widehat\Psi_\ell^{(X)}+\sum_{I=1}^\infty\delta\Psi_{\ell,I}^{(X)}\right)-\tr\ln\widehat\Psi_\ell^{(X)}\right],
\end{align}
from which we obtain
\begin{align}
  s_1^{(X)}&=\lim_{r\to\infty}\sum_{\ell=0}^\infty \frac{(\ell+1)^2}{2}\left[\tr[\widehat\Psi_\ell^{(X)}]^{-1}\delta\Psi_{\ell,1}^{(X)}\right],
  \label{s1(ell)}
  \\
 s_2^{(X)}&=\lim_{r\to\infty}\sum_{\ell=0}^\infty \frac{(\ell+1)^2}{2}\left[\tr[\widehat\Psi_\ell^{(X)}]^{-1}\delta\Psi_{\ell,2}^{(X)}-\frac{1}{2}\tr[\widehat\Psi_\ell^{(X)}]^{-1}\delta\Psi_{\ell,1}^{(X)}[\widehat\Psi_\ell^{(X)}]^{-1}\delta\Psi_{\ell,1}^{(X)}\right].
 \label{s2(ell)}
\end{align}

In order to perform the $\overline{\rm MS}$ subtraction from the divergent
part, we can use
\begin{align}
  \ln\mathcal A^{(X)} &=
  -\frac{1}{2}\tr\left[\widehat{\mathcal M}^{(X)}\right]^{-1}\delta\mathcal M
  +\frac{1}{4}\tr\left[\widehat{\mathcal M}^{(X)}\right]^{-1}\delta\mathcal M\left[\widehat{\mathcal M}^{(X)}\right]^{-1}\delta\mathcal M
  +\mathcal O(\delta \mathcal M^3),
\end{align}
to obtain
\begin{align}
 s_1^{(X)}&=\frac{1}{2}\tr\left[\widehat{\mathcal M}^{(X)}\right]^{-1}\delta\mathcal M,\\
 s_2^{(X)}&=-\frac{1}{4}\tr\left[\widehat{\mathcal M}^{(X)}\right]^{-1}\delta\mathcal M\left[\widehat{\mathcal M}^{(X)}\right]^{-1}\delta\mathcal M.
\end{align}
These quantities can be calculated by using the ordinary diagrammatic
method.  We calculate them adopting dimensional regularization, and
perform $\overline{\rm MS}$ subtraction to obtain the renormalized
quantities, $s_{1,\overline{\rm MS}}^{(X)}$ and $s_{2,\overline{\rm
    MS}}^{(X)}$; their explicit forms are given in
Appendix~\ref{apx_msbar}.\footnote
{There was an error in the calculation of the counterterms in
  \cite{Endo:2017tsz}, which is corrected in the expression in
  Appendix~\ref{apx_msbar}.}

The prefactor after the $\overline{\rm MS}$ subtraction can be
obtained as
\begin{align}
  \ln\mathcal A_{\overline{\rm MS}}^{(X)}
  = \left[ \ln\mathcal A+s_1^{(X)}+s_2^{(X)} \right]
  -s_{1,\overline{\rm MS}}^{(X)}-s_{2,\overline{\rm MS}}^{(X)},
\end{align}
where $s_1^{(X)}$ and $s_2^{(X)}$ in the square bracket should be
understood as those given in Eqs.\ \eqref{s1(ell)} and
\eqref{s2(ell)}, respectively.  Notice that the quantity in the square
bracket can be evaluated order by order in $\ell$.  At a large $\ell$,
the contribution behaves as $\ell^{-2}$ and the sum over $\ell$
converges.  One may truncate higher $\ell$, or fit the terms with
large enough $\ell$ and sum them up for better precision.

\section{Conclusions and Discussion}\label{sec_conclusions}
\setcounter{equation}{0}

In this paper, we have studied the decay rate of the false vacuum in gauge
theory for the case with a multi-field bounce.  If more than one scalar
fields contribute to the bounce, the mixing among the gauge fields and
the scalars becomes $r$-dependent, making it complicated to calculate
the one-loop contributions to the decay rate.  We have extended the
results of \cite{Endo:2017gal,Endo:2017tsz} to perform a gauge invariant
calculation of the decay rate at the one-loop level.  The one-loop
contributions to the decay rate, which are denoted as ${\cal A}$ in this
article, are given by the ratio of the functional determinants of the
fluctuation operators around the bounce and those around the false
vacuum (see Eqs.\ \eqref{eq_A_ccbar} and \eqref{eq_A_Aphi}).  Using the
fact that the functional determinant of the fluctuation operator, ${\cal
M}^{(X)}$, can be related to the asymptotic behavior of the function,
$\psi^{(X)}$, obeying ${\cal M}^{(X)}\psi^{(X)}=0$, we have derived an
expression of the decay rate at the one-loop level, which is manifestly
independent of the gauge parameter, $\xi$.  In particular, we have
discussed how the functional determinant of $\mathcal M^{(A_\mu
\varphi)}$, {\it i.e.} the fluctuation operator for the gauge and scalar
fields, can be evaluated in a gauge invariant way.  Our main results are
summarized in Section \ref{sec_results}.

In our study, we have worked in both the Fermi gauge and the background
gauge.  We have shown that the Fermi gauge has an advantage to show the
gauge invariance of the result especially when there exist gauge zero
modes.  The path integral over the gauge zero modes should be replaced
by the integration over the gauge volume; the rule of the replacement
can be straightforwardly obtained in the Fermi gauge.  The background
gauge has, however, an advantage in numerically calculating the decay
rate because of the better convergence.  In order to utilize the
calculation in the background gauge, we derived a prescription to
translate the result in the background gauge to that in the Fermi
gauge.

We have also discussed how we can remove ultraviolet divergences from
the decay rate.  We have shown a procedure to subtract the divergences
from the one-loop results and to perform the renormalization in the
$\overline{\rm MS}$ scheme (see Section \ref{sec_renormalization}).

Our results apply to various types of models with many scalar fields
having gauge charges; in such models, it is often the case
that there shows up a true vacuum whose energy density is lower than
that of the electroweak vacuum.  Our results can be used to calculate
the decay rate of the electroweak vacuum in such models.
Phenomenological application of our results may be discussed elsewhere.

\section*{Acknowledgements}

The work of SC is supported by JSPS KAKENHI grant.
The work of TM is supported by JSPS KAKENHI grant Nos.\ 16H06490 and 18K03608.
The work of YS is supported by JSPS KAKENHI grant No.\ 16H06492.

\appendix

\section{Evaluation of Determinants}\label{apx_calc}
\setcounter{equation}{0}

\subsection{Alternative fluctuation operators}

One of the important steps in our calculation is to rewrite
$\det\Psi^{(SL\varphi)}_\ell$ using $\eta$ and $\lambda$ at
$r\sim\infty$. Since the fluctuation matrices approach those at the
false vacuum as $r\to\infty$, one may expect that the behavior of
$\Psi^{(SL\varphi)}_\ell$ at $r\sim\infty$ is the same as that of the
solution at the false vacuum,
$\widehat{\Psi}^{(SL\varphi)}_\ell$. However, it is not always the
case because they are related as
\begin{align}
 \Psi^{(SL\varphi)}_\ell(r\to\infty)=\widehat{\Psi}^{(SL\varphi)}_\ell(r)G(r),
\end{align}
where $G$ is a square matrix. Since the elements of $\widehat
\Psi^{(SL\varphi)}_\ell$ generally become hierarchical as $r\to\infty$,
larger elements easily contaminate smaller elements due to the mixing
induced by $G$. The crucial problem here is that the contamination
cannot be avoided just by taking a linear combination of the solutions
since $G$ is $r$-dependent. Such contamination leads to a severe
cancellation when we calculate the determinant of
$\Psi^{(SL\varphi)}_\ell$. In the following, we discuss a way to avoid
this problem.

We first define the projection operators as
\begin{align}
 P_H&=V_HV_H^T,\\
 P_G&=M(M^TM)^{-1}M^T,
\end{align}
which satisfy $P_H+P_G=I_\varphi$ and $P_HP_G=P_GP_H=0$.  Then, we
define $r$-dependent orthogonal matrices $O_{\varphi}(r)$ and $O_G(r)$
having the following properties:
\begin{align}
 O_{\varphi}^T(r)M(r)O_G(r)&=
\begin{pmatrix}
 \eqmakebox[tagW]{$W(r)$}\\
 \coolunder{tagW}{n_G}{0}
\end{pmatrix}
\begin{matrix*}[l]
    \coolright{n_G}\\
    \coolright{n_H}
\end{matrix*}=
\begin{pmatrix}
 \eqmakebox[tagWB]{$W_B(r)$}&0\\
 0&\eqmakebox[tagWB]{$W_U(r)$}\\
 \coolunder{tagWB}{n_B}{0}&\coolunder{tagWU}{n_U}{0}
\end{pmatrix}
\begin{matrix*}[l]
    \coolright{n_B}\\
    \coolright{n_U}\\
    \coolright{n_H}
\end{matrix*},\label{eq_rot_M}\\
\nonumber\\
 O_{\varphi}^T(r)P_H(r)\Omega(r)P_H(r)O_{\varphi}(r)&=
\begin{pmatrix}
 0&0\\
 \coolunder{tag0}{n_G}{0}&\coolunder{tagm2}{n_H}{m^2(r)}
\end{pmatrix}
\begin{matrix*}[l]
    \coolright{n_G}\\
    \coolright{n_H}
\end{matrix*},\label{eq_rot_Omega}\\
\nonumber
\end{align}
where $m^2$, $W$, $W_B$ and $W_U$ are diagonal matrices. Since we work
in the field basis given in Eqs.~\eqref{eq_basis_M} and
\eqref{eq_basis_Omega}, we can take
\begin{align}
 \lim_{r\to\infty}O_\varphi(r)=I_\varphi,~\lim_{r\to\infty}O_G(r)=I_G.\label{eq_lim_O}
\end{align}

The basic idea is to use the following matrices instead of $M$ and $\Omega$.
\begin{align}
  M_{\rm alt}&=
\begin{cases}
  M,&r \ll r_*\\
  O_{\varphi}^TMO_G,& r \gg r_*
\end{cases},\\
\Omega_{\rm alt}&=
 \begin{cases}
   \Omega,& r \ll r_*\\
   O_{\varphi}^TP_H\Omega P_HO_{\varphi}+O_{\varphi}^TP_G\Omega P_GO_{\varphi},&
   r \gg r_*
\end{cases}.
\end{align}
Here, $r_*$ is taken to be much larger than the typical scale of the bounce.
Then, the behavior of the solution at $r\to\infty$ can be understood
easily since the fluctuation operators become block-diagonalized. The
behavior at $r\sim r_*$ will be defined in Appendix \ref{apx_error}.

The decomposed solution in Section~\ref{sec_solutions} cannot
be used as it is since $M_{\rm alt}$ does not satisfy
Eqs.~\eqref{eq_M_eom} and \eqref{eq_M_deriv} for $r\gtrsim r_*$.
However, we find that the solution becomes exact if we add extra scalar
mass terms to the fluctuation operators as
\begin{align}
 \mathcal M_{\ell,{\rm alt}}^{(SL\varphi)}&=\left.\mathcal M_{\ell}^{(SL\varphi)}\right|_{M\to M_{\rm alt},\Omega\to\Omega_{\rm alt}+\delta\Omega},\label{eq_M_alt_ell}\\
 \mathcal M_{0,{\rm alt}}^{(S\varphi)}&=\left.\mathcal M_{0}^{(S\varphi)}\right|_{M\to M_{\rm alt},\Omega\to\Omega_{\rm alt}+\delta\Omega},\label{eq_M_alt_0}
\end{align}
where
\begin{align}
 \delta\Omega=-M_{\rm alt}(M_{\rm alt}^TM_{\rm alt})^{-1}\varepsilon^T,
\end{align}
with
\begin{align}
 \varepsilon=-\Delta_0M_{\rm alt}+\Omega_{\rm alt}M_{\rm alt}.
\end{align}
Then, the solutions of
\begin{align}
  \mathcal M_{\ell,{\rm alt}}^{(SL\varphi)}\Psi_{\ell,\rm alt}^{(SL\varphi)}&=0,
\end{align}
and
\begin{align}
  \mathcal M_{0,{\rm alt}}^{(S\varphi)}\Psi_{0,{\rm alt}}^{(S\varphi)}=0,
\end{align}
are given by Eqs.~\eqref{eq_decomp} -- \eqref{eq_lambda} and
\eqref{eq_decomp_0} with replacing $M\to M_{\rm alt}$ and
$\Omega\to \Omega_{\rm alt}$. Since the fluctuation operators are not
modified for $r\lesssim r_*$, we take
\begin{align}
  \Psi_{\ell,\rm alt}^{(SL\varphi)}(r\lesssim r_*) = &\,
  \Psi_{\ell}^{(SL\varphi)}(r\lesssim r_*),
  \\
  \Psi_{0,{\rm alt}}^{(S\varphi)}(r\lesssim r_*) = &\,
  \Psi_{0}^{(S\varphi)}(r\lesssim r_*).
\end{align}
Similarly, we define the solutions, $\Psi_{\ell,\rm alt}^{(c\bar c)}$
and $\Psi_{\ell,\rm alt}^{(\eta\lambda)}$, which satisfy the same
differential equations as $\Psi_{\ell}^{(c\bar{c})}$ and
$\Psi^{(\eta\lambda)}_\ell$ but with $M\to M_{\rm
  alt}$ and $\Omega\to\Omega_{\rm alt}$.

For later convenience, we also define
\begin{align}
 P_{G,\rm alt}&=M_{\rm alt}(M_{\rm alt}^TM_{\rm alt})^{-1}M_{\rm alt}^T,\\
 P_{H,\rm alt}&=I_\varphi-P_{G,\rm alt}.
\end{align}

The deformation at $r\sim r_*$ should be smooth enough and the
decomposition of the solutions should be well-defined for the entire
region of $r$. In addition, we need to ensure the use of these
alternative fluctuation operators does not affect the final results. We
will discuss these issues in Appendix~\ref{apx_error}.

\subsection{Evaluation of solutions for $\ell>0$}

\subsubsection{Behavior at infinity}

Here, we evaluate $\Psi^{(SL\varphi)}_{\ell,{\rm alt}}$ at
$r\to\infty$.  For this purpose, we can use the fact that
$\Psi_{\ell,\rm alt}^{(c\bar c)}$ and $\Psi_{\ell,\rm
  alt}^{(\eta\lambda)}$ behave as follows:
\begin{align}
  \Psi_{\ell,\rm alt}^{(c\bar c)}(r\to\infty) \simeq &\,
  \begin{pmatrix}
    \frac{e^{\sqrt{\xi}W_Br}}{r^{3/2}}I_B & 0 \\
    0 & r^\ell I_U
  \end{pmatrix}
  \mathcal T_\ell^{(c\bar c)},
\end{align}
and
\begin{align}
  \Psi_{\ell,\rm alt}^{(\eta\lambda)}(r\to\infty)
  \simeq &\,
  \begin{pmatrix}
    \frac{e^{W_Br}}{r^{3/2}}I_B & 0 & 0 \\
    0 & \frac{1}{r^2}I_U & 0 \\
    0 & 0 & \frac{e^{mr}}{r^{3/2}} I_H
  \end{pmatrix} \mathcal T_\ell^{(\eta\lambda)},
  \label{eq_eta_inf}
\end{align}
where $\mathcal T_\ell^{(c\bar c)}$ and $\mathcal
T_\ell^{(\eta\lambda)}$ are constant square matrices.

We examine the behavior of each type of the solutions at $r\to\infty$
below.

\begin{itemize}
 \item Type 1

In the Fermi gauge, $\chi^{(1)}=r^\ell I_G$.  Then, from
Eq.~\eqref{eq_decomp},
\begin{align}
 \Psi_{\ell,{\rm alt}}^{(SL\varphi)(1)}(r\to\infty)&\simeq
\begin{pmatrix}
 \eqmakebox[tagPhi1infF]{$\frac{\ell}{r}I_G$}\\
 \eqmakebox[tagPhi1infF]{$\frac{L}{r}I_G$}\\
 \eqmakebox[tagPhi1infF]{$W$}\\
 \coolunder{tagPhi1infF}{n_G}{0}
\end{pmatrix}
\begin{matrix*}[l]
    \coolright{n_G}\\
    \coolright{n_G}\\
    \coolright{n_G}\\
    \coolright{n_H}
\end{matrix*}\hspace{2ex}r^\ell.\\
\nonumber
\end{align}

In the background gauge, $\chi^{(1)}=\Psi_{\ell,\rm alt}^{(c\bar
  c)}$.  Using Eq.~\eqref{eq_decomp},
\begin{align}
 \Psi_{\ell,\rm BG,alt}^{(SL\varphi)(1)}(r\to\infty)&\simeq
\begin{pmatrix}
 \eqmakebox[tagPhi1infB]{$\sqrt{\xi} W_B$}&\eqmakebox[tagPhi1infB2]{$0$}\\
 \eqmakebox[tagPhi1infB]{$0$}&\eqmakebox[tagPhi1infB2]{$\frac{\ell}{r}I_U$}\\
 \eqmakebox[tagPhi1infB]{$\frac{L}{r}I_B$}&\eqmakebox[tagPhi1infB2]{$0$}\\
 \eqmakebox[tagPhi1infB]{$0$}&\eqmakebox[tagPhi1infB2]{$\frac{L}{r}I_U$}\\
 \eqmakebox[tagPhi1infB]{$W_B$}&\eqmakebox[tagPhi1infB2]{$0$}\\
 \eqmakebox[tagPhi1infB]{$0$}&\eqmakebox[tagPhi1infB2]{$W_U$}\\
 \coolunder{tagPhi1infB}{n_B}{0}&\coolunder{tagPhi1infB2}{n_U}{0}
\end{pmatrix}
\begin{matrix*}[l]
    \coolright{n_B}\\
    \coolright{n_U}\\
    \coolright{n_B}\\
    \coolright{n_U}\\
    \coolright{n_B}\\
    \coolright{n_U}\\
    \coolright{n_H}
\end{matrix*}\hspace{2ex}\Psi_{\ell,\rm alt}^{(c\bar c)}.\\
\nonumber
\end{align}

 \item Type 2

Since $P_{H,{\rm alt}}M'_{\rm alt}=0$ for $r\gtrsim r_*$,
$\lambda^{(2)}=0$ from Eq.~\eqref{eq_lambda}.
In the Fermi gauge, $\zeta^{(2)}=r^\ell$, and
\begin{align}
 \eta^{(2)} (r\rightarrow\infty) =
\begin{pmatrix}
 \eqmakebox[tageta]{$\mathcal O(r^{-1}W'_B)$}&0\\
 \coolunder{tageta}{n_B}{0}&\coolunder{tageta2}{n_U}{\frac{L}{\ell+2}+\mathcal O(W_U^2)}
\end{pmatrix}
\begin{matrix*}[l]
    \coolright{n_B}\\
    \coolright{n_U}
\end{matrix*}\hspace{2ex}r^\ell.\\
\nonumber
\end{align}
Then, from Eqs.~\eqref{eq_mid} -- \eqref{eq_bot},
\begin{align}
 \Psi_{\ell,{\rm alt}}^{(SL\varphi)(2)}(r\to\infty)=
\begin{pmatrix}
 \eqmakebox[tagPhi2infF]{$-\frac{\xi}{4}rI_B$}&\eqmakebox[tagPhi2infF2]{$0$}\\
 \eqmakebox[tagPhi2infF]{$0$}&\eqmakebox[tagPhi2infF2]{$\frac{\ell-(\ell+2)\xi}{4(\ell+2)}rI_U$}\\
 \eqmakebox[tagPhi2infF]{$-\frac{\xi\ell}{4L}rI_B$}&\eqmakebox[tagPhi2infF2]{$0$}\\
 \eqmakebox[tagPhi2infF]{$0$}&\eqmakebox[tagPhi2infF2]{$\frac{\ell[(\ell+4)-(\ell+2)\xi]}{4L(\ell+2)} rI_U$}\\
 \eqmakebox[tagPhi2infF]{$-\frac{\xi}{4(\ell+2)}r^2W_B$}&\eqmakebox[tagPhi2infF2]{$0$}\\
 \eqmakebox[tagPhi2infF]{$0$}&\eqmakebox[tagPhi2infF2]{$\mathcal O(r^2W_U)$}\\
 \coolunder{tagPhi2infF}{n_B}{0}&\coolunder{tagPhi2infF2}{n_U}{0}
\end{pmatrix}
\begin{matrix*}[l]
    \coolright{n_B}\\
    \coolright{n_U}\\
    \coolright{n_B}\\
    \coolright{n_U}\\
    \coolright{n_B}\\
    \coolright{n_U}\\
    \coolright{n_H}
\end{matrix*}\hspace{2ex}r^\ell.\\
\nonumber
\end{align}
In the background gauge, $\zeta^{(2)}=\Psi_{\ell,\rm alt}^{(c\bar
  c)}$.  Then,
\begin{align}
 \eta^{(2)} (r\rightarrow\infty) =
\begin{pmatrix}
 \eqmakebox[tageta]{$\mathcal O(r^{-1}W'_B)$}&0\\
 \coolunder{tageta}{n_B}{0}&\coolunder{tageta2}{n_U}{\frac{L}{\ell+2}+\mathcal O(W_U^2)}
\end{pmatrix}
\begin{matrix*}[l]
    \coolright{n_B}\\
    \coolright{n_U}
\end{matrix*}\hspace{2ex}\Psi_{\ell,\rm alt}^{(c\bar c)},\\
\nonumber
\end{align}
and
\begin{align}
 \Psi_{\ell,\rm BG,alt}^{(SL\varphi)(2)}(r\to\infty)=
\begin{pmatrix}
 \eqmakebox[tagPhi2infB]{$\mathcal O(W'_B)$}&\eqmakebox[tagPhi2infB2]{$0$}\\
 \eqmakebox[tagPhi2infB]{$0$}&\eqmakebox[tagPhi2infB2]{$\frac{\ell-(\ell+2)\xi}{4(\ell+2)}rI_U$}\\
 \eqmakebox[tagPhi2infB]{$\mathcal O(W'_B)$}&\eqmakebox[tagPhi2infB2]{$0$}\\
 \eqmakebox[tagPhi2infB]{$0$}&\eqmakebox[tagPhi2infB2]{$\frac{\ell[(\ell+4)-(\ell+2)\xi]}{4L(\ell+2)} rI_U$}\\
 \eqmakebox[tagPhi2infB]{$W_B^{-1}$}&\eqmakebox[tagPhi2infB2]{$0$}\\
 \eqmakebox[tagPhi2infB]{$0$}&\eqmakebox[tagPhi2infB2]{$\mathcal O(r^2W_U)$}\\
 \coolunder{tagPhi2infB}{n_B}{0}&\coolunder{tagPhi2infB2}{n_U}{0}
\end{pmatrix}
\begin{matrix*}[l]
    \coolright{n_B}\\
    \coolright{n_U}\\
    \coolright{n_B}\\
    \coolright{n_U}\\
    \coolright{n_B}\\
    \coolright{n_U}\\
    \coolright{n_H}
\end{matrix*}\hspace{2ex}\Psi_{\ell,\rm alt}^{(c\bar c)}.\\
\nonumber
\end{align}

 \item Type 3

   When gauge bosons become massive at the false vacuum, it is convenient
   to use the following relations:
   \begin{align}
     \Delta_\ell\left(r\Psi^{({\rm top})}_\ell\right)&=L\eta-\frac{\xi}{r}\partial_rr^2Y,\\
     \Psi^{({\rm mid})}_\ell&=\frac{1}{Lr^2}\partial_rr^2\left(r\Psi^{({\rm top})}_\ell\right)+\frac{r\xi}{L}Y,\\
     \Psi^{({\rm bot})}_\ell&=M\chi+\lambda+M(M^TM)^{-1}\zeta,
   \end{align}
   where
   \begin{align}
     Y=
     \begin{cases}
       \zeta & \mbox{: Fermi}\\
       -M^TM\chi & \mbox{: BG}
     \end{cases}.\label{eq_y}
   \end{align}
   Using Eqs.~\eqref{eq_chi} and \eqref{eq_mid} -- \eqref{eq_bot} as
   well as the above relations,
   \begin{align}
     \Psi_{\ell,{\rm alt}}^{(SL\varphi)(3)}(r\to\infty)=
     \begin{pmatrix}
       \eqmakebox[tagPhi3inf]{$\frac{L}{r}W_B^{-2}$}&\eqmakebox[tagPhi3inf2]{$0$}&0\\
       \eqmakebox[tagPhi3inf]{$0$}&\eqmakebox[tagPhi3inf2]{$-\frac{r}{L}I_U$}&0\\
       \eqmakebox[tagPhi3inf]{$W_B^{-1}$}&\eqmakebox[tagPhi3inf2]{$0$}&0\\
       \eqmakebox[tagPhi3inf]{$0$}&\eqmakebox[tagPhi2infF2]{$-\frac{2r}{L^2}I_U$}&0\\
       \eqmakebox[tagPhi3inf]{$\mathcal O(r^{-1}W'_B)$}&\eqmakebox[tagPhi3inf2]{$0$}&0\\
       \eqmakebox[tagPhi3inf]{$0$}&\eqmakebox[tagPhi3inf2]{$\frac{L}{2r}(W'_U)^{-1}$}&0\\
       \coolunder{tagPhi3inf}{n_B}{0}&\coolunder{tagPhi3inf2}{n_U}{0}&\coolunder{tagPhi3inf3}{n_H}{I_H}
     \end{pmatrix}
     \begin{matrix*}[l]
       \coolright{n_B}\\
       \coolright{n_U}\\
       \coolright{n_B}\\
       \coolright{n_U}\\
       \coolright{n_B}\\
       \coolright{n_U}\\
       \coolright{n_H}
     \end{matrix*}\hspace{2ex}\Psi_{\ell,\rm alt}^{(\eta\lambda)}.\\
     \nonumber
   \end{align}
\end{itemize}

\subsubsection{Translational zero modes}
Here, we discuss the translational zero modes. The following discussion
applies to both the Fermi gauge and the background gauge.  With the
alternative fluctuation operator, $\mathcal M^{(SL\varphi)}_{1,\rm
alt}$, the existence of the translational zero modes is not guaranteed;
$\det\Psi_{1,\rm alt}^{(SL\varphi)}(r\to\infty)$ can be
non-vanishing. However, it does not cause a problem since it approaches
zero as $r_*\to\infty$ and we always take $\nu\to0$ after $r_*\to\infty$.

Let us start with the original fluctuation matrix (the one without
``alt''). To construct the solution explicitly, we regularize the
translational zero modes using
\begin{align}
  \mathcal M_{1,\rm reg}^{(SL\varphi)}=\mathcal M_{1}^{(SL\varphi)}+\nu
  \begin{pmatrix}
    \eqmakebox[tagReg1A]{$0$}&\eqmakebox[tagReg1B]{$0$}\\
    \coolunder{tagReg1A}{2n_G}{0}&\coolunder{tagReg1B}{n_\varphi}{P_H}
  \end{pmatrix}
  \begin{matrix*}[l]
    \coolright{2n_G}\\
    \coolright{n_\varphi}
  \end{matrix*},\\
  \nonumber
\end{align}
instead of $\mathcal M_1^{(SL\varphi)}+\nu$.
Since
\begin{equation}
 \mathcal M_{1,\rm reg}^{(SL\varphi)}\psi^{({\rm tr})}=\nu\psi^{({\rm tr})},
\end{equation}
we can use the same discussion for the zero modes as in Section
\ref{sec_zeromodes}. Then, we define its alternative as
\begin{equation}
 \mathcal M^{(SL\varphi)}_{1,\rm reg,alt}=\left.\mathcal M^{(SL\varphi)}_{1,\rm reg}\right|_{M\to M_{\rm alt},~\Omega\to\Omega_{\rm alt}+\delta\Omega,~P_H\to P_{H,\rm alt}}.
\end{equation}
Since $\psi^{({\rm tr})}$ is included in $\Psi_{1}^{(SL\varphi)(3)}$, it is enough to calculate
\begin{equation}
 \mathcal M^{(SL\varphi)}_{1,\rm reg,alt}\check\Psi_{1,\rm alt}^{(SL\varphi)(3)}=-\Psi_{1,\rm alt}^{(SL\varphi)(3)}.
\end{equation}
The solution is given by
\begin{equation}
 \check\Psi_{1,\rm alt}^{(SL\varphi)(3)}=\left.\Psi_{1,\rm alt}^{(SL\varphi)(3)}\right|_{\Psi_{1,\rm alt}^{(\eta\lambda)}\to\check\Psi_{1,\rm alt}^{(\eta\lambda)}},
\end{equation}
where
\begin{equation}
 \check\Psi_{1,\rm alt}^{(\eta\lambda)}=
\begin{pmatrix}
 \check \eta\\
 V_{H,\rm alt}^T\check\lambda
\end{pmatrix}.
\end{equation}
Here,
$\check\eta$ and $\check\lambda$ satisfy
\begin{align}
 \Delta_1\check\eta&=\left.\Delta_1\eta\right|_{\eta\to\check\eta,\zeta\to0,\lambda\to\check\lambda,M\to M_{\rm alt},\Omega\to\Omega_{\rm alt}},\\
 \Delta_1\check\lambda&=\left.\Delta_1\lambda\right|_{\eta\to\check\eta,\zeta\to0,\lambda\to\check\lambda,M\to M_{\rm alt},\Omega\to\Omega_{\rm alt}}+
\begin{pmatrix}
  0&0&V_{H,\rm alt}
 \end{pmatrix}\Psi_{1,\rm alt}^{(SL\varphi)}.
\end{align}
\subsubsection{Functional determinant ($\ell>1$)}
Here, we calculate the determinant of the solutions obtained above.

In the Fermi gauge,
\begin{align}
 \det \Psi_{\ell,{\rm alt}}^{(SL\varphi)}(r\to\infty) = &\, \det
\begin{pmatrix}
 \Psi_{\ell,{\rm alt}}^{(SL\varphi)(1)}&\Psi_{\ell,{\rm alt}}^{(SL\varphi)(2)}&\Psi_{\ell,{\rm alt}}^{(SL\varphi)(3)}
\end{pmatrix}
\nonumber\\
\simeq &\,
\left(-\frac{\xi r}{2(\ell+2)}\right)^{n_B}\left(\frac{\ell[\ell+\xi(\ell+2)]}{4(\ell+2)r}\right)^{n_U}r^{2\ell n_G}\frac{\det\Psi_{\ell,\rm alt}^{(\eta\lambda)}}{\det W'_U},
\end{align}
and
\begin{align}
 \det\widehat\Psi_{\ell}^{(SL\varphi)}(r\to\infty)&= \det\widehat\Psi_\ell^{(B)}\det\widehat\Psi_\ell^{(U)}\det\widehat\Psi_\ell^{(\lambda)}\nonumber\\
 &\simeq\left(-\frac{\xi r}{2(\ell+2)}\right)^{n_B}r^{2\ell n_B}\det\widehat\Psi_\ell^{(\eta)}\left(-\frac{\ell+\xi(\ell+2)}{2(\ell+2)}r^{2\ell}\right)^{n_U}\det\widehat\Psi_\ell^{(\lambda)}.
\end{align}
Thus,
\begin{align}
 \frac{\det\Psi_{\ell,{\rm alt}}^{(SL\varphi)}(r\to\infty)}{\det\widehat\Psi_{\ell}^{(SL\varphi)}(r\to\infty)}&=
\left(-\frac{\ell}{2r_\infty}\right)^{n_U}\frac{\det\Psi_{\ell,\rm alt}^{(\eta\lambda)}}{\det\widehat\Psi^{(\eta)}_\ell\det\widehat\Psi^{(\lambda)}_\ell\det W'_U}.
\end{align}
Keeping only the leading terms in $r$, we can interchange $r\to\infty$
and $r_*\to\infty$ and
\begin{align}
 \frac{\det\Psi_{\ell}^{(SL\varphi)}(r\to\infty)}{\det\widehat\Psi_{\ell}^{(SL\varphi)}(r\to\infty)}&=\left(-\frac{\ell}{2r_\infty}\right)^{n_U}\frac{\det\Psi_{\ell}^{(\eta\lambda)}}{\det\widehat\Psi^{(\eta)}_\ell\det\widehat\Psi^{(\lambda)}_\ell\det W'_U}.
\end{align}

In the background gauge,
\begin{align}
  &\det\Psi_{\ell,{\rm BG},{\rm alt}}^{(SL\varphi)}(r\to\infty)=\det
  \begin{pmatrix}
    \Psi_{\ell,\rm BG,alt}^{(SL\varphi)(1)}&\Psi_{\ell,\rm BG,alt}^{(SL\varphi)(2)}&\Psi_{\ell,\rm BG,alt}^{(SL\varphi)(3)}
  \end{pmatrix}\nonumber\\&~~~~
  \simeq \left(-\sqrt{\xi}\right)^{n_B}\frac{1}{\det \widehat{W}}\left(\frac{\ell[\ell+\xi(\ell+2)]}{4(\ell+2)r}\right)^{n_U}\frac{\left[\det \Psi_{\ell,\rm alt}^{(c\bar c)}\right]^2\det\Psi_{\ell,\rm alt}^{(\eta\lambda)}}{\det W'_U},
\end{align}
and
\begin{align}
  &\det\widehat\Psi_{\ell,{\rm BG}}^{(SL\varphi)}(r\to\infty)= \det\widehat\Psi_\ell^{(B)}\det\widehat\Psi_\ell^{(U)}\det\widehat\Psi_\ell^{(\lambda)}\nonumber\\ &~~~~
 \simeq\left(-\sqrt{\xi}\right)^{n_B}\frac{1}{\det\widehat W}\left[r^{-\ell n_U}\det\widehat \Psi_\ell^{(c\bar c)}\right]^2\det\widehat\Psi_\ell^{(\eta)}\left(-\frac{\ell+\xi(\ell+2)}{2(\ell+2)}r^{2\ell}\right)^{n_U}\det\widehat\Psi_\ell^{(\lambda)}.
\end{align}
Here, we have used $W_B(r\to\infty)=\widehat W$.
Thus,
\begin{align}
  &\frac{\det\Psi_{\ell,{\rm BG},{\rm alt}}^{(SL\varphi)}(r\to\infty)}{\det\widehat\Psi_{\ell,{\rm BG}}^{(SL\varphi)}(r\to\infty)}
  =
  \lim_{r\to\infty}
\left(-\frac{\ell}{2r_\infty}\right)^{n_U}\frac{\det\Psi_{\ell,\rm alt}^{(\eta\lambda)}}{\det\widehat\Psi_\ell^{(\eta)}\det\widehat\Psi_\ell^{(\lambda)}\det W'_U}\left(\frac{\det \Psi_{\ell,\rm alt}^{(c\bar c)}}{\det\widehat \Psi_\ell^{(c\bar c)}}\right)^2,
\end{align}
and hence
\begin{align}
 \frac{\det\Psi_{\ell,{\rm BG}}^{(SL\varphi)}(r\to\infty)}{\det\widehat\Psi_{\ell,{\rm BG}}^{(SL\varphi)}(r\to\infty)}&=\left(-\frac{\ell}{2r_\infty}\right)^{n_U}\frac{\det\Psi_{\ell}^{(\eta\lambda)}}{\det\widehat\Psi^{(\eta)}_\ell\det\widehat\Psi^{(\lambda)}_\ell \det W'_U}\left(\frac{\det \Psi_\ell^{(c\bar c)}}{\det\widehat \Psi_\ell^{(c\bar c)}}\right)^2.
\end{align}

From Eqs.~\eqref{BC_slphi1}, \eqref{BC_slphi2} and \eqref{BC_slphi3},
\begin{align}
  \lim_{r\to0}\frac{\det\Psi_\ell^{(SL\varphi)}(r)}{\det\widehat\Psi_\ell^{(SL\varphi)}(r)}
  =\frac{\det \widehat W}{\sqrt{\det M^T_0 M_0}},
\end{align}
independently of the choice of the gauge fixing.  Taking the ratio
between the determinants with $r\to0$ and $r\to\infty$, we get
Eq.~\eqref{det_l>1}.

\subsubsection{Functional determinant ($\ell=1$)}


In the Fermi gauge,
\begin{align}
 \det \Psi_{1,{\rm reg},{\rm alt}}^{(SL\varphi)}(\nu;r\to\infty) = &\, \det
\begin{pmatrix}
 \Psi_{1,{\rm alt}}^{(SL\varphi)(1)}&\Psi_{1,{\rm alt}}^{(SL\varphi)(2)}&\Psi_{1,{\rm alt}}^{(SL\varphi)(3)}+\nu\check\Psi_{1,{\rm alt}}^{(SL\varphi)(3)}
\end{pmatrix}
\nonumber\\
\simeq &\,
\left(-\frac{\xi r}{6}\right)^{n_B}\left(\frac{1+3\xi}{12r}\right)^{n_U}r^{2 n_G}\frac{\det\left[\Psi_{1,\rm alt}^{(\eta\lambda)}+\nu\check\Psi_{1,\rm alt}^{(\eta\lambda)}\right]}{\det W'_U}
\nonumber\\ &\,
+\mathcal O(\nu^2),
\end{align}
and
\begin{align}
 \det\widehat\Psi_{1,{\rm reg}}^{(SL\varphi)}(\nu;r\to\infty)&\simeq\left(-\frac{\xi r}{6}\right)^{n_B}r^{2 n_B}\det\widehat\Psi_1^{(\eta)}\left(-\frac{1+3\xi}{6}r^{2}\right)^{n_U}\det\widehat\Psi_1^{(\lambda)}+\mathcal O(\nu).
\end{align}
Thus,
\begin{align}
 \frac{\det\Psi_{1,{\rm reg},{\rm alt}}^{(SL\varphi)}(\nu;r\to\infty)}{\det\widehat\Psi_{1,{\rm reg}}^{(SL\varphi)}(\nu;r\to\infty)}&=
\left(-\frac{1}{2r_\infty}\right)^{n_U}\frac{\det\left[\Psi_{1,\rm alt}^{(\eta\lambda)}+\nu\check\Psi_{1,\rm alt}^{(\eta\lambda)}\right]}{\det\widehat\Psi^{(\eta)}_1\det\widehat\Psi^{(\lambda)}_1\det W'_U}+\mathcal O(\nu^2,\mathcal E\nu),
\end{align}
where
\begin{equation}
 \mathcal E=\lim_{\nu\to0}\frac{\det\Psi_{1,{\rm reg},{\rm alt}}^{(SL\varphi)}(\nu;r\to\infty)}{\det\widehat\Psi_{1,{\rm reg}}^{(SL\varphi)}(\nu;r\to\infty)}.
\end{equation}
Notice that $\mathcal E$ can be non-vanishing when the translational zero
modes disappear in the alternative fluctuation operator. As
$r_*\to\infty$, $\mathcal E$ should approach zero. Keeping only the
leading terms in $r$, we can interchange $r\to\infty$ and
$r_*\to\infty$. Then,
\begin{align}
 \lim_{\nu\to0}\frac{1}{\nu}\frac{\det\Psi_{1}^{(SL\varphi)}(\nu;r\to\infty)}{\det\widehat\Psi_{1}^{(SL\varphi)}(\nu;r\to\infty)}&=\lim_{\nu\to0}\frac{1}{\nu}\left(-\frac{1}{2r_\infty}\right)^{n_U}\frac{\det\left[\Psi_{1}^{(\eta\lambda)}+\nu\check\Psi_{1}^{(\eta\lambda)}\right]}{\det\widehat\Psi^{(\eta)}_1\det\widehat\Psi^{(\lambda)}_1\det W'_U}.
\end{align}

In the background gauge,
\begin{align}
  &\det\Psi_{1,{\rm reg},{\rm BG},{\rm alt}}^{(SL\varphi)}(\nu;r\to\infty)=\det
  \begin{pmatrix}
    \Psi_{1,\rm BG,alt}^{(SL\varphi)(1)}&\Psi_{1,\rm BG,alt}^{(SL\varphi)(2)}&\Psi_{1,\rm BG,alt}^{(SL\varphi)(3)}+\nu\check\Psi_{1,\rm BG,alt}^{(SL\varphi)(3)}
  \end{pmatrix}\nonumber\\&~~~~
  \simeq \left(-\sqrt{\xi}\right)^{n_B}\frac{1}{\det \widehat{W}}\left(\frac{1+3\xi}{12r}\right)^{n_U}\frac{\left[\det \Psi_{1,\rm alt}^{(c\bar c)}\right]^2\det\left[\Psi_{1,\rm alt}^{(\eta\lambda)}+\nu\check\Psi_{1,\rm alt}^{(\eta\lambda)}\right]}{\det W'_U}+\mathcal O(\nu^2),
\end{align}
and
\begin{align}
  &\det\widehat\Psi_{1,{\rm reg},{\rm BG}}^{(SL\varphi)}(\nu;r\to\infty)= \det\widehat\Psi_1^{(B)}\det\widehat\Psi_1^{(U)}\det\widehat\Psi_1^{(\lambda)}\nonumber\\ &~~~~
 \simeq\left(-\sqrt{\xi}\right)^{n_B}\frac{1}{\det\widehat W}\left[r^{-1 n_U}\det\widehat \Psi_1^{(c\bar c)}\right]^2\det\widehat\Psi_1^{(\eta)}\left(-\frac{1+3\xi}{6}r^{2}\right)^{n_U}\det\widehat\Psi_1^{(\lambda)}+\mathcal O(\nu).
\end{align}
Here, we have used $W_B(r\to\infty)=\widehat W$.
Thus,
\begin{align}
  &\lim_{\nu\to0}\frac{1}{\nu}\frac{\det\Psi_{1,{\rm reg},{\rm BG},{\rm alt}}^{(SL\varphi)}(\nu;r\to\infty)}{\det\widehat\Psi_{1,{\rm reg},{\rm BG}}^{(SL\varphi)}(\nu;r\to\infty)}
  \nonumber \\ &\, ~~~~ =
  \lim_{\nu\to0}\lim_{r\to\infty}\frac{1}{\nu}
\left(-\frac{1}{2r_\infty}\right)^{n_U}\frac{\det\left[\Psi_{1,\rm alt}^{(\eta\lambda)}+\nu\check\Psi_{1,\rm alt}^{(\eta\lambda)}\right]}{\det\widehat\Psi_1^{(\eta)}\det\widehat\Psi_1^{(\lambda)}\det W'_U}\left(\frac{\det \Psi_{1,\rm alt}^{(c\bar c)}}{\det\widehat \Psi_1^{(c\bar c)}}\right)^2,
\end{align}
and hence
\begin{align}
 \lim_{\nu\to0}\frac{1}{\nu}\frac{\det\Psi_{1,{\rm reg},{\rm BG}}^{(SL\varphi)}(\nu;r\to\infty)}{\det\widehat\Psi_{1,{\rm reg},{\rm BG}}^{(SL\varphi)}(\nu;r\to\infty)}&=\lim_{\nu\to0}\frac{1}{\nu}\left(-\frac{1}{2r_\infty}\right)^{n_U}\frac{\det\left[\Psi_{1}^{(\eta\lambda)}+\nu\check\Psi_{1}^{(\eta\lambda)}\right]}{\det\widehat\Psi^{(\eta)}_1\det\widehat\Psi^{(\lambda)}_1 \det W'_U}\left(\frac{\det \Psi_1^{(c\bar c)}}{\det\widehat \Psi_1^{(c\bar c)}}\right)^2.
\end{align}
Taking the ratio
between the determinants with $r\to0$ and $r\to\infty$, we obtain
Eq.~\eqref{trzeromode}.

\subsection{Evaluation of solutions for $\ell=0$}

\subsubsection{Behavior at infinity}

The behavior of $\Psi_{0,\rm alt}^{(c\bar c)}$ and $\Psi_{0,\rm
alt}^{(\lambda)}$ is given by
\begin{align}
  \Psi_{0,\rm alt}^{(c\bar c)}(r\to\infty)&\simeq
  \begin{pmatrix}
    \frac{e^{\sqrt{\xi}W_Br}}{r^{3/2}}I_B & 0 \\
    0 & I_U+\mathcal O(r^{-2})
  \end{pmatrix}
  \mathcal T_0^{(c\bar c)},
\end{align}
and
\begin{align}
  \Psi_{0,\rm alt}^{(\lambda)}(r\to\infty) \simeq
  \frac{e^{mr}}{r^{3/2}} I_H \mathcal T_0^{(\lambda)},
\end{align}
where $\mathcal T_0^{(c\bar c)}$ and $\mathcal T_0^{(\lambda)}$ are
constant square matrices.

Next, we consider the behavior of $\Psi^{(SL\varphi)}_{0,{\rm alt}}$
at $r\to\infty$.
\begin{itemize}
 \item Type 1

In the Fermi gauge, $\chi^{(1)}=I_G$.
Then, from Eq.~\eqref{eq_decomp_0},
\begin{align}
 \Psi_{0,\rm alt}^{(S\varphi)(1)}(r\to\infty)\simeq
\begin{pmatrix}
 \eqmakebox[tagPhi1infL0F]{$0$}&\eqmakebox[tagPhi1infL0F2]{$0$}\\
 \eqmakebox[tagPhi1infL0F]{$0$}&\eqmakebox[tagPhi1infL0F2]{$0$}\\
 \eqmakebox[tagPhi1infL0F]{$W_B$}&\eqmakebox[tagPhi1infL0F2]{$0$}\\
 \eqmakebox[tagPhi1infL0F]{$0$}&\eqmakebox[tagPhi1infL0F2]{$W_U$}\\
 \coolunder{tagPhi1infL0F}{n_B}{0}&\coolunder{tagPhi1infL0F}{n_U}{0}
\end{pmatrix}
\begin{matrix*}[l]
    \coolright{n_B}\\
    \coolright{n_U}\\
    \coolright{n_B}\\
    \coolright{n_U}\\
    \coolright{n_H}
\end{matrix*}.\label{eq_psi_0_infty_Fermi}\\
\nonumber
\end{align}

In the background gauge, $\chi^{(1)}=\Psi_{0,\rm alt}^{(c\bar c)}$, and
\begin{align}
 \Psi_{0,\rm BG,alt}^{(S\varphi)(1)}(r\to\infty)\simeq
\begin{pmatrix}
 \eqmakebox[tagPhi1infL0B]{$\sqrt{\xi}W_B$}&\eqmakebox[tagPhi1infL0B2]{$0$}\\
 \eqmakebox[tagPhi1infL0B]{$0$}&\eqmakebox[tagPhi1infL0B2]{$I_U\partial_r$}\\
 \eqmakebox[tagPhi1infL0B]{$W_B$}&\eqmakebox[tagPhi1infL0B2]{$0$}\\
 \eqmakebox[tagPhi1infL0B]{$0$}&\eqmakebox[tagPhi1infL0B2]{$W_U$}\\
 \coolunder{tagPhi1infL0B}{n_B}{0}&\coolunder{tagPhi1infL0B2}{n_U}{0}
\end{pmatrix}
\begin{matrix*}[l]
    \coolright{n_B}\\
    \coolright{n_U}\\
    \coolright{n_B}\\
    \coolright{n_U}\\
    \coolright{n_H}
\end{matrix*}\hspace{2ex}\Psi_{0,\rm alt}^{(c\bar c)}.\label{eq_psi_0_infty_BG}\\
\nonumber
\end{align}

 \item Type 2

There are some useful relations:
\begin{align}
 \frac{1}{r^3}\partial_rr^3\Psi^{(\rm top)}_0&=-\xi Y,\label{eq_top_0}\\
 \Psi^{(\rm bot)}_0&=\lambda+M\Gamma,\label{eq_bot_0}
\end{align}
where $Y$ is given in Eq.~\eqref{eq_y} and $\Gamma$ is a function obeying
\begin{align}
 \partial_r\Gamma&=\Psi^{(\rm top)}_0+(M^TM)^{-1}[\partial_r\zeta+2(M')^T\lambda].
\end{align}
In the Fermi gauge,  $\zeta^{(2)}=I_G$.
From Eqs.~\eqref{eq_top_0} and \eqref{eq_bot_0}, we get
\begin{align}
 \Psi_{0,\rm alt}^{(S\varphi)(2)}(r\to\infty)&\simeq
\begin{pmatrix}
 \eqmakebox[tagPhi2infL0F]{$-\frac{\xi}{4}rI_G$}\\
 \eqmakebox[tagPhi2infL0F]{$-\frac{\xi}{8}r^2W$}\\
 \coolunder{tagPhi2infL0F}{n_G}{0}
\end{pmatrix}
\begin{matrix*}[l]
    \coolright{n_G}\\
    \coolright{n_G}\\
    \coolright{n_H}
\end{matrix*}.\\
\nonumber
\end{align}

In the background gauge,  $\zeta^{(2)}=\Psi_{0,\rm alt}^{(c\bar c)}$, and
\begin{align}
 \chi^{(2)}(r\to\infty)\simeq
\begin{pmatrix}
 \eqmakebox[tagchi2B]{$\mathcal O(W'_B)$}&0\\
 \coolunder{tagchi2B}{n_B}{0}&\coolunder{tagchi2B2}{n_U}{-W_U^{-2}}
\end{pmatrix}
\begin{matrix*}[l]
    \coolright{n_B}\\
    \coolright{n_U}
\end{matrix*}\hspace{2ex}\Psi_{0,\rm alt}^{(c\bar c)}.\\
\nonumber
\end{align}
The behavior of $W_U$ is given by
\begin{equation}
 W_U(r\to\infty)\propto \frac{e^{W'_U(W_U)^{-1}r}}{r^{3/2}}.
\end{equation}

From Eqs.~\eqref{eq_top_0} and \eqref{eq_bot_0}, we get
\begin{align}
\Psi_{\ell,\rm BG,alt}^{(S\varphi)(2)}(r\to\infty)&\simeq
\begin{pmatrix}
 I_G&0\\
 0&O_\varphi
\end{pmatrix}
\begin{pmatrix}
 \eqmakebox[tagPhi2infL0B]{$\mathcal O(W'_B)$}&\eqmakebox[tagPhi2infL0B2]{$0$}\\
 \eqmakebox[tagPhi2infL0B]{$0$}&\eqmakebox[tagPhi2infL0B2]{$-\frac{\xi}{4}rI_U$}\\
 \eqmakebox[tagPhi2infL0B]{$W_B(W_B^TW_B)^{-1}$}&\eqmakebox[tagPhi2infL0B2]{$0$}\\
 \eqmakebox[tagPhi2infL0B]{$0$}&\eqmakebox[tagPhi2infL0B2]{$-\frac{1}{2}\frac{\xi}{r^3}(W'_U)^{-1}\mathcal K_{{\rm alt}}$}\\
 \coolunder{tagPhi2infL0B}{n_B}{0}&\coolunder{tagPhi2infL0B2}{n_U}{0}
\end{pmatrix}
\begin{matrix*}[l]
    \coolright{n_B}\\
    \coolright{n_U}\\
    \coolright{n_B}\\
    \coolright{n_U}\\
    \coolright{n_H}
\end{matrix*}\hspace{2ex}\Psi_{0,\rm alt}^{(c\bar c)},\\
\nonumber
\end{align}
in the background gauge.  Here, $\mathcal K_{{\rm alt}}$ is the
$n_U\times n_U$ matrix defined by
\begin{align}
 \mathcal K_{{\rm alt}}
=\lim_{r\to\infty}\frac{r^3}{\xi}
\begin{pmatrix}
 0&I_U
\end{pmatrix}\left(\partial_r\Psi_{0,\rm alt}^{(c\bar c)}\right)\left(\Psi_{0,\rm alt}^{(c\bar c)}\right)^{-1}
\begin{pmatrix}
 0\\
 I_U
\end{pmatrix}.
\end{align}
Notice that its elements are finite and its determinant is non-vanishing.

\item Type 3

We can take $\zeta^{(3)}=0$ and $\lambda^{(3)}=\Psi^{(\lambda)}_{0,\rm alt}$.
From Eq.~\eqref{eq_chi}, $\chi^{(3)}(r)=0$ for $r\gtrsim r_*$.  Then,
from Eq.~\eqref{eq_decomp_0},
\begin{align}
 \Psi_{0,\rm alt}^{(S\varphi)(3)}(r\to\infty)\simeq
\begin{pmatrix}
 0\\
 0\\
 \coolunder{tagPhi3infL0}{n_H}{I_H}
\end{pmatrix}
\begin{matrix*}[l]
    \coolright{n_G}\\
    \coolright{n_G}\\
    \coolright{n_H}
\end{matrix*}\hspace{2ex}\Psi_{0,\rm alt}^{(\lambda)}.\\
\nonumber
\end{align}
\end{itemize}

\subsubsection{Gauge zero modes}
Here, we discuss the gauge zero modes.  Even with the alternative
fluctuation operator, there appear gauge zero modes, which are given by
\begin{equation}
 \psi_{\rm alt}^{(A)}=
\begin{pmatrix}
 0\\
 M_{\rm alt}
\end{pmatrix}\mathcal U^{(A)}.
\end{equation}

To subtract these zero modes, we need to solve
\begin{equation}
 \mathcal M_{0,\rm alt}^{(S\varphi)}\check\psi_{\rm alt}^{(A)}=-\psi_{\rm alt}^{(A)}.
\end{equation}
The solutions can be obtained by using Eqs.~\eqref{decomp_gauge_zero} --
\eqref{Delta0checklambda} with the replacement of $M\to M_{\rm alt}$
and $\Omega\to \Omega_{\rm alt}$.

We define
\begin{align}
\begin{pmatrix}
 Z_B^{(A)}\\
 Z_U^{(A)}
\end{pmatrix}
\begin{matrix*}[l]
    \coolright{n_B}\\
    \coolright{n_U}
\end{matrix*}=\int drr^3M_{\rm alt}^TM_{\rm alt}
\mathcal U^{(A)},
\end{align}
where $Z_B^{(A)}$ and $Z_U^{(A)}$ become constant at $r\to\infty$.
Then, from Eqs.~\eqref{eq_top_0} and \eqref{eq_bot_0}, we obtain
\begin{align}
 \check\Psi_{0,{\rm alt}}^{(S\varphi)(1U)}(r\to\infty)\simeq
\begin{pmatrix}
 \eqmakebox[tagPhi1infL0FZ2]{$\frac{\xi}{4r}Z_B^{(A)}$}\\
 \eqmakebox[tagPhi1infL0FZ2]{$\frac{\xi}{4r}Z_U^{(A)}$}\\
 \eqmakebox[tagPhi1infL0FZ2]{$\frac{\xi}{4} W_BZ_B^{(A)}\ln r$}\\
 \eqmakebox[tagPhi1infL0FZ2]{$-\frac{1}{2r^3}(W'_U)^{-1}Z_U^{(A)}$}\\
 \coolunder{tagPhi1infL0FZ2}{n_U}{0}
\end{pmatrix}
\begin{matrix*}[l]
    \coolright{n_B}\\
    \coolright{n_U}\\
    \coolright{n_B}\\
    \coolright{n_U}\\
    \coolright{n_H}
\end{matrix*},\\
\nonumber
\end{align}
with which we define
\begin{equation}
 \check\Psi_{0,\rm alt}^{(S\varphi)(1U)}=
\begin{pmatrix}
 \check\psi_{\rm alt}^{(1)} ~~\cdots~~ \check\psi_{\rm alt}^{(n_U)}
\end{pmatrix}.
\end{equation}

Let us move on to the background gauge.
The gauge zero modes are given by
\begin{equation}
 \psi_{\rm BG,alt}^{(A)}=
\begin{pmatrix}
 I_G\partial_r\\
 M_{\rm alt}
\end{pmatrix}\Psi_{0,\rm alt}^{(c\bar c)}\mathcal U_{\rm BG}^{(A)},
\end{equation}
where
\begin{equation}
 \mathcal U_{\rm BG}^{(A)}=\left[\mathcal T_0^{(c\bar c)}\right]^{-1}\mathcal U^{(A)}.
\end{equation}
We need to solve
\begin{align}
 \mathcal M_{0,{\rm BG,alt}}^{(S\varphi)}\check\psi_{0,{\rm BG,alt}}^{(A)}=-\psi_{0,{\rm BG,alt}}^{(A)}.
\end{align}
Notably, $\check\psi_{0,{\rm BG,alt}}^{(A)}$ can be expressed as
Eq.~\eqref{decomp_gauge_zero} with $\check\zeta=\check\lambda=0$, while
$\check{\chi}^{(A)}$ satisfies
\begin{align}
  \Delta_0 \check{\chi}^{(A)} =&\,
  \left. \Delta_0 \chi \right|_{
    \chi\to\check{\chi}^{(A)},
    \zeta\to0,
    \lambda\to0,M\to M_{\rm alt},\Omega\to\Omega_{\rm alt}
  }+\Psi_{0,\rm alt}^{(c\bar c)}\mathcal U_{\rm BG}^{(A)}.\label{eq_chi_zero}
\end{align}
The following relation also holds.
\begin{align}
 &\frac{1}{r^3}\partial_rr^3\check\Psi^{(\rm top)(A)}_0=\xi M^TM\check\chi^{(A)}+\Psi_0^{(c\bar c)}\mathcal U_{\rm BG}^{(A)}.\label{eq_top_zero}
\end{align}

From Eq.~\eqref{eq_chi_zero},
\begin{align}
 \check\chi^{(A)}(r\to\infty)=
\begin{pmatrix}
 \eqmakebox[tagchi1BZ]{$0$}&0\\
 \coolunder{tagchi1BZ}{n_B}{0}&\coolunder{tagchi1BZ2}{n_U}{\frac{r^2}{8}I_U}
\end{pmatrix}
\begin{matrix*}[l]
    \coolright{n_B}\\
    \coolright{n_U}
\end{matrix*}\hspace{2ex}\Psi_{0,\rm alt}^{(c\bar c)}\mathcal U_{\rm BG}^{(A)}.\\
\nonumber
\end{align}
From Eqs.~\eqref{eq_top_zero} and \eqref{eq_bot_0}, we obtain
\begin{align}
 \check\psi_{0,{\rm BG,alt}}^{(A)}(r\to\infty)\simeq
\begin{pmatrix}
 \eqmakebox[tagPhi1infL0BZ]{$0$}&\eqmakebox[tagPhi1infL0BZ2]{$0$}\\
 \eqmakebox[tagPhi1infL0BZ]{$0$}&\eqmakebox[tagPhi1infL0BZ2]{$\frac{r}{4}I_U$}\\
 \eqmakebox[tagPhi1infL0BZ]{$0$}&\eqmakebox[tagPhi1infL0BZ2]{$0$}\\
 \eqmakebox[tagPhi1infL0BZ]{$0$}&\eqmakebox[tagPhi1infL0BZ2]{$\frac{r^2}{8}W_U$}\\
 \coolunder{tagPhi1infL0BZ}{n_B}{0}&\coolunder{tagPhi1infL0BZ2}{n_U}{0}
\end{pmatrix}
\begin{matrix*}[l]
    \coolright{n_B}\\
    \coolright{n_U}\\
    \coolright{n_B}\\
    \coolright{n_U}\\
    \coolright{n_H}
\end{matrix*}\hspace{2ex}\Psi_{0,\rm alt}^{(c\bar c)}\mathcal U_{\rm BG}^{(A)},\\
\nonumber
\end{align}
with which we define
\begin{equation}
 \check\Psi_{0,\rm BG,alt}^{(S\varphi)(1U)}=\sum_{B}\check\psi_{0,{\rm BG,alt}}^{(B)}\mathcal U^{(B)T}\mathcal T_0^{(c\bar c)}.
\end{equation}

\subsubsection{Functional determinant}
Here, we calculate the determinant of the solutions obtained above.

In the Fermi gauge,
\begin{align}
 \det\Psi^{(S\varphi)}_{0,{\rm reg},{\rm alt}}(\nu;r\to\infty) = &\, \det
\begin{pmatrix}
 \Psi_{0,\rm alt}^{(S\varphi)(1B)}&\Psi_{0,\rm alt}^{(S\varphi)(1U)}+\nu\check\Psi_{0,\rm alt}^{(S\varphi)(1U)}&\Psi_{0,\rm alt}^{(S\varphi)(2)}&\Psi_{0,\rm alt}^{(S\varphi)(3)}
\end{pmatrix}
\nonumber \\ &\,
+\mathcal O(\nu^{n_U+1})\nonumber\\
\simeq &\, \nu^{n_U}\left(\frac{\xi}{4}r\right)^{n_B}\det W_B\left(-\frac{\xi}{8r^2}\right)^{n_U}\frac{\det Z_U}{\det W'_U}\det\Psi_{0,\rm alt}^{(\lambda)}+\mathcal O(\nu^{n_U+1}),
\end{align}
and
\begin{align}
 \det\widehat\Psi_{0,{\rm reg}}^{(S\varphi)}(\nu;r\to\infty)&= \det\Psi_0^{(B)}\det\Psi_0^{(U)}\det\Psi_0^{(\lambda)}\nonumber\\
 &\simeq\left(\frac{\xi}{4}r\right)^{n_B}\det \widehat W\left(\frac{\xi}{4}r\right)^{n_U}\det\widehat\Psi_0^{(\lambda)}+\mathcal O(\nu).
\end{align}

Thus,
\begin{align}
 \frac{1}{\nu^{n_U}}\frac{\det\Psi^{(S\varphi)}_{0,{\rm reg},{\rm alt}}(\nu;r\to\infty)}{\det\widehat\Psi_{0,{\rm reg}}^{(S\varphi)}(\nu;r\to\infty)}&=\left(-\frac{1}{2r_\infty^3}\right)^{n_U}\frac{\det Z_U}{\det W'_U}\frac{\det\Psi_{0,\rm alt}^{(\lambda)}}{\det\widehat\Psi_0^{(\lambda)}}+\mathcal O(\nu),
\end{align}
and
\begin{align}
 \lim_{\nu\to0}\frac{1}{\nu^{n_U}}\frac{\det\Psi^{(S\varphi)}_{0,{\rm reg}}(\nu;r\to\infty)}{\det\widehat\Psi_{0,{\rm reg}}^{(S\varphi)}(\nu;r\to\infty)}&=\left(-\frac{1}{2r_\infty^3}\right)^{n_U}\frac{\det\mathcal X_U}{\det W'_U}\frac{\det\Psi^{(\lambda)}_0}{\det\widehat\Psi^{(\lambda)}_0}.
\end{align}

In the background gauge,
\begin{align}
 \det\Psi^{(SL\varphi)}_{0,{\rm reg},{\rm BG},{\rm alt}}(\nu;r\to\infty)&=\det
\begin{pmatrix}
  \Psi_{0,\rm BG,alt}^{(S\varphi)(1)}+\nu\check\Psi_{0,\rm BG,alt}^{(S\varphi)(1)}&\Psi_{0,\rm BG,alt}^{(S\varphi)(2)}&\Psi_{0,\rm BG,alt}^{(S\varphi)(3)}
\end{pmatrix}+\mathcal O(\nu^{n_U+1})\nonumber\\
 &\simeq \nu^{n_U}\left(\sqrt{\xi}\right)^{n_B}\left(-\frac{\xi}{8r^2}\right)^{n_U}\frac{\det \mathcal K_{{\rm alt}}}{\det W'_U}\left[\det \Psi_{0,\rm alt}^{(c\bar c)}\right]^2\det\Psi_{\rm alt}^{(\lambda)}+\mathcal O(\nu^{n_U+1}),
\end{align}
and
\begin{align}
 \det\widehat\Psi_{0,{\rm reg},{\rm BG}}^{(SL\varphi)}(\nu;r\to\infty)&= \det\Psi_0^{(B)}\det\Psi_0^{(U)}\det\Psi_0^{(\lambda)}\nonumber\\
 &\simeq\left(\sqrt{\xi}\right)^{n_B}\left[\det\widehat \Psi_0^{(c\bar c)}\right]^2\left(\frac{\xi}{4}r\right)^{n_U}\det\widehat\Psi_0^{(\lambda)}+\mathcal O(\nu).
\end{align}
Thus,
\begin{align}
 \frac{1}{\nu^{n_U}}\frac{\det\Psi^{(S\varphi)}_{0,{\rm reg},{\rm BG},{\rm alt}}(\nu;r\to\infty)}{\det\widehat\Psi_{0,{\rm reg},{\rm BG}}^{(S\varphi)}(\nu;r\to\infty)}&=\left(-\frac{1}{2r_\infty^3}\right)^{n_U}\frac{\det \mathcal K_{{\rm alt}}}{\det W'_U}\frac{\det\Psi_{0,\rm alt}^{(\lambda)}}{\det\widehat\Psi_0^{(\lambda)}}\left(\frac{\det \Psi_{0,\rm alt}^{(c\bar c)}}{\det\widehat \Psi_0^{(c\bar c)}}\right)^2+\mathcal O(\nu),
\end{align}
and
\begin{align}
 \lim_{\nu\to0}\frac{1}{\nu^{n_U}}\frac{\det\Psi^{(S\varphi)}_{0,{\rm reg},{\rm BG}}(\nu;r\to\infty)}{\det\widehat\Psi_{0,{\rm reg},{\rm BG}}^{(S\varphi)}(\nu;r\to\infty)}&=\left(-\frac{1}{2r_\infty^3}\right)^{n_U}\frac{\det \mathcal K}{\det W'_U}\frac{\det\Psi^{(\lambda)}_0}{\det\widehat\Psi^{(\lambda)}_0}\left(\frac{\det \Psi_0^{(c\bar c)}}{\det\widehat \Psi_0^{(c\bar c)}}\right)^2.
\end{align}

From Eqs.~\eqref{psi1(l=0)}, \eqref{psi2(l=0)} and \eqref{psi3(l=0)},
\begin{align}
  \lim_{r\to0}\frac{\det\Psi_0^{(S\varphi)}(r)}{\det\widehat\Psi_0^{(S\varphi)}(r)}
  =\frac{\sqrt{\det M^T_0 M_0}}{\det \widehat W},
\end{align}
independently of the choice of the guage fixing.  Based on the above
results, Eq.~\eqref{gaugezeromode} is obtained.

\section{Use of Alternative Fluctuation Opeartors}
\label{apx_error}
\setcounter{equation}{0}

In this appendix, we justify the use of the alternative fluctuation
operators in the evaluation of the determinants at $r\to\infty$.

\subsection{General discussion}

\subsubsection{Setup}

We compare the determinants of the two functions, $\mathfrak F$ and
$\mathfrak F_{\rm alt}$, which satisfy
\begin{align}
 \left[-\partial_r^2-\Lambda(r)\partial_r-\delta\Lambda(r)\partial_r+\Xi(r)+\delta\Xi(r)\right]\mathfrak F(r)&=0,\label{eq_orig_deq}\\
 \left[-\partial_r^2-\Lambda(r)\partial_r+\Xi(r)\right]\mathfrak F_{\rm alt}(r)&=0,
\end{align}
where $\Lambda,\delta\Lambda,\Xi,\delta\Xi,\mathfrak F$ and $\mathfrak F_{\rm alt}$
are $n\times n$ matrices. The support of $\delta\Lambda(r)$ and
$\delta\Xi(r)$ is $r>r_*$ and we take
\begin{align}
 \mathfrak F(r)=\mathfrak F_{\rm alt}(r) ~~:~~ r<r_* .
\end{align}

We assume that there exist constants, $C_\mathfrak F$, $C_{\delta
  \Lambda}$ and $C_{\delta \Xi}$, such that
\begin{align}
 ||\mathfrak F_{\rm alt}'(r)\mathfrak F_{\rm alt}^{-1}(r)||<C_\mathfrak F,~~~
 ||\delta\Lambda(r)||<\frac{C_{\delta \Lambda}}{r^3},~~~
 ||\delta\Xi(r)||<\frac{C_{\delta \Xi}}{r^3},\label{eq_cond_abs}
\end{align}
where $||A||$ is the induced (spectral) norm of the real matrix $A$,
which is defined as
\begin{equation}
  ||A||=\max_{x\in {\mathbb R}^n, |x|\neq0}
  \frac{|Ax|}{|x|},\label{eq_induced_norm}
\end{equation}
with $|x|=\sqrt{\sum_ix_i^2}$.
In addition, we assume
\begin{equation}
  \mu\left(-\mathfrak F_{\rm alt}'(r)\mathfrak F_{\rm alt}^{-1}(r)\right)\leq0,~~~
  \mu\left(-\Lambda(r)\right)<-\frac{k}{r},\label{eq_cond_mu}
\end{equation}
where $k>2$ is a non-integer constant and $\mu(A)$ is the logarithmic norm
of matrix $A$, which is defined as
\begin{equation}
 \mu(A)=\lim_{h\to0^+}\frac{||I+hA||-1}{h}.
\end{equation}
Notice that $\mu(A)$ gives the largest eigenvalue of $\frac{1}{2}(A+A^T)$.

The goal of this subsection is to show that the following quantity has
an upper bound:
\begin{align*}
 \lim_{r\to\infty}\left|\ln\frac{\det \mathfrak F(r)}{\det \mathfrak F_{\rm alt}(r)}\right|.
\end{align*}

\subsubsection{Recursive formula}

Let us construct a formal solution of Eq.~\eqref{eq_orig_deq} using
$\mathfrak F_{\rm alt}$.  We express $\mathfrak F$ as
\begin{align}
  \mathfrak F(r)= \mathfrak F_{\rm alt}(r)\mathcal P\exp\left[\int_{r_*}^rds\mathfrak F_{\rm alt}^{-1}(s)\Theta(s)\mathfrak F_{\rm alt}(s)\right],
\end{align}
where $\Theta$ is a function satisfying
\begin{align}
  \Theta'+\left(\mathfrak F_{\rm alt}'\mathfrak F_{\rm alt}^{-1}+\Lambda\right)\Theta+\Theta \mathfrak F_{\rm alt}'\mathfrak F_{\rm alt}^{-1}=\delta\Xi-\delta\Lambda \mathfrak F_{\rm alt}'\mathfrak F_{\rm alt}^{-1}-\Theta^2-\delta\Lambda\Theta.
\end{align}
Here, the path-ordered exponential of matrix $A(s)$ is defined as\footnote
{One may also write the path-ordered exponential as
  \begin{align*}
    \mathcal P \exp \left[ \int_{r_{\rm i}}^{r_{\rm f}} ds A(s) \right] =
    \lim_{N\rightarrow\infty}
    e^{A(r_{\rm f})\delta s}
    e^{A(r_{\rm f}-\delta s)\delta s}
    e^{A(r_{\rm f}-2\delta s)\delta s}
    \cdots
    e^{A(r_{\rm f}-N\delta s)\delta s},
  \end{align*}
  with
  \begin{align*}
    \delta s=\frac{r_{\rm f}-r_{\rm i}}{N}.
  \end{align*}
}
\begin{align}
  \mathcal P \exp \left[ \int_{r_{\rm i}}^{r_{\rm f}} ds A(s) \right] =
  1 +
  \sum_{p=1}^\infty
  \int_{r_{\rm i}}^{r_{\rm f}} ds_p
  \int_{r_{\rm i}}^{s_p} ds_{p-1} \cdots
  \int_{r_{\rm i}}^{s_2} ds_1
  A(s_p) A(s_{p-1}) \cdots A(s_1).
\end{align}

Treating $\delta\Lambda$ and $\delta\Xi$ as perturbations,
we expand $\Theta$ as
\begin{align}
 \Theta=\Theta^{(1)}+\Theta^{(2)}+\cdots,
\end{align}
where $\Theta^{(p)}$ is the $p$-th order term with respect to
fluctuations.  Formally, we obtain
\begin{align}
 \Theta^{(p)}(r)=\int_{r_*}^r ds\mathcal P\exp\left[\int^r_sdt\left(-\mathfrak F_{\rm alt}'(t)\mathfrak F_{\rm alt}^{-1}(t)-\Lambda(t)\right)\right]\Delta^{(p)}(s)\mathcal P\exp\left[\int^s_rdt\mathfrak F_{\rm alt}'(t)\mathfrak F_{\rm alt}^{-1}(t)\right],\label{eq_formal_sol}
\end{align}
where
\begin{align}
  \Delta^{(p)}=
\begin{cases}
 \delta\Xi-\delta\Lambda \mathfrak F_{\rm alt}'\mathfrak F_{\rm alt}^{-1} & :\ p=1\\
 -\delta\Lambda\Theta^{(p-1)}-\sum_{i=1}^{p-1}\Theta^{(i)}\Theta^{(p-i)} & :\ p>1
\end{cases}.
\end{align}
This recursive formula can be solved order by order and we can obtain
$\Theta$ if the sum over $p$ converges.

\subsubsection{Error evaluation formula}

Let us first evaluate the right path-ordered exponential in
Eq.~\eqref{eq_formal_sol}.  Using Eq.~\eqref{eq_induced_norm} as well as
\begin{equation}
 \mathcal P\exp\left[\int^s_rdt\mathfrak F_{\rm alt}'(t)\mathfrak F_{\rm alt}^{-1}(t)\right]y_R(r)=y_R(s),
\end{equation}
the following relation holds:
\begin{align}
 \left|\left|\mathcal P\exp\left[\int^s_rdt\mathfrak F_{\rm alt}'(t)\mathfrak F_{\rm alt}^{-1}(t)\right]\right|\right|&=\max_{|y_R(r)|\neq0}\frac{|y_R(s)|}{|y_R(r)|},
\end{align}
where $y_R(t)$ satisfies
\begin{equation}
  y_R' (t)=\mathfrak F'_{\rm alt}(t)\mathfrak F_{\rm alt}^{-1}(t)y_R(t).
\end{equation}
Then,
\begin{align}
 D^-|y_R(t)|&=\lim_{h\to0^+}\frac{|y_R(t)|-|y_R(t-h)|}{h}\nonumber\\
 &=\lim_{h\to0^+}\frac{|y_R(t)|-|y_R(t)-h\mathfrak F'_{\rm alt}(t)\mathfrak F_{\rm alt}^{-1}(t)y_R(t)|}{h}\nonumber\\
 &\geq-\mu\left(-\mathfrak F'_{\rm alt}(t)\mathfrak F_{\rm alt}^{-1}(t)\right)|y_R(t)|,
\end{align}
where $D^-$ is the left-hand derivative. After integration, we obtain
\begin{align}
 \ln\frac{|y_R(r)|}{|y_R(s)|}&\geq-\int_s^rdt\mu\left(-\mathfrak F'_{\rm alt}(t)\mathfrak F_{\rm alt}^{-1}(t)\right)\geq0,
\end{align}
resulting in
\begin{align}
 \left|\left|\mathcal P\exp\left[\int^s_rdt\mathfrak F_{\rm alt}'(t)\mathfrak F_{\rm alt}^{-1}(t)\right]\right|\right|&\leq1.\label{eq_res_R}
\end{align}

Similarly, we evaluate the left path-ordered exponential in
Eq.~\eqref{eq_formal_sol}.  From Eq.~\eqref{eq_induced_norm},
\begin{align}
 \left|\left|\mathcal P\exp\left[\int^r_sdt\left(-\mathfrak F_{\rm alt}'(t)\mathfrak F_{\rm alt}^{-1}(t)-\Lambda(t)\right)\right]\right|\right|&=\max_{|y_L(s)|\neq0}\frac{|y_L(r)|}{|y_L(s)|},
\end{align}
with
\begin{equation}
  y_L' (t)=\left(-\mathfrak F_{\rm alt}'(t)\mathfrak F_{\rm alt}^{-1}(t)-\Lambda(t)\right)y_L(t).
\end{equation}
Then, we obtain
\begin{align}
  D^+|y_L(t)|&=\lim_{h\to0^+}\frac{|y_L(t+h)|-|y_L(t)|}{h}\nonumber\\
  &=\lim_{h\to0^+}\frac{|y_L(t)+h\left(-\mathfrak F_{\rm alt}'(t)\mathfrak F_{\rm alt}^{-1}(t)-\Lambda(t)\right)y_L(t)|-|y_L(t)|}{h}\nonumber\\
  &\leq\mu\left(-\mathfrak F_{\rm alt}'(t)\mathfrak F_{\rm alt}^{-1}(t)-\Lambda(t)\right)|y_L(t)|,
\end{align}
where $D^+$ is the right-hand derivative, based on which we find
\begin{align}
  \ln\frac{|y_L(r)|}{|y_L(s)|} & \leq
  \int_r^sdt\mu\left(-\mathfrak F_{\rm alt}'(t)\mathfrak F_{\rm alt}^{-1}(t)\right)+\int_r^sdt\mu\left(-\Lambda(t)\right)
  <
  -k\ln\left(\frac{r}{s}\right).
\end{align}
Thus, the following inequality holds:
\begin{align}
 \left|\left|\mathcal P\exp\left[\int^r_sdt\left(-\mathfrak F_{\rm alt}'(t)\mathfrak F_{\rm alt}^{-1}(t)-\Lambda(t)\right)\right]\right|\right|&<\left(\frac{s}{r}\right)^k.\label{eq_res_L}
\end{align}

Using Eqs.~\eqref{eq_res_R} and \eqref{eq_res_L}, we obtain
\begin{align}
 ||\Theta^{(p)}(r)||<\int_{r_*}^r ds\left(\frac{s}{r}\right)^k||\Delta^{(p)}(s)||.
\end{align}
Now, we show that $||\Theta^{(p)}||$ is bounded from above as
\begin{align}
  ||\Theta^{(p)}(r)|| < C^{(p)}\frac{r_*}{r^2},
  \label{ThetaBound}
\end{align}
where $C^{(p)}$'s are positive constants.  Indeed, it is the case for
$p=1$ as
\begin{align}
  ||\Theta^{(1)}(r)||&<\int_{r_*}^r ds\left(\frac{s}{r}\right)^k||\Delta^{(1)}(s)||\nonumber\\
  &<(C_{\delta\Xi}+C_{\delta\Lambda}C_\mathfrak F)\int_{r_*}^r ds\left(\frac{s}{r}\right)^k\frac{1}{s^3}\nonumber\\
  &=\frac{C_{\delta\Xi}+C_{\delta\Lambda}C_\mathfrak F}{k-2}\frac{1}{r^2}\left(1-\frac{r_*^{k-2}}{r^{k-2}}\right)\nonumber\\
  &<C^{(1)}\frac{r_*}{r^2},
\end{align}
where
\begin{equation}
  C^{(1)}=\frac{C_{\delta\Xi}+C_{\delta\Lambda}C_\mathfrak F}{(k-2)r_*}.
\end{equation}
Next, let us assume $||\Theta^{(q)}(r)||<C^{(q)}\frac{r_*}{r^2}$ for
all $q<p$, and show that the inequality \eqref{ThetaBound} holds.  Indeed,
\begin{align}
 ||\Theta^{(p)}(r)||&<\int_{r_*}^r ds\left(\frac{s}{r}\right)^k||\Delta^{(p)}(s)||\nonumber\\
 &<\int_{r_*}^r ds\left(\frac{s}{r}\right)^k\left(r_*\frac{C_{\delta\Lambda}C^{(p-1)}}{s^{5}}+\sum_{i=1}^{p-1}r_*^2\frac{C^{(i)}C^{(p-i)}}{s^{4}}\right)\nonumber\\
 &=\frac{C_{\delta\Lambda}C^{(p-1)}}{k-4}r_*\left(\frac{1}{r^{4}}-\frac{r_*^{k-4}}{r^k}\right)+\sum_{i=1}^{p-1}\frac{C^{(i)}C^{(p-i)}}{k-3}r_*^2\left(\frac{1}{r^{3}}-\frac{r_*^{k-3}}{r^k}\right)\nonumber\\
 &<C^{(p)}\frac{r_*}{r^{2}},
\end{align}
where
\begin{align}
 C^{(p)}=\left(\frac{C_{\delta\Lambda}}{r_*^2}\frac{C^{(p-1)}}{|k-4|}+\sum_{i=1}^{p-1}\frac{C^{(i)}C^{(p-i)}}{|k-3|}\right).
\end{align}
Thus, the inequality \eqref{ThetaBound} is valid for all $p$.

\begin{figure}
 \begin{center}
  \includegraphics[width=0.5\linewidth]{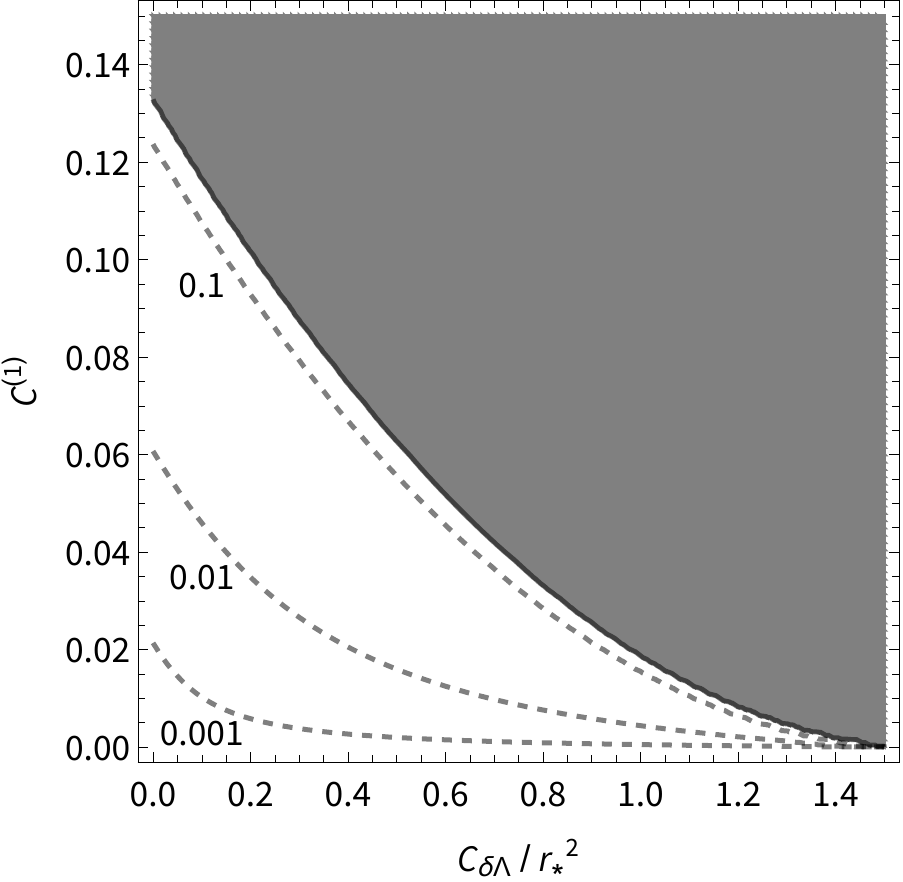}
 \end{center}
 \caption{Contours of constant $C^{\rm tot}$, taking $k=2.5$. The
   dashed lines indicate $C^{\rm tot}=0.1$, $0.01$, and $0.001$.
   In the gray region, $C^{\rm tot}$ diverges.}
 \label{fig_conv}
\end{figure}

Finally, we evaluate an upper bound on $||\Theta(r)||$.  Let us define
\begin{align}
 C^{\rm tot}=\sum_{p=1}^{\infty}C^{(p)}.
\end{align}
Then, if it converges, we get
\begin{align}
 ||\Theta(r)||<C^{\rm tot}\frac{r_*}{r^2}.
\end{align}
In Fig.~\ref{fig_conv}, we show the contours of constant $C^{\rm tot}$, taking
$k=2.5$.  As we can see from the figure, $C^{\rm tot}$ is actually
convergent for  small enough $C^{(1)}$ and $C_{\delta\Lambda}/r_*^2$,
and $C^{{\rm tot}}$ approaches zero as $C^{(1)}$ and
$C_{\delta\Lambda}/r_*^2$ go to zero.

Thus, if $C^{\rm tot}$ converges, we obtain
\begin{align}
  \lim_{r\to\infty}\left|\ln\frac{\det \mathfrak F(r)}{\det \mathfrak F_{\rm alt}(r)}\right|
  =\left|\int_{r_*}^\infty ds\tr \Theta(s)\right|
  \leq n\int_{r_*}^\infty ds||\Theta(s)||
  =nC^{\rm tot}.
\end{align}

\subsection{Alternative fluctuation operators}

Let us consider a one-parameter family of differential equations with
different transition point, $r_*$, and we denote the solution as
$\mathfrak F_{\rm alt}(r_*;r)$. Then,
$C_{\delta\Xi},C_{\delta\Lambda}$ and $C_\mathfrak F$ are dependent on
$r_*$. From the above discussion, sufficient conditions for
\begin{equation}
 \lim_{r_*\to\infty}\lim_{r\to\infty}\left|\ln\frac{\det\mathfrak F(r)}{\det\mathfrak F_{\rm alt}(r_*;r)}\right|=1,
\end{equation}
are given by
\begin{align}
 \lim_{r_*\to\infty}\frac{C_{\delta\Xi}(r_*)+C_{\delta\Lambda}(r_*)C_\mathfrak F(r_*)}{r_*}&=0,\label{eq_cond_C_1}\\
 \lim_{r_*\to\infty}\frac{C_{\delta\Lambda}(r_*)}{r_*^2}&=0.\label{eq_cond_C_2}
\end{align}
In this subsection, we give several general remarks.  Then, in the
following subsections, for $(SL\varphi)$ modes, $(c\bar c)$ modes and
$(\eta\lambda)$ modes, we construct $\mathfrak F_{\rm alt}(r_*;r)$
explicitly and show that the above conditions as well as
Eq.~\eqref{eq_cond_mu} are satisfied.

\subsubsection{Extended fluctuation operators}

Since the conditions \eqref{eq_cond_abs} are sensitive to the
behavior of the fluctuation operators around $r\sim r_*$, we need to
specify their deformation explicitly.

In particular, the differential equations for $\Psi_{\rm
alt}^{(\eta\lambda)}$ should be well-defined for all $r$, which requires
that the decomposition of solutions should be exact for all $r$. The
decomposition with the alternative fluctuation operators becomes
exact when the following relations are
satisfied;
\begin{align}
 &(M_{\rm alt}^T)'M_{\rm alt}=M_{\rm alt}^TM'_{\rm alt},\label{eq_deriv_ext}\\
 &\Omega_{\rm alt}^T=\Omega_{\rm alt},\label{eq_Omega_ext}\\
 &P_{H,{\rm alt}}(-\Delta_0M_{\rm alt}+\Omega_{\rm alt}M_{\rm alt})=0.\label{eq_M_eom_ext}
\end{align}
However, it is generally difficult to deform the fluctuation operators
keeping these relations. In the following, we show a way to overcome it
by extending the fluctuation operators.

We introduce $n_G$ spectator scalars and extend $M$ and $\Omega$ as
\begin{align}
 M^{\rm ext}&=
\begin{pmatrix}
 \eqmakebox[tagMext]{$M$}\\
 \coolunder{tagMext}{n_G}{0}
\end{pmatrix}
\begin{matrix*}[l]
    \coolright{n_G+n_H}\\
    \coolright{n_G}
\end{matrix*},\\%
\nonumber\\
 \Omega^{\rm ext}&=
\begin{pmatrix}
 \eqmakebox[tagOmegaext1]{$\Omega$}&\eqmakebox[tagOmegaext2]{$0$}\\
 \coolunder{tagOmegaext1}{n_G+n_H}{0}&\coolunder{tagOmegaext2}{n_G}{\Omega_{\rm sp}}
\end{pmatrix}
\begin{matrix*}[l]
    \coolright{n_G+n_H}\\
    \coolright{n_G}
\end{matrix*},\\%
\nonumber
\end{align}
where $\Omega_{\rm sp}$ is an $n_G\times n_G$ matrix, which will be
determined later. We take the same $\Omega_{\rm sp}$ for the false
vacuum. Thus, the spectator scalars do not affect the ratio of the
determinants. We also introduce $P_{G,{\rm alt}}^{\rm ext}$ and $P_{H,{\rm alt}}^{\rm ext}$,
which are defined as
\begin{align}
 P_{G,{\rm alt}}^{\rm ext}&=M^{\rm ext}_{\rm alt}(M^{{\rm ext}T}_{\rm alt}M^{\rm ext}_{\rm alt})^2M^{{\rm ext}T}_{\rm alt},\\
 P_{H,{\rm alt}}^{\rm ext}&=I_\varphi^{\rm ext}-P_{G,{\rm alt}}^{\rm ext},
\end{align}
where $I_\varphi^{\rm ext}$ is the
$(n_\varphi+n_G)\times(n_\varphi+n_G)$ identity matrix.

We assume that the deformation begins at $r=r_*=r_1$ and ends at
$r=r_5$ with $r_5-r_1\ll r_*$. We construct the alternative matrices,
$M^{\rm ext}_{\rm alt}$ and $\Omega^{\rm ext}_{\rm alt}$, which
satisfy $M^{\rm ext}_{\rm alt}(r)=M^{\rm ext}(r)$ and $\Omega^{\rm
  ext}_{\rm alt}(r)=\Omega^{\rm ext}(r)$ for $r<r_1$. For $r_1<r<r_5$,
we deform them in the following way.
\begin{enumerate}
 \item We express $M_{\rm alt}^{\rm ext}$ as
\begin{equation}
 M_{\rm alt}^{\rm ext}(r)=
\begin{pmatrix}
 M_{\rm alt}(r)\\
 \omega(r)
\end{pmatrix},
\end{equation}
where $\omega$ expresses the elements for the spectator scalars and
$\omega(r_1)=0$.
Keeping the following relations:
\begin{align}
 \Omega_{\rm alt}^{\rm ext}&=\Omega^{\rm ext}-P_{H,\rm alt}^{\rm ext}\left[-\Delta_0M_{\rm alt}^{\rm ext}+\Omega^{\rm ext} M_{\rm alt}^{\rm ext}\right](M_{\rm alt}^{{\rm ext}T}M_{\rm alt}^{\rm ext})^{-1}M_{\rm alt}^{{\rm ext}T}\nonumber\\
 &\hspace{3ex}-M_{\rm alt}^{\rm ext}(M_{\rm alt}^{{\rm ext}T}M_{\rm alt}^{\rm ext})^{-1}\left[-\Delta_0M_{\rm alt}^{{\rm ext}T}+M_{\rm alt}^{{\rm ext}T}\Omega^{\rm ext} \right]P_{H,\rm alt}^{\rm ext},\label{eq_deform_Omega}\\
 \Omega_{\rm sp}&=(\Delta_0\omega)\omega^{-1},
\end{align}
we deform $M_{\rm alt}^{\rm ext}$ in the following steps.
\begin{enumerate}
 \item $r_1<r<r_2$: Turn on $\omega$ keeping $\omega\propto
       \omega(r_2)$.
 \item $r_2<r<r_3$: Rotate $M_{\rm alt}$ so that $M_{\rm
       alt}^T(r_3)=(W,0)$, where $W$ is a diagonal matrix. During the
       rotation, we change $\omega$ so that
\begin{align}
 \omega^T\omega'-\omega'^T\omega=M'^T_{\rm alt}M_{\rm alt}-M_{\rm alt}^TM'_{\rm alt}.
\end{align}
 \item $r_3<r<r_4$: Turn off $\omega$ keeping
       $\omega\propto\omega(r_3)$.
\end{enumerate}
 \item $r_4<r<r_5$: Diagonalize $P_{H,{\rm alt}}^{\rm ext}\Omega_{\rm
       alt}^{\rm ext}P_{H,{\rm alt}}^{\rm ext}$.
\end{enumerate}
Here, we take $r_{p+1}-r_p$ independently of $r_*$.  One can easily
check that Eqs.~\eqref{eq_deriv_ext}, \eqref{eq_Omega_ext} and
\eqref{eq_M_eom_ext} hold at each step.

\subsubsection{Linear approximation}

To evaluate the effect of the deformation on the functional
determinants, we construct each step of the deformation more
concretely. Since the difference between the original matrices and the
alternative matrices are expected to be very small, we can work in the
linear approximation. As in the previous sections, we choose the basis
of the fields so that
\begin{align}
 \lim_{r\to\infty}M(r) W^{-1}(r)=
\begin{pmatrix}
 I_G\\
 0
\end{pmatrix},\label{eq_M_basis}\\
 \lim_{r\to\infty}\Omega(r)=
\begin{pmatrix}
 0&0\\
 0&\widehat m^2
\end{pmatrix}.
\end{align}
Notice that the first condition is somewhat stronger than
Eq.~\eqref{eq_lim_O} since some of the elements of $W^{-1}(r)$ diverge
as $r\to\infty$. The existence of such a field basis is guaranteed by
the one-to-one correspondence between the massive gauge bosons and the
NG bosons.

Since $O_\varphi$ and $O_G$ are very close to the identity matrices for
a large enough $r_*$, we approximate them as
\begin{align}
 O_\varphi(r_*)&\simeq I_{\varphi}+\delta_\varphi,\\
 O_G(r_*)&\simeq I_G+\delta_G,
\end{align}
where $\delta_\varphi$ and $\delta_G$ are constant anti-symmetric
matrices and satisfy
\begin{equation}
 ||\delta_\varphi||_{\max}<\delta_{\max}^2,~||\delta_G||_{\max}<\delta_{\max}^2,\label{eq_delta_max}
\end{equation}
with $||\cdots ||_{\max}$ being the max norm and $\delta_{\max}\ll1$.

We also approximate that $M(r)$ and
$\Omega(r)$ are almost constant during the
deformation and express them as
\begin{align}
 M(r)&\simeq M_*=
(I_{\varphi}+\delta_\varphi)
\begin{pmatrix}
 W_*\\
 0
\end{pmatrix}(I_G-\delta_G),\\
 \Omega(r)&\simeq
(I_{\varphi}+\delta_\varphi)
\begin{pmatrix}
 0&0\\
 0&m^2(r_*)
\end{pmatrix}(I_{\varphi}-\delta_\varphi),
\end{align}
where $W_*=W(r_*)$. From Eq.\eqref{eq_M_basis}, $\delta_G$ should
satisfy
\begin{equation}
 ||W_*\delta_G W_*^{-1}||_{\max}\ll 1,
\end{equation}
for a large enough $r_*$.

We define sigmoid functions that have the following features:
\begin{align}
 &\varsigma_{p}(r)\simeq
\begin{cases}
 0&r\lesssim r_p\\
 1&r\gtrsim r_{p+1}
\end{cases},\\
 &\max_{r}|\partial_r^n\varsigma_p(r)|\sim (r_{p+1}-r_p)^{-n}.
\end{align}
Then, we construct the deformation as follows.
\begin{itemize}
 \item $r_1<r<r_2$:
 We turn on $\omega$ as
 \begin{align}
  \omega(r)&=\varsigma_1(r)\delta_{\max}^{1/2}W_*.
 \end{align}
 In this step, $M_{\rm alt}$ and $\Omega_{\rm alt}^{\rm ext}$ are
 constant.

 \item $r_2<r<r_3$:
We rotate $M_{\rm alt}$ and $\omega$ as
\begin{align}
  M_{\rm alt}(r)&=(I_\varphi-\varsigma_2(r)\delta_\varphi)M_*(I_G+\varsigma_2(r)\delta_G),\\
 \omega(r)&=\sqrt{\delta_{\max}}(I_\varphi+\varepsilon_\varphi(r))W_*(1-\varepsilon_G(r)),
\end{align}
where
\begin{align}
 \varepsilon_\varphi(r)&=\frac{\varsigma_2(r)}{\delta_{\max}}
\begin{pmatrix}
 I_G&0
\end{pmatrix}\delta_\varphi
\begin{pmatrix}
 I_G\\
 0
\end{pmatrix},\\
 \varepsilon_G(r)&=\frac{\varsigma_2(r)}{\delta_{\max}}\delta_G.
\end{align}
Notice that the elements of $\varepsilon_\varphi$ and $\varepsilon_G$
       are much smaller than one due to Eq.~\eqref{eq_delta_max}. In
       this step, $\Omega_{\rm alt}^{\rm ext}$ is changed according to
       Eq.~\eqref{eq_deform_Omega}.

 \item $r_3<r<r_4$: We turn off $\omega$ as
\begin{align}
 \omega(r)=(1-\varsigma_3(r))\omega(r_3).
\end{align}
In this step, $M_{\rm alt}$ and $\Omega_{\rm alt}^{\rm ext}$ are
 constant.

 \item $r_4<r<r_5$: Finally, we rotate $P_{H,\rm alt}^{\rm
       ext}\Omega_{\rm alt}^{\rm ext}P_{H,\rm alt}^{\rm ext}$ as
\begin{align}
 P_{H,\rm alt}^{\rm ext}\Omega_{\rm alt}^{\rm ext}P_{H,\rm alt}^{\rm ext}&=P_{H,\rm alt}^{\rm ext}
\begin{pmatrix}
 I_\varphi-\varsigma_4(r)\delta_\varphi&0\\
 0&I_G
\end{pmatrix}\Omega_{\rm alt}^{\rm ext}
\begin{pmatrix}
 I_\varphi+\varsigma_4(r)\delta_\varphi&0\\
 0&I_G
\end{pmatrix}P_{H,\rm alt}^{\rm ext}.
\end{align}
Notice that we can still use $\delta_\varphi$ to diagonalize $\Omega$ since
\begin{equation}
 P_{H,\rm alt}^{\rm ext}(r_4)\Omega_{\rm alt}^{\rm ext}(r_4)P_{H,\rm alt}^{\rm ext}(r_4)=P_{H,\rm alt}^{\rm ext}(r_4)\Omega^{\rm ext}(r_4)P_{H,\rm alt}^{\rm ext}(r_4).
\end{equation}
In this step, $M_{\rm alt}^{\rm ext}$ is constant.
\end{itemize}

Finally, let us discuss the deformation of $V_H^{\rm ext}$, which is
defined as
\begin{equation}
 V_H^{\rm ext}=
 \begin{pmatrix}
  V_H&0\\
  0&I_G
 \end{pmatrix}\equiv
\begin{pmatrix}
 u_1^{\rm ext}&\cdots&u_{n_H+n_G}^{\rm ext}
\end{pmatrix}.
\end{equation}
Its alternative can be defined in the following way.  Since the
difference between $M^{\rm ext}$ and $M_{\rm alt}^{\rm ext}$ is small, we can
construct $(n_H+n_G)$ independent vectors as
\begin{equation}
 \tilde u^{\rm ext}_{p,\rm alt}=\frac{P^{\rm ext}_{H,\rm alt}u_p^{\rm ext}}{|P^{\rm ext}_{H,\rm alt}u_p^{\rm ext}|}.
\end{equation}
Using Gram-Schmidt orthogonalization, we get an orthonormal basis, $u^{\rm ext}_{p,\rm alt}$, and define
\begin{equation}
 V^{\rm ext}_{H,\rm alt}=
\begin{pmatrix}
 u_{1,\rm alt}^{\rm ext}&\cdots&u_{n_H+n_G,\rm alt}^{\rm ext}
\end{pmatrix}.
\end{equation}

\subsection{$(SL\varphi)$ modes}

In the rest of this appendix, we confirm Eqs.~\eqref{eq_cond_C_1} and
\eqref{eq_cond_C_2} for each mode.  We start with the $(SL\varphi)$
modes.

We consider the functions
$\mathfrak F^{(SL\varphi)}$ and $\mathfrak F^{(SL\varphi)}_{\rm alt}$
which obey
\begin{align}
 &\begin{pmatrix}
  \sqrt{\xi}I_G&0&0\\
  0&I_G &0\\
  0&0 &I_\varphi^{\rm ext}
 \end{pmatrix}
\mathcal M_{\ell}^{(SL\varphi){\rm ext}}
 \begin{pmatrix}
  \sqrt{\xi}I_G&0&0\\
  0&I_G &0\\
  0&0 &I_\varphi^{\rm ext}
 \end{pmatrix}\mathfrak F^{(SL\varphi)}=0,\\
 &\begin{pmatrix}
  \sqrt{\xi}I_G&0&0\\
  0&I_G &0\\
  0&0 &I_\varphi^{\rm ext}
 \end{pmatrix}
\mathcal M_{\ell,\rm alt}^{(SL\varphi){\rm ext}}
 \begin{pmatrix}
  \sqrt{\xi}I_G&0&0\\
  0&I_G &0\\
  0&0 &I_\varphi^{\rm ext}
 \end{pmatrix}\mathfrak F^{(SL\varphi)}_{\rm alt}=0,
\end{align}
where the superscript ``${\rm ext}$'' indicates that the $n_G$ spectator
scalars are included. Then,
\begin{equation}
 \lim_{r\to\infty}\left|\ln\frac{\det\Psi^{(SL\varphi)}_{\ell}(r)}{\det\Psi^{(SL\varphi)}_{\ell,{\rm alt}}(r)}\right|=\lim_{r\to\infty}\left|\ln\frac{\det\mathfrak F^{(SL\varphi)}(r)}{\det \mathfrak F^{(SL\varphi)}_{\rm alt}(r)}\right|.
\end{equation}
Here, we have ignored $\omega$, which is non-vanishing only around
$r\sim r_*$. One can show that its effect disappears as
$r_*\to\infty$.

Then, we can show the following.
\begin{itemize}
 \item All of the eigenvalues of $(\Lambda+\Lambda^T)/2$ are $3/r$ and
       hence $\mu(-\Lambda)=-3/r$.
 \item The fluctuation operator at $r\to\infty$ has the same number of
   increasing solutions and deceasing solutions. If there are no zero
   modes, $\mathfrak F^{(SL\varphi)}_{\rm alt}$ contains all of the
   increasing solutions and  $\mu(-{\mathfrak
     F^{(SL\varphi)}_{\rm alt}}'{\mathfrak F^{(SL\varphi)}_{\rm
       alt}}^{-1})<0$.  When there are zero modes, $\mathfrak
   F^{(SL\varphi)}_{\rm alt}$ fails to contain some of the increasing
   solutions and $\mu(-{\mathfrak
     F^{(SL\varphi)}_{\rm alt}}'{\mathfrak F^{(SL\varphi)}_{\rm
       alt}}^{-1})$ becomes positive. However, it is harmless since we
   use the fluctuation operator with regulator, with which all the
   solutions are increasing.
 \item Since the growth of $\mathfrak F^{(SL\varphi)}_{\rm alt}$ is
   either exponential or power at $r\rightarrow\infty$, $C_\mathfrak
   F$ is finite.
 \item Since we can take a smaller $\delta_{\max}$ for a larger $r_*$,
       it is easy to find $C_{\delta\Lambda}$ that
satisfies Eq.~\eqref{eq_cond_C_1}. Notice that the max norm and the
       spectral norm of an $n\times n$ matrix, $A$, are related through
\begin{equation}
 ||A||\leq n||A||_{\rm max}.
\end{equation}
 \item As for $\delta\Xi$, we need a careful
treatment since it includes $(W^TW)^{-1}$.  It appears in $\Omega^{\rm
       ext}-\Omega_{\rm alt}^{\rm ext}$ and
\begin{align}
 \delta\Omega^{\rm ext}&=-M_{\rm alt}^{\rm ext}(M_{\rm alt}^{{\rm ext},T}M_{\rm alt}^{\rm ext})^{-1}\left[-\Delta_0M_{\rm alt}^{{\rm ext}T}+M_{\rm alt}^{{\rm ext}T}\Omega^{\rm ext}_{\rm alt}\right].
\end{align}
From Eq.~\eqref{eq_deform_Omega}, it can be rewritten as
\begin{equation}
 \delta\Omega^{\rm ext}=-M_{\rm alt}^{\rm ext}(M_{\rm alt}^{{\rm ext},T}M_{\rm alt}^{\rm ext})^{-1}\left[-\Delta_0M_{\rm alt}^{{\rm ext}T}+M_{\rm alt}^{{\rm ext}T}\Omega^{\rm ext}\right]P_{G,{\rm alt}},
\end{equation}
for $r_1<r<r_4$.
It is non-vanishing only when $r_2<r<r_3$, where
\begin{align}
 \left[-\Delta_0M_{\rm alt}^{\rm ext}+\Omega^{\rm ext} M_{\rm alt}^{\rm ext}\right](M_{\rm alt}^{{\rm ext}T}M_{\rm alt}^{\rm ext})^{-1}M_{\rm alt}^{{\rm ext}T}\simeq
\begin{pmatrix}
 \varsigma''_2(r)\left[-\delta_\varphi+W\delta_GW^{-1}\right]\\
 0
\end{pmatrix},
\end{align}
and $\Omega^{\rm ext}-\Omega_{\rm alt}^{\rm ext}$ and $\delta\Omega^{\rm
       ext}$ are obtained by appropriate projections.
Since its elements decrease as $\delta_{\max}$ becomes smaller, we can
       find $C_{\delta\Xi}$ that
satisfies Eq.~\eqref{eq_cond_C_2}.
\end{itemize}

The same discussion applies to $\ell=0$.

\subsection{$(c\bar c)$ modes}

For the ghosts, we introduce
$\mathfrak F^{(c\bar c)}$ and $\mathfrak F^{(c\bar c)}_{\rm alt}$, obeying
\begin{align}
 \mathcal M_\ell^{(c\bar c){\rm ext}}\mathfrak F^{(c\bar c)}=0,\\
 \mathcal M_{\ell,\rm alt}^{(c\bar c){\rm ext}}\mathfrak F^{(c\bar c)}_{\rm alt}=0.
\end{align}
The discussion is almost the same as in the case of the $(SL\varphi)$
modes, but is much simpler because, for the ghosts, $\delta
\Lambda=0$ and
\begin{equation}
 \delta\Xi=\xi(M^{{\rm
ext}T}M^{{\rm ext}}-M^{{\rm ext}T}_{\rm alt}M^{{\rm ext}}_{\rm alt}).
\end{equation}

\subsection{$(\eta\lambda)$ modes}

Let us define
\begin{align}
  & \mathcal M^{(\eta\lambda)}_\ell =
  \nonumber \\ &~~~~
\begin{pmatrix}
 W^{-1}\left[-\Delta_\ell^{(M^TM)} + 2M^TM'(M^TM)^{-1} \frac{1}{r^2} \partial_r r^2 \right]W&- \frac{2L}{r} W^{-1}M'^T V_H\\
 \frac{2L}{r} V_H^TM' (M^TM)^{-1}W&V_H^T\left[-\Delta_\ell^{(\Omega)}
  - 4 M' (M^TM)^{-1} M'^T \right]V_H
\end{pmatrix},
\nonumber \\
\end{align}
where
$\Delta_\ell^{(M^TM)}\equiv\Delta_\ell- M^TM$ and
$\Delta_\ell^{(\Omega)}\equiv\Delta_\ell-\Omega$.
Then, we consider the solution of the following equation:
\begin{align}
 \mathcal M^{(\eta\lambda)}_\ell\tilde\Psi_\ell^{(\eta\lambda)}=0.
\end{align}
Here, the solution, $\tilde\Psi_\ell^{(\eta\lambda)}$, with
appropriate boundary conditions is related to
$\Psi_\ell^{(\eta\lambda)}$ as
\begin{align}
 \begin{pmatrix}
 W&0\\
 0&I_H
\end{pmatrix}\tilde\Psi_\ell^{(\eta\lambda)}=\Psi_\ell^{(\eta\lambda)}.
\end{align}

We also define its alternative version and its extension to include
the spectator scalars.  We introduce $\mathfrak F^{(\eta\lambda)}$ and
$\mathfrak F^{(\eta\lambda)}_{\rm alt}$ which obey
\begin{align}
 \mathcal M_\ell^{(\eta\lambda){\rm ext}}\mathfrak F^{(\eta\lambda)}=0,\\
 \mathcal M_{\ell,\rm alt}^{(\eta\lambda){\rm ext}}\mathfrak F_{\rm alt}^{(\eta\lambda)}=0.
\end{align}
Then,
\begin{align}
 \lim_{r\to\infty}\left|\ln\frac{\det\Psi^{(\eta\lambda)}_{\ell}(r)}{\det\Psi^{(\eta\lambda)}_{\ell,{\rm alt}}(r)}\right|=\lim_{r\to\infty}\left|\ln\frac{\det\mathfrak F^{(\eta\lambda)}(r)}{\det \mathfrak F^{(\eta\lambda)}_{\rm alt}(r)}\right|.
\end{align}

We can show the following.
\begin{itemize}
 \item Since $\Lambda$ is diagonal with the diagonal components $3/r$,
   $\mu(-\Lambda)=-3/r$.
 \item From Eq.~\eqref{eq_eta_inf} and the divergent behavior of
   $W_U^{-1}$, $\mathfrak F^{(\eta\lambda)}_{\rm alt}$ contains only
   the increasing solutions. Thus, $\mu(-\mathfrak
   {F^{(\eta\lambda)}}'_{\rm alt}{\mathfrak F_{\rm
     alt}^{(\eta\lambda)}}^{-1})<0$.
 \item Since $\mathfrak F^{(\eta\lambda)}_{\rm alt}$ grows
   exponentially, $C_\mathfrak F$ is finite.
 \item Just like the $(SL\varphi)$ modes, substituting the formulas of
       the linear approximation,
we can find $C_{\delta\Lambda}$ and $C_{\delta\Xi}$ that
satisfy Eqs.~\eqref{eq_cond_C_1} and \eqref{eq_cond_C_2}.
\end{itemize}

The same discussion applies to $\ell=0$.
\section{$\overline{\rm MS}$ Counterterms}\label{apx_msbar}
\setcounter{equation}{0}

In this appendix, we give the explicit form of the counterterms in the
$\overline{\rm MS}$ scheme.

We define mass squared matrices of gauge bosons and scalars in the
false vacuum as
\begin{align}
 \widehat m_A^2 \equiv &\, \widehat M^T\widehat M,\\
 \widehat m_\varphi^2 \equiv &\, \widehat\Omega+\widehat M\widehat M^T,
\end{align}
both of which are diagonal, and the Fourier transformation of mass squared difference
\begin{align}
 \widetilde{\delta m_A^2}(k)&=\int d^4x[M^T(r)M(r)-\widehat m_A^2]e^{-ikx},\\
 \widetilde{\delta m_\varphi^2}(k)&=\int d^4x[\Omega(r)+M(r)M^T(r)-m_\varphi^2]e^{-ikx},\\
 \widetilde{M}(k)&=\int d^4xM(r)e^{-ikx}.
\end{align}
Note that $M$ and $\Omega$ only depend on $r=\sqrt{x_\mu x_\mu}$ due to the spherical symmetry of the bounce.
Accordingly, all of the Fourier transformation given above are functions only of $k\equiv\sqrt{k_\mu k_\mu}$.
Furthermore, $\widetilde{\delta m_A^2}$ and $\widetilde{\delta m_\varphi^2}$ are symmetric matrices.

After a straightforward calculation, we obtain
\begin{align}
 s^{(A_\mu\varphi)}_{1,\overline{\rm MS}}&=2\mathcal I_1(\widehat m_A^2,\widetilde{\delta m_A^2}) + \frac{1}{2} \mathcal I_1(\widehat m_\varphi^2,\widetilde{\delta m_\varphi^2}) + \frac{1}{16\pi^2}\tr\left[\widehat m_A^2\widetilde{\delta m_A^2}(0)\right],\\
 s^{(c\bar c)}_{1,\overline{\rm MS}}&=-\mathcal I_1(\widehat m_A^2,\widetilde{\delta m_A^2}),\\
 s^{(A_\mu\varphi)}_{2,\overline{\rm MS}}&=-\mathcal I_2(\widehat m_A^2,\widehat m_A^2,\widetilde{\delta m_A^2})-2\mathcal I_2(\widehat m_A^2,\widehat m_\varphi^2,k\widetilde{M})-\frac{1}{4}\mathcal I_2(\widehat m_\varphi^2,\widehat m_\varphi^2,\widetilde{\delta m_\varphi^2})\nonumber\\
 &\hspace{3ex}+\frac{1}{256\pi^4}\int_0^\infty dk\,k^3\tr\left[\widetilde{\delta m_A^2}(k)\widetilde{\delta m_A^2}(k)\right],\\
 s^{(c\bar c)}_{2,\overline{\rm MS}}&=\frac{1}{2}\mathcal I_2(\widehat m_A^2,\widehat m_A^2,\widetilde{\delta m_A^2}),
\end{align}
where
\begin{align}
  \mathcal I_1(m^2,\widetilde{\delta m^2})&=-\sum_i[\widetilde{\delta m^2}(0)]_{ii}\frac{1}{16\pi^2} A_{0,\overline{\text{MS}}}([m^2]_{ii}),\\
  \mathcal I_2(m_1^2,m_2^2,\widetilde{\delta  m^2})&=\sum_{ij}\frac{1}{128\pi^4}\int_0^\infty dk\,k^3\left[\widetilde{\delta m^2}(k)\right]_{ij}\left[\widetilde{\delta m^2}(k)\right]_{ij} B_{0,\overline{\text{MS}}}(k, [m_1^2]_{ii}, [m_2^2]_{jj}).
\end{align}
Renormalizing the divergence in the $\overline{\text{MS}}$ scheme, we can evaluate the one-point function $A_{0,\overline{\text{MS}}}$ and the two-point function $B_{0,\overline{\text{MS}}}$ as \cite{Passarino:1978jh}
\begin{align}
  A_{0,\overline{\text{MS}}}(m^2) &= \lim_{\delta\to +0} m^2\left(1+\ln\frac{\mu^2}{m^2-i\delta}\right),\\
  B_{0,\overline{\text{MS}}}(k, m^2, M^2) &= \lim_{\delta\to +0}
  \left[ 2+x_1(\delta)\ln\frac{x_1(\delta)-1}{x_1(\delta)}+x_2(\delta)\ln\frac{x_2(\delta)-1}{x_2(\delta)}
  \right.\nonumber\\
  &\qquad\quad\left.-\ln(1-x_1(\delta))-\ln(x_2(\delta)-1)+\ln\left(\frac{\mu^2}{k^2}
  \right) \right],
\end{align}
where $\mu$ is the renormalization scale while $x_1(\delta)$ and
$x_2(\delta)$ are solutions of the following quadratic equation:
\begin{align}
  -k^2 x^2 + (k^2 - m^2 + M^2)x + m^2 - i\delta = 0.
\end{align}

\bibliographystyle{unsrt}
\bibliography{multifield}

\end{document}